\documentclass{aa}  

\usepackage{graphicx}
\usepackage{txfonts}
\usepackage{multirow}
\usepackage{booktabs}
\def\rearth{{\rm\,R_\oplus}}
\newcommand*\rot{\rotatebox{90}}

\usepackage{natbib}

\begin{document}

   \title{Formation of planetary systems by pebble accretion and migration}

  \subtitle{Hot super-Earth systems from breaking compact resonant chains}
   \titlerunning{Breaking the resonant chains}

\author{André Izidoro\inst{1},  Bertram Bitsch\inst{2}, Sean N. Raymond\inst{3}, Anders Johansen\inst{4},
\\ Alessandro Morbidelli\inst{5}, Michiel Lambrechts\inst{4}, \and Seth A. Jacobson\inst{6} }
\authorrunning{Izidoro et al.}
\institute{UNESP, Univ. Estadual Paulista - Grupo de Din{\^a}mica Orbital \& Planetologia, Guaratinguet{\'a}, CEP 12516-410 S{\~a}o Paulo, Brazil \\ \email{E-mail:izidoro.costa@gmail.com} \and
Max-Planck-Institut für Astronomie, Königstuhl 17, 69117 Heidelberg, Germany \and
Laboratoire d'astrophysique de Bordeaux, Univ. Bordeaux, CNRS, B18N, allée Geoffroy Saint-Hilaire, 33615 Pessac, France \and
Lund Observatory, Department of Astronomy and Theoretical Physics, Lund University, Box 43, 22100 Lund, Sweden \and 
Laboratoire Lagrange, UMR7293, Université Côte d’Azur, CNRS, Observatoire de la Côte d’Azur, Boulevard de l’Observatoire, \\ 06304 Nice Cedex 4, France \and
Department of Earth and Environmental Sciences, Michigan State University, East Lansing, MI, USA
} 

\date{Accepted XXX. Received YYY; in original form ZZZ}

 
  \abstract
{At least 30\% of main sequence stars host planets with sizes of between 1 and 4 Earth radii and orbital periods of less than 100 days. We use N-body simulations including a model for gas-assisted pebble accretion and disk--planet tidal interaction to study the formation of  super-Earth systems. We show that the integrated pebble mass reservoir creates a bifurcation between hot super-Earths or hot-Neptunes ($\lesssim15M_{\oplus}$) and super-massive planetary cores potentially able to become gas giant planets ($\gtrsim15M_{\oplus}$). Simulations with moderate pebble fluxes grow multiple super-Earth-mass planets that migrate inwards and pile up at the inner edge of the disk forming long resonant chains. We follow the long-term dynamical evolution of these systems and use the period ratio distribution of observed planet-pairs to constrain our model. Up to $\sim$95\% of resonant chains become  dynamically unstable after the gas disk dispersal, leading to a phase of late collisions that breaks the original resonant configurations. Our simulations naturally match observations  when they produce a dominant fraction ($\gtrsim95\%$) of unstable systems with a sprinkling ($\lesssim5\%$) of stable resonant chains (the Trappist-1 system represents one such example). Our results demonstrate that super-Earth systems are inherently multiple (${\rm N\geq2}$) and that the observed excess of single-planet transits is a consequence of the mutual inclinations excited by the planet--planet instability. In simulations in which planetary seeds are initially distributed in the inner and outer disk, close-in super-Earths are systematically ice rich. This contrasts with the interpretation that most super-Earths are rocky based on bulk-density measurements of super-Earths and photo-evaporation modeling of their bimodal radius distribution. We investigate the conditions needed to form rocky super-Earths. The formation of rocky super-Earths requires special circumstances, such as far more efficient planetesimal formation  well inside the snow line, or  much faster planetary growth by pebble accretion in the inner disk. Intriguingly, the necessary conditions to match the bulk of hot super-Earths are at odds with the conditions needed to match the Solar System.}


   \maketitle
%

\section{Introduction}

Exoplanet systems present a diversity of architectures compared with the structure of our home planetary system. Planets with sizes between those of Earth and Neptune, that is, with between 1 and 4 Earth radii,  have been found in compact multi-planet systems orbiting their host stars at orbital periods shorter than 100 days \citep{lissauertal11a,marcyetal14,fabryckyetal14}. These systems are typically referred to as hot super-Earth systems and their high abundance \citep[e.g.,][]{mayoretal11,batalhaetal13,
howardetal13,fressinetal13} is one of the greatest surprises in exoplanet science. Observations  and planet occurrence studies suggest that at least 30\% of the FGK-type stars in the galaxy host hot super-Earths with a period of less than 100 days \citep{mayoretal11,howardetal12,fressinetal13,petiguraetal13,zhuetal18,mulders18,muldersetal18}. Hot super-Earths are found by inference to have orbits with low orbital eccentricities and mutual inclinations \citep{mayoretal11,lissauertal11a,johansenetal12,fangmargot12,xieetal16,zhuetal18}.  

Our understanding of the origins of hot super-Earths remains incomplete. Many models have been proposed in the last decade or so, but these scenarios have been gradually refined by observational constraints and simulations and many have already  been discarded \cite[see discussions in][]{raymondetal08b,raymondetal14,raymondcossou14,schlichting14,morbyraymond16,ogiharaetal15a,chatterjeetan15,izidoroetal17}. We now briefly discuss three scenarios: (a) in-situ accretion ; (b) drift-then-assembly; and (c) migration. 

The in-situ scenario for the formation of hot super-Earths proposes that hot super-Earths formed more or less where they are seen today. This scenario requires (i) large amounts of mass in solids in the inner protoplanetary disk in order to form massive planets and (ii) the late formation of super-Earths in gas-poor or gas-free environments in order to avoid gas-driven migration \citep{raymondetal08b,chianglaughlin13,raymondcossou14,schlichting14,schlaufman14,dawsonetal15,dawsonetal16}. The required high density of solids in the protoplanetary disk is problematic if one imposes the  canonical dust/gas ratio, because it would imply that the disk is gravitationally unstable. Thus,  in-situ formation models often assume an enhanced dust/gas ratio~\citep[e.g.,][]{hansenmurray12,hansenmurray13,hansen14,dawsonetal15,dawsonetal16,leechiang16}, but without modeling how such an enhancement occurred or its properties (e.g., its radial extent). The late formation of super-Earths is sometimes justified by invoking that, as long as there is a significant amount of gas in the disk, gas dynamical friction damps the eccentricities of growing planets and prevents mutual orbit crossing. However, numerical simulations including all dynamical effects show that, whenever the density of planetesimals is large, planet growth is inexorably fast \citep{ogiharaetal15a}; thus a ``late formation'' seems implausible. An alternative possibility is that  planetesimal formation itself occurred late, but, given the assumed large dust/gas ratio, this also appears implausible because the streaming instability becomes effective as soon as the metallicity is larger than a few percent \citep{yangetal17}. If super-Earths form too early in the gas-rich inner parts of the  disk,  fast inward type-I migration  leads to very compact systems of hot super-Earths that fail to match observations~\citep{ogiharaetal15a}. Effects of disk winds have been  invoked  to avoid large-scale  fast inward type-I migration~\citep{ogiharaetal15b}. Inward type-I migration is only completely suppressed if the the corotation torque is assumed to be fully unsaturated, which may not be realistic for Earth-mass planets~\citep{ogiharaetal18}. This suggests that some level of inward gas-driven migration is difficult to avoid.


The drift-then-assembly model proposes that small particles such as pebbles or planetesimals drift inwards by gas drag and pile up at the pressure bump that may form at the transition between the magnetorotational instability-active inner regions and the exterior dead zone \citep{chatterjeetan14,boleyford13,boleyetal14,chatterjeetan15,huetal16}. This collection of particles forms a ring of solid material that eventually becomes gravitationally unstable and collapses to form a planet. The newly created planet induces another pressure bump outside its orbits and the process repeats. Although this idea is interesting, the model remains to be further developed. For instance, it has been shown that Hall shear instability may  dominate near the disk inner edge  when the magnetic field is aligned with  the spin vector of the disk. This instability may  prevent the local formation of close-in super-Earths via the processes envisioned  in the drift-then-assembly model \citep{mohantyetal18}.  \cite{uedaetal19} also showed that the accumulation of dust at the disk inner edge occurs only if the dust grains are large enough and the viscosity in the dead-zone is sufficiently low.





The migration model proposes that super-Earths or their constituent planetary embryos formed in gas-rich disks and  migrated inward from outside their current orbits by planet--disk gravitational interaction \citep{terquempapaloizou07,idalin08,idalin10,mcneilnelson10,hellarynelson12,cossouetal14,colemannelson14,colemannelson16,izidoroetal17,ogiharaetal18,raymondetal18,carreraetal18}. Simulations modeling planet--disk interaction predict that hot super-Earths  typically  migrate inwards  and pile up at the disk inner edge forming long chains of first-order mean motion resonances \citep{terquempapaloizou07,raymondetal08b,mcneilnelson10,rein12,reinetal12,hornetal12,ogiharakobayashi13,cossouetal14,raymondcossou14,
ogiharaetal15a,liuetal15,liuetal16,izidoroetal17,ormeletal17,unterbornetal18,ogiharaetal18}. During the gas disk phase, the orbital eccentricities and inclinations of super-Earths are tidally damped by the gaseous disk  \citep{papaloizoularwood00,goldreichsari03,tanakaward04,cresswellnelson08,bitschkley10,bitschetal11,teyssandierterquem14}. As the disk evolves and loses mass, eccentricity and inclination damping become less efficient.  Once the gas dissipates, these effects vanish. If eccentricities and orbital inclinations of planets in the chain grow because of mutual interactions, the orbits of the planets may eventually cross each other leading to collisions and scattering events \cite[e.g.,][]{kominami04,iwasakietal06,matsumotoetal12,morbidelli18}. This evolution typically breaks the resonant configurations established during the  gas disk phase, leading to a phase of giant impacts (for a detailed discussion on the onset of dynamical instabilities in resonant systems, see \citet[][and references therein]{meyerwisdom08,goldreichschlichting14,deckbatygin15,xuetal18,pichierietal18,pichierimorbidelli20}. The final (post-instability) configuration of such systems is nonresonant. Thus, the fact that most super-Earths are not found in resonant systems~\citep{lissaueretal11a,fabryckyetal14} should not be used as an argument against the migration model. In fact, the current distributions of super-Earths are consistent with all systems emerging from resonant chains. Systems like Kepler-223 \citep{millsetal16} and  TRAPPIST-1 \citep{gillonetal17,lugeretal17} have multiple-planet resonant chains that are naturally produced by migration. These resonant chains represent the small fraction of systems that did not become unstable after gas dispersal \citep{cossouetal14,izidoroetal17,ogiharaetal18}.  

The migration model can match the period ratio distribution of Kepler planets by combining a fraction of unstable and stable systems, typically $\lesssim10\%$ of stable and $\gtrsim90\%$ of unstable systems \citep{izidoroetal17}. The model also suggests that the large number of Kepler systems with single transiting planets versus multiple transiting planets -- known as the Kepler dichotomy \citep{johansenetal12} -- is a consequence of the dispersion of orbital inclinations of super-Earths rather than a true dichotomy in planetary multiplicity.  Dynamical instabilities excite the orbital inclinations of planets such that observations are likely to miss transits of mutually inclined planets \citep{winn10}, raising the number of single-transiting systems. Although some studies claim that the Kepler dichotomy is real  ~\citep[e.g.,][]{fangmargot12,johansenetal12,moriartyballard16}, the migration-instability model is arguably the simplest explanation for an {apparent} dichotomy, predicting an insignificant number of real single-planet systems in the Kepler sample.

To date, a downside of the migration model has been that simulations have assumed from the beginning that several Earth-mass  planetary embryos  formed in different parts of the disk. However, it remains unclear as to whether such distributions of embryos  could really have arisen naturally, or as to how the details of initial conditions affect the final systems. It is crucial to evaluate the legitimacy of these assumptions and more importantly to assess whether the migration model remains viable when a more self-consistent approach is used. 

The goal of this paper is to revisit the migration model \citep{cossouetal14,izidoroetal17} and to build a comprehensive  scenario for the origins of super-Earths  that is consistent with a broad picture of planet formation. This is the main upgrade of this study relative to \citet{izidoroetal17}. A key new ingredient in our scenario is pebble accretion. 
Pebble accretion plays a role after the formation of planetesimals, which are themselves thought to form by clumping of drifting pebbles via the so-called streaming instability ~\citep{youdingoodman05,johansenetal09,carreraetal15,simonetal16,carreraetal17}. Planetesimals then grow by mutual collisions, and once they approach lunar mass, they start to accrete pebbles efficiently ~\citep{johansenlacerda10,ormelklahr10,lambrechtsjohansen12,johansenetal15,xuetal17}.
Pebble accretion can explain the rapid growth of the building blocks of terrestrial planets, super-Earths, and ice giants \cite[e.g.,][]{lambrechtsjohansen12,levisonetal15,levisonetal15b,bitschetal15b,bitschjohansen16,nduguetal18,chambers16,johansenetal15,johansenlambrechts17,lambrechtsetal18}. However, what sets the destiny of planets in becoming either hot super-Earths or a different class of planet is not entirely clear.

This paper is part of a trilogy that develops a unified model to explain the formation of rocky Earth-like planets, hot super-Earths, and giant planets from pebble accretion and migration. This paper is dedicated to the formation pathways, dynamical evolution, and compositions of hot super-Earths.  The other two companion papers of this trilogy focus on (1) the formation of terrestrial planets and super-Earths inside the snow line, highlighting the role of the pebble flux \citep[][hereafter referred to as Paper I]{lambrechtsetal18}, and  (2) understanding the conditions required for gas giant planet formation in the face of orbital migration \citep[][a companion paper of this series, hereafter referred to as Paper III]{bitschetal18c}

Paper I models planetary growth exclusively inside the snow line. It shows that sufficiently low pebble fluxes -- integrated pebble fluxes of $114{\rm M_{\oplus}}$ -- lead to the slow growth of protoplanetary embryos that do not migrate substantially during the gas disk lifetime. These embryos are typically the same  mass as Mars at the end of the gas disk phase. An increased pebble flux by a simple factor of two bifurcates the evolution of these systems  inducing the formation of more massive rocky planetary embryos that migrate inwards and pile up at the inner edge of the disk.  These different growth histories separate the formation of truly Earth-like planets from that of rocky super-Earths. The long-term dynamical evolution of these systems reveals that dynamical instabilities after gas dispersal finally set the architecture of these systems. Dynamical instabilities among small Mars-mass 
planetary embryos result in collisions that lead to the formation of Earth-like planets of no more than 4 Earth masses. Instabilities among large rocky planetary embryos near the disk inner edge assemble rocky hot super-Earth systems.  An extensive  analysis of the formation of hot super-Earths,  also accounting for their possible origins beyond the snow line, is not performed in Paper I, but is shown here in the present work. Finally, Paper III shows that if the pebble flux is large enough (e.g., integrated pebble flux of $350-480M_{\oplus}$) super-Earths  turn into gas giant planets. Paper III presents a self-consistent model of the growth and migration of gas giant planets. It also highlights the role of migration in the formation of gas giants, an aspect that is typically ignored in  simulations  modeling the formation of the Solar System from pebble accretion \citep{levisonetal15,chambers16}.

This trilogy of papers is designed to provide a comprehensive view of planet formation and evolution, revealing the possible broad diversity of planetary systems produced from  pebble accretion, disk evolution, migration, and long-term dynamical evolution of planetary systems.

The current paper is structured as follows. Section \ref{sec:2}  describes the methods, namely our gas disk model, and pebble accretion prescription. Section \ref{rolepebbleflux} presents simulations designed to understand the role of the pebble flux in the formation and early  evolution of the system. We also tested the role of the initial distribution of protoplanetary embryos and pebble sizes inside the snow line. In Section \ref{sec:sec4} we discuss our results in light of those of Paper I. In Section \ref{sec:super-Earth} we present the results of our simulations dedicated to modeling the growth and long-term dynamical evolution of hot super-Earth systems. In Section \ref{sec:observations} we lay out observational constraints that we use to evaluate the success of different models in matching the true super-Earth distributions.  We focus on the period ratio distribution, the Kepler ``dichotomy'' and the compositions (rocky vs. icy) of super-Earths. In Section \ref{sec:solarsystem} we place the Solar System in the context of our model. In Section \ref{sec:7} we  present our  conclusions. In Appendix \ref{sec:AppendixA}  we detail our prescription for gas-driven migration.

\section{Method}
\label{sec:2}

Our simulations are performed with FLINTSTONE, our new N-body code built on MERCURY \citep{chambers99}. In addition to the standard orbital integration, we include modules to consistently compute the following:
\begin{itemize}
\item the evolution of an underlying, gas-dominated protoplanetary disk;
\item the gas-assisted accretion of pebbles onto growing planetary cores; and
\item the orbital migration and tidal damping of cores.
\end{itemize}
Other aspects of the MERCURY code were left intact. For instance, collisions are modeled as inelastic merging events that conserve linear momentum. We also do not model volatile loss during giant impacts.
Thus, the final water and ice content of planets in our simulations should be interpreted as upper limits. We also do not perform modeling of planetary interior. In this work we use the term ``ice''  indiscriminately as a proxy for all physical states of water in our planets. In all our simulations, planetary objects are ejected from the system if they reach heliocentric distances of greater than 100~AU. 

We compared FLINTSTONE with the code used in our companion paper (Paper I) by \citet{lambrechtsetal18} which is built on Symba \citep{duncanetal98} and found similar results for test problems regarding planet migration, pebble accretion, and damping of eccentricity and inclination. 

Below we describe our disk model and our prescriptions for pebble accretion, gas-driven migration, inclination, and eccentricity gas-tidal damping.

\subsection{Gas disk model}


Our underlying disk is represented by 1D radial  profiles derived from 3D hydrodynamical simulations modeling gas disk evolution \citep{bitschetal15}.\footnote{We note that the gas disk model considered in this work is more sophisticated than the one from in Paper I} The hydrodynamical model accounts for the effects of viscous heating, stellar irradiation, and radial diffusion.

In the standard alpha-disk paradigm, the gas accretion rate on the young star is written as
\begin{equation}
\dot{M}_\text{gas} =  3\pi\alpha h^2 r^2 \Omega_{{\rm k}} \Sigma_{{\rm gas}},
\label{eq:1}
\end{equation}
where ${ h = H/r}$ is the disk aspect ratio with H being the disk scale height, $\alpha$  the dimensionless viscosity parameter \citep{shakurasunyaev73}, r   the distance to the star,  ${ \Omega_k = \sqrt{GM_{\odot}/r^3}}$ the orbital the keplerian frequency, and $\Sigma_{gas}$ the gas surface density.

Following \cite{hartmannetal98} and \cite{bitschetal15}, the relationship between the disk and star age and the gas accretion rate onto the star is given by
\begin{equation}
{{\rm log}{\left( \frac{\dot{M}_{{\rm gas}}}{M_{\odot}/\text{yr}}\right) }~=~-8 - 1.4\log\left(\frac{\text{t}_{\text{disk}} + 10^5~\text{yr}}{10^6~\text{yr}}\right)}.
\label{eq:2}
\end{equation}
Finally, the hydrostatic equilibrium yields 
\begin{equation}
T =  h^2 \frac{G M_{\odot}}{r}\frac{\mu}{\mathcal{R}},
\label{eq:3}
\end{equation}
where ${\rm t_{disk}}$ is the disk age, $T$ is the temperature at the gas disk midplane, ${\mu}$ is the gas mean molecular weight set  to ${\rm 2.3~gmol^{-1}}$, $G$ is the gravitational constant,  ${M_{\odot}}$ is the mass of the star set equal to 1 solar mass,  and $\mathcal{R}$ is the ideal gas constant. We use ${t_{{\rm start}}}$ to represent the disk age at the starting time of our simulations.

From Eqs. \ref{eq:1}, \ref{eq:2}, and  \ref{eq:3} we obtain the gas surface density ($\Sigma_{{\rm gas}}$) of the disk by using the fits of the disk temperature  at the midplane provided in \cite{bitschetal15}. The disk metallicity and ${\rm \alpha}$-viscosity parameter are set to 1\% and to $\alpha = 5.4 \times 10^{-3}$, respectively. The main source of viscosity in protoplanetary disks is relatively unknown \citep[e.g.,][]{turneretal14}. Recent ALMA observations \citep[e.g.,][]{pinteetal16,dongetal17,zhangetal18} have suggested disk viscosities that can be  one order of magnitude lower than the value assumed in our simulations (however, see \citet{dullemondetal18} for a substantially higher radial diffusion). Lower disk viscosities could increase the efficiency of pebble accretion because the pebble scale height depends on viscosity. At the same time, in sufficiently low-viscosity disks lower mass planets would be able to open up gaps in the disk. It would be interesting to test the impact of the disk viscosity on our model but this remains a subject for future investigations.

As the disk evolves, we recalculate the disk structure every 500 years rather than at every time-step to save computation time. This does not significantly  impact the quality of our approach because the disk structure only significantly changes on  long timescales ($\gg$500 years).

Protoplanetary disks are also expected to have inner cavities in their gas density distribution created due to the magnetic star--disk interaction \citep{koenigl91}. The dipolar magnetosphere of a rotating young star may disrupt the very inner parts of the protoplanetary disk  dictating the gas accretion flow onto the stellar surface \citep{romanovaetal03,bouvieretal07,flocketal17}. The truncation radius is probably at a few stellar radii, inside the co-rotation radius with  the star and where the magnetic field pressure balances the pressure of the accreting disk. In standard disks with typical magnetized young stars the disk truncation radius is expected to be around $\sim$0.03-0.2~AU \citep[e.g.,][]{bouvieretal13}.  

The inner cavity of a protoplanetary disk is expected to have an important impact on planet formation because it is likely to act as an efficient planet trap, preventing  inwardly migrating planets from simply falling onto the star \citep{massetetal06,romanovalovelace06,romanovaetal18}. The innermost planets in several Kepler systems indeed have orbital periods corresponding to the expected truncation radius of  disks, corroborating the existence of disk inner edges \cite[e.g.,][]{muldersetal18}.

To account for an inner cavity, our disk inner edge is fixed at $\sim0.1$~AU in our nominal simulations unless stated otherwise.  In order to represent the drop in surface density at this location we multiply the gas density by the following factor
\begin{equation}
\Re = \tanh\left( \frac{r-r_{{\rm in}}}{0.05r_{{\rm in}}}\right),
\end{equation}
where $r$ is the heliocentric distance  and $r_{{\rm in}}$ is the location of the disk inner edge. This  approach has  also been used in  previous works \citep{cossouetal14,izidoroetal17}.

In Appendix \ref{sec:AppendixA} we describe how planet migration is modeled in our simulations. We emphasize that the migration prescription considered in this work is more sophisticated than that of Paper I. Unlike Paper I, here we take into account corotation (entropy and vortensity driven) effects in order to obtain a more quantitatively realistic model.

\subsection{Model setup}\label{sec:setupmodels}

Our simulations start with a distribution of small planetary embryos with masses randomly chosen between 0.005 and 0.015 Earth masses. Each simulation starts with a slightly different distribution of planetary seed masses. For this mass regime, pebble accretion typically dominates over planetesimal accretion \citep{johansenlambrechts17}. In all our simulations, the initial orbital inclination of planetary embryos is randomly selected between 0 and 0.5 degrees. The orbital eccentricities of all planets are set initially to $10^{-4}$. Other orbital angles were randomly selected between 0 and 360 degrees.

The birth location of the first planetesimals is unconstrained by observations. Some models suggest that planetesimal formation is likely to initiate just outside the snow line \citep{drazkowskadullemond14,armitageetal16,drazkowskaalibert17,carreraetal17,schoonenbergormel17,drazkowskadullemond18}.  Other models suggest planetesimal formation takes place just inside the snow line \citep{idaguillot16} or even around 1~AU \citep{drazkowskaetal16}. In our simulations we test different distributions of seeds where they naturally account for planetesimal formation: only throughout the inner disk (inside the snow line), only throughout the outer disk (outside the snow line), and both inside and outside the snow line (throughout the disk). In our disk model the snow line moves inward as the disk evolves; at 0.5~Myr, it is around 3~AU but by 3~Myr is already around 0.7~AU. The details of our different initial distributions of protoplanetary embryos are presented in Table \ref{tab:model}. In simulations of Models I and II, planetary seeds are initially distributed past 0.7~AU. Because at ${\rm t_{disk}=3~Myr}$ the snow line is at 0.7~AU, simulations with ${\rm t_{start}=3~Myr}$ correspond to scenarios where planetary seeds formed only outside the snow line.

\begin{table}
\scriptsize
\caption{Initial conditions of our simulations. From left to right, columns mark the simulation model, the region where planetary seeds are initially distributed in the disk, their initial mutual radial separation, and the disk age at the start of the simulation. The mutual separation is also given in units of mutual hill radii, ${\rm Rhm=\left( \frac{m_1+m_2}{3M_\oplus}\right)^{1/3}\frac{a_1+a_2}{2}}$, where  ${\rm m_1}$ and ${\rm m_2}$ are the masses and ${\rm a_1}$ and ${\rm a_2}$ are the semi-major axes of any two adjacent planetary embryos.}
   \centering 
\begin{tabular}{@{}lcccc@{}}
  \hline
&        &  Radial                  & Mutual       &  $\text{t}_{\text{start}}$    \\
&        &  Distribution            & separation   &  \\
\hline\hline\\[-1.4mm]
\multirow{7}{*}{{\rot{\rlap{{\small Model}}}}} & I         &  0.7 to ${\rm \sim20~AU}$    &  0.25~AU  (${\rm \sim5-100~Rhm}$)        & 0.5 and 3~Myr\\
\cmidrule{2-5}
                          & II        & 0.7 to ${\rm \sim60~AU}$     &   ${\rm \sim 0.01-3~AU}$ (${\rm \sim10-30~Rhm}$)  & 0.5 and 3~Myr  \\
\cmidrule{2-5}
                          & III       &  ${\rm 0.2~(or~0.3)}$ to ${\rm \sim2~AU}$     &  0.025~AU (${\rm \sim15-55~Rhm}$)      & 0.5 and 3~Myr \\
\hline
\label{tab:model}
\end{tabular}
\end{table}

The formation time of the first planetesimals is also poorly constrained. It has been proposed that in our inner Solar System at least two distinct generations of planetesimals were born: one forming early at about 0.5~Myr after the so-called calcium--aluminium-rich inclusions \citep[CAIs;][]{villeneuveetal09,kruijeretal12} and others late at about 3~Myr after CAIs.  However, it is not clear whether this scenario is a generic outcome of planet formation. Given our limited understanding of the timing of planetesimal formation, we explore in our simulations two end-member scenarios. For simplicity, we consider a single generation of planetesimals for all our simulations. Our first scenario corresponds to the case where planetesimals form very early. In this case our simulations start with ${\rm t_{start}=0.5~Myr}$. In the second scenario, planetesimals are assumed to form late and our simulations start with ${\rm t_{start}=3.0~Myr}$ (see also Paper III). We note that different ${t_{{\rm start}}}$ imply different disk structures at the beginning of our simulations and this has an important impact on the system evolution both in terms of planet migration and pebble accretion \citep{bitschetal15b}.

 In  our simulations, planetary seeds starting initially inside the snow line are assumed to be rocky while those outside are considered to have 50\% of their mass as ice. Rocky and icy planetary seeds have bulk densities of 5.5${\rm~g/cm^3}$ and 2${\rm ~g/cm^3}$, respectively.

\subsection{Pebble accretion}

\label{subsec:accretion}

As in Paper I we do not model drifting pebbles as individual particles because of the high computational cost \cite[but see][]{kretkelevisonetal14,levisonetal15}. Instead, pebbles in the disk are modeled as a background field, which depends on the  flux and Stokes number of the pebbles, which evolve over time. Our modeling of the pebble flux is more sophisticated than that from Paper I, where the number of pebbles in the disk decays exponentially during the disk lifetime and the Stokes number of the pebbles is fixed. Here, the pebble flux and Stokes number are computed in the context of dust coagulation models \citep{birnstieletal12,lambrechtsjohansen12,lambrechtsjohansen14}, as summarized below. We follow the prescription of pebble accretion by protoplanets from \citet{johansenetal15}, which can also account for reduced accretion rates for eccentric and inclined bodies. 

The accretion rate onto the planetary core is given as
\begin{equation}
 {\dot M}_{\rm core} = \pi R_{\rm acc}^2 \rho_{\rm p,mid} \bar{S} \delta v \ ,
\end{equation}
where $\rho_{\rm p,mid}$ is the midplane density of pebbles, related to the pebble surface density layer $\Sigma_{\rm peb}$ via
\begin{equation}
 \rho_{\rm p,mid} = \Sigma_{\rm peb} /(\sqrt{2\pi} H_{\rm peb}) \ ,
\end{equation}
where $H_{\rm peb} = H_{\rm gas} \sqrt{\alpha_{\rm set} / \tau_{\rm f}}$ is the pebble scale height \citep{youdinlithwick07}, $\tau_{\rm f}$ is the Stokes number, and $\alpha_{set}$ is the dimensionless viscosity parameter representing disk midplane turbulence  which determines the vertical settling of pebbles. In our nominal simulations, $\alpha_{\rm set}=\alpha$, although some studies argue for  $\alpha_{\rm set}\ll\alpha$ \citep{zhustone14}. $R_{\rm acc}$ denotes the accretion radius of the planet, $\delta v = v_{\rm rel} + \Omega R_{\rm acc}$ with $v_{\rm rel}$ being the relative velocity between pebbles and planets. To calculate the mass accretion rate, we first need the accretion radius and the pebble density averaged over the accretion radius. The stratification integral $\bar{S}$ is defined as the mean pebble density normalized by the pebble density in the midplane, namely:
\begin{equation}
 \bar{S} = \frac{1}{\pi R_{\rm acc}^2} \int_{z_0 - R_{\rm acc}}^{z_0 + R_{\rm acc}} 2 \exp [-z^2 / (2H_{\rm peb}^2)] \sqrt{R_{\rm acc}^2 - (z -z_0)^2} {\rm dz},
 \label{eq:strat}
\end{equation}
where $z_0$ is the height over the disk midplane where the protoplanet is located along its orbit. This equation is derived by considering layers of  constant $z$ and consequently constant pebble density, but there is no analytical solution to this integral and the solution has to be obtained numerically. However, \citet{johansenetal15} used a square approximation that integrates the pebble density over a square instead of a circle rendering the integral analytically solvable, which we use here. The exact solution is shown in the Appendix of \citet{johansenetal15}. 

The pebble flux  is calculated as
\begin{equation}
  \label{eq:Mdotpebble}
 \dot{M}_{\rm peb} = 2 \pi r_{\rm g} \frac{{\rm d}r_{\rm g}}{{\rm d}t} Z \Sigma_{\rm g} (r_{\rm g}) \ .
\end{equation}
Here, $r_{\rm g}$ represents the heliocentric distance at which dust particles have  grown to pebble sizes and start drifting inwards by gas drag, and $\Sigma_{\rm g} (r_{\rm g})$ is the gas surface density at the  pebble production line location $r_{\rm g}$ (for more details see \cite{lambrechtsjohansen14}). The quantity $Z$ represents the dust to gas ratio in the disk that can be converted into pebbles at the pebble production line $r_{\rm g}$ at time $t$. In our nominal model, we take $Z=1\%$ . \citet{lambrechtsjohansen14} derived the time-dependent radial location of the pebble production line as
\begin{equation}
 r_{\rm g} = \left(\frac{3}{16}\right)^{1/3} (GM_\star)^{1/3} (\epsilon_{\rm D} Z)^{2/3} t^{2/3} \ ,
\end{equation}
and thus
\begin{equation}
 \frac{{\rm d}r_{\rm g}}{{\rm d}t} = \frac{2}{3} \left(\frac{3}{16}\right)^{1/3} (GM_\star)^{1/3} (\epsilon_{\rm D} Z)^{2/3} t^{-1/3} \ ,
\end{equation}
where $M_\star$ is the stellar mass, which we set to $1 M_\odot$, $G$ is the gravitational constant, and $\epsilon_{\rm D} = 0.05$ is associated with the logarithmic growth range from dust grains to pebble sizes. In our model, at 0.5~Myr, 3~Myr, and 5~Myr the pebble production line is at ${\sim77{\rm ~AU}}$, ${\sim250{\rm ~AU}}$, and ${\sim350{\rm ~AU}}$, respectively\footnote{Our disk is probably on the large side of the spectrum of typical radial disk sizes ($\sim$100~AU), but some disks are as large as 500-1000~AU \cite[e.g.,][]{hughesetal08}}. We note that in the drift-limited pebble-growth  model, the pebble flux depends on the gas surface density at the pebble production line  (Eq. \ref{eq:Mdotpebble}).  As the pebble production line moves beyond $\sim$50~AU, the  gas disk and drift-limited pebble-growth  models assumed in this work can strongly underestimate the pebble column density \citep{bitschetal18a} compared to observations \citep{wilneretal05,carrascoetal16}.  This  is a consequence of the low gas surface density in the remote regions of our evolving protoplanetary disk. In our nominal disk with ${S_{{\rm peb}}=1}$, the total mass in pebbles between 50 and 90~AU is only about 10${{\rm ~M_{\oplus}}}$ at 1~Myr.  Observations suggest that the total mass in dust and pebbles in  the outer parts (50-90au) of the HL Tau disk, which is about 1~Myr old, is between 80 and 280 ${{\rm ~M_{\oplus}}}$~\citep{carrascoetal16}. On the other hand, some observed disks seem not to be massive enough to form the observed exoplanet population~\citep{manaraetal18,mulders18}. To account for a possible range of  disk masses, we rescale the gas surface density at the pebble production line by a factor ${S_{{\rm peb}}}$in order to increase the pebble surface density (or decrease in some extreme cases; see eq. \ref{eq:SigmaPeb}). We note that $S_{{\rm peb}}$ is treated as a free-parameter in our model.

Particles in the gas disk grow and drift, so the local Stokes numbers of the particles are limited by drift \citep{birnstieletal12,lambrechtsjohansen12}. By equating the pebble growth timescale with the drift timescale  (i.e., valid for the drift-limited dominant pebble size; see section 2.4 in \cite{lambrechtsjohansen14}) the stokes number may be written as
\begin{equation}\label{stokesnumber}
 \tau_{\rm f} = \frac{\sqrt{3}}{8} \frac{\epsilon_{\rm P}}{\eta} \frac{\Sigma_{\rm peb}(r_P)}{\Sigma_{\rm g}(r_P)} \ ,
\end{equation}
where $\Sigma_{\rm g} (r_{\rm P})$ and $\Sigma_{\rm peb} (r_{\rm P})$ are the gas and pebble surface densities at the  location of the planets $r_{\rm P}$.  We represent the radial pressure support of the disk through the dimensionless parameter
\begin{equation}
\label{eq:eta}
 \eta = - \frac{1}{2} \left( \frac{H}{r} \right)^2 \frac{\partial \ln P}{\partial \ln r} \ .
\end{equation}

\begin{figure}
\centering
\includegraphics[scale=.4]{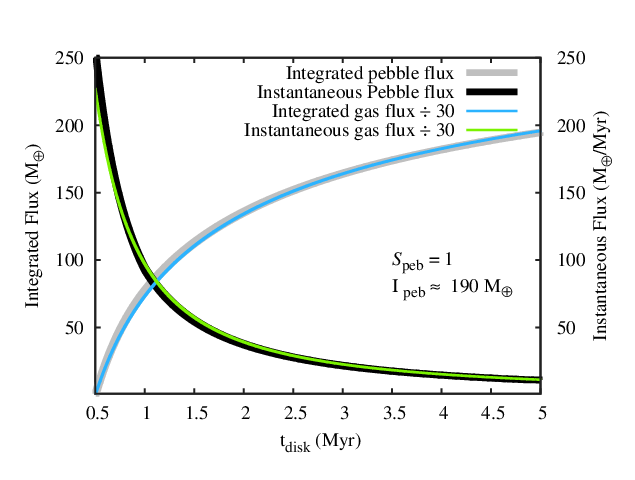}
\caption{Pebble flux in our nominal disk model.  The integrated mass in drifted pebbles (${\rm I_{{\rm peb}}}$) is shown on the left-hand vertical axis as a function of disk age (${\rm t_{disk}}$). Fluxes are calculated at 5~AU. The right-hand vertical axis shows the instantaneous pebble flux decreasing as the disk dissipates.  Simulations of Paper I and III also feature the decay of the pebble flux. All our simulations start with a ${\rm t_{disk}}$ of at least 0.5 Myr, thus the pebble production line is already past the initial positions of our outermost seeds ($\sim$60~AU; see Table \ref{tab:model}).  This plot is generated considering  ${S_{{\rm peb}} = 1}$ in Eq. \ref{eq:SigmaPeb} but a simple rescaling of these curves accounts for other considered pebble fluxes. Both curves correspond to the pebble flux beyond the snow line. The pebble flux inside the snow line is reduced by a factor of two because of the mass sublimation of the ice component of the pebbles when they cross the snow line.  The  gas flux and the integrated gas flux in our gas disk model are shown by the green and blue lines, respectively. We note that in the plot we rescaled the gas flux by a factor of 1/30, for clear comparison with the pebble flux.}
    \label{fig:flux}
\end{figure}

The pebble flux decreases in time as the disk evolves \citep{birnstieletal12,lambrechtsetal14,bitschetal18a}. The (evolving) pebble surface density $\Sigma_{\rm peb}$ can be calculated using the pebble flux and Stokes number (Eqs. \ref{eq:Mdotpebble} and \ref{stokesnumber}) as
\begin{equation}
 \label{eq:SigmaPeb}
 \Sigma_{\rm peb} = \sqrt{\frac{2 S_{\text{peb}} \dot{M}_{\rm peb} \Sigma_{\rm g} }{\sqrt{3} \pi \epsilon_{\rm P} r_{\rm P} v_{\rm K}}} \ ,
\end{equation}
where $r_{\rm P}$ denotes the semi-major axis of the planet, and $v_{\rm K} = \Omega_{\rm K} r_{\rm P}$.  ${S_{{\rm peb}}}$ is a  nondimensional linear scaling factor of the pebble flux (see Eq. \ref{eq:Mdotpebble}). In Eq. \ref{eq:SigmaPeb},  $\epsilon_{\rm P} =0.5$ represents the pebble sticking efficiency under the assumption of near-perfect sticking \citep{lambrechtsjohansen14}. ${S_{{\rm peb}}=1}$ corresponds to an integrated pebble flux ${\rm I_{{\rm peb}} \approx 194~M_{\oplus}}$ beyond the snow line (see Figure \ref{fig:flux}).   It remains possible that a disk with a higher(lower) pebble flux could  be simply  associated with a disk with higher(lower)\ metallicity (see Eq. 11 of \cite{lambrechtsjohansen14}). For simplicity, we assume a unique  disk metallicity  set equal to  1\% during the entire disk lifetime to model planet migration in all simulations. The same approach was taken in our companion paper by \citet{bitschetal18c}.

We assume that the water component of pebbles crossing the snow line sublimates, releasing the rocky or silicate counterpart in the form of smaller dust grains. As the snow line moves inwards, the boundary between big (icy) and small (rocky) pebbles moves with the snow line. The Stokes number of icy pebbles is given by Eq. \ref{stokesnumber} in the drift-limited solution (dominant pebble size). We do not use Eq. \ref{stokesnumber} to calculate the Stokes number of silicate pebbles in the inner disk. Instead,  we set the size of silicate pebbles to sizes of 1~mm which correspond to the typical chondrule sizes  of ordinary chondrites \citep{friedrichetla15} in our nominal simulations. We then calculate the stokes number of 1mm silicate pebbles in the inner disk via the classical definition of Stokes number (instead of using Eq. \ref{stokesnumber})  as 
\begin{equation}
 \tau_{\rm f,rock} = \frac{\rho_{\rm rock}  {\rm Rocky}_{\rm peb}}{ \Sigma_{\rm g}},
\end{equation}
\citep[e.g.,][]{lambrechtsjohansen12} where ${\rm Rocky_{{\rm peb}}}$ and  $\rho_{\rm rock}$ define the size  and bulk density of silicate pebbles, respectively. The bulk density of silicate pebbles is assumed to be 5.5${\rm g/cm^3}$ in all simulations. Inside the snow line, the pebble flux is reduced by a factor of two to account for the sublimation of the water component of the pebbles ($\dot{M}_{\rm peb,rock} = 0.5\dot{M}_{\rm peb}$). We calculate the  pebble surface density inside the snow line using the following equation.
\begin{equation}
  \label{eq:Mdotpebblerock}
 \Sigma_{\rm peb,rock} =  \frac{\dot{M}_{\rm peb,rock} } {2 \pi r v_{\rm peb,rock}} \,
,\end{equation}
where the radial velocity of rocky pebbles is given by \citep[e.g.,][]{ormelklahr10,bitschetal18a}
\begin{equation}
  \label{eq:vrad_rock}
 v_{{r}_{\rm peb,rock}} = -\frac{ 2.0  \eta v_{\rm k}}{  \tau_{\rm f,rock} +  \tau_{\rm f,rock}^{-1}} - \frac{\nu}{r} \,
,\end{equation}
where  ${\rm \nu = \alpha H^2 \Omega_k}$ is the gas kinematic viscosity.

\begin{figure}
\centering
\includegraphics[scale=.22]{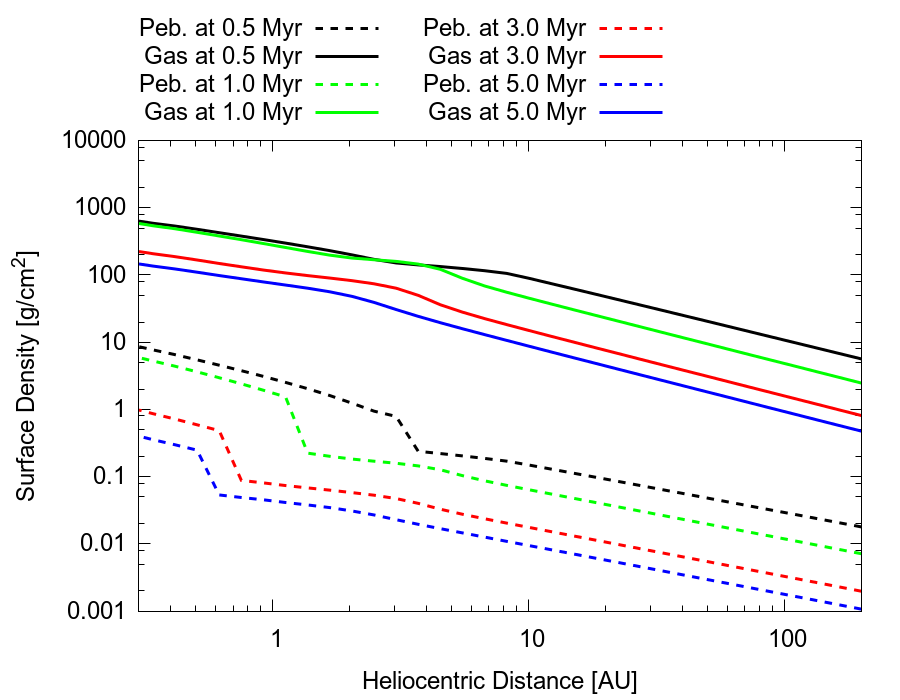}
\includegraphics[scale=.26]{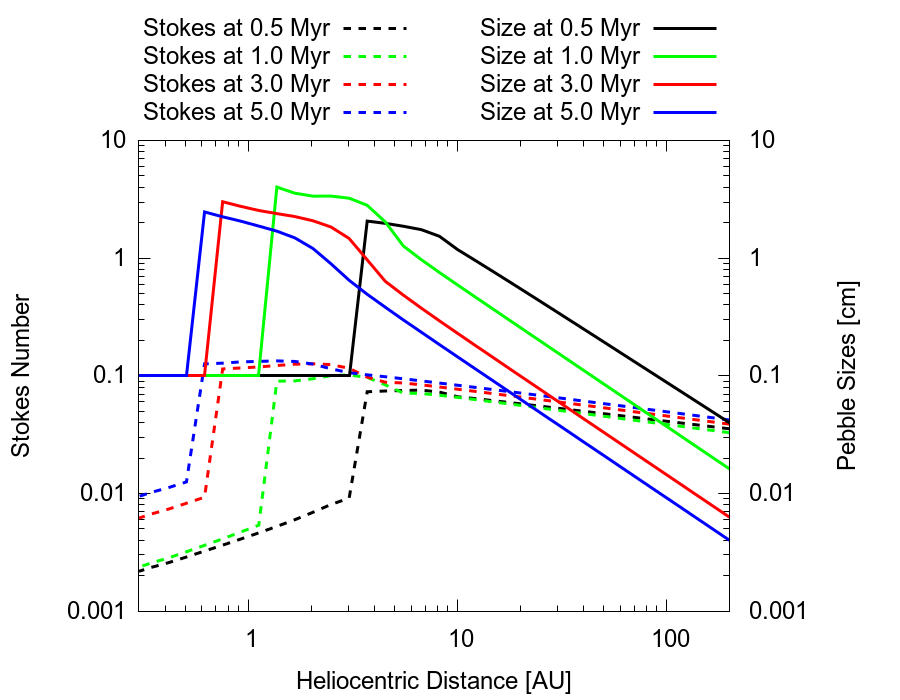}
\caption{\textbf{Top:} Gas and pebble surface densities as a function of orbital distance at different disk ages. \textbf{Bottom:} Pebble sizes and Stokes number as a  function of orbital distance at different disk ages. In both panels, ${\rm Rocky}_{\rm peb}$~=~1mm and ${S_{{\rm peb}}=1}$, which correspond to our nominal values. The sharp drop in the Stokes number in the inner regions of the disk is due to the snow line transition where we set the size of silicate pebbles to 1~mm.  } 
    \label{fig:pebbles_surf_size}
\end{figure}

In our nominal simulations, we set ${\rm Rocky_{{\rm peb}}}$=1~mm which corresponds to the size of chondrules, but we also test the effects of larger silicate pebbles in our models. These pebble sizes were already used in \citet{morbidellietal15b} to reproduce the different growth speeds of the terrestrial planets compared to the gas giants in the Solar System. On the other hand, although  laboratory  experiments \citep[e.g.,][]{windmarketal12,blum2018}  and numerical models \citep{birnstieletal10,banzattietal15} suggest that the growth of silicate pebbles beyond milimeter size is challenging, centimeter-sized chondrule clusters are found in unequilibrated ordinary chondrites \citep{metzleretal12}. Silicate centimeter(cm)-sized pebbles probably also existed in our early inner Solar System at least to some (small) level. Thus, in our simulations we also analyze the effects of considering  1cm   pebbles inside the snow line. We have verified that in our disk model, at the very early stages of the disk (${\rm t_{disk}\approx0.0~Myr}$),  1cm (1mm) pebbles inside 0.5~AU would be in the Stokes (Epstein) regime of gas--particle coupling \citep{epstein24}. However, as our simulations start with ${\rm t_{start}=t_{disk}=0.5~Myr}$ or 3~Myr, both 1mm and 1cm silicate pebbles  beyond $\sim$0.2~AU (the starting location of our innermost seeds) are in the  Epstein regime of gas-particle coupling. The top panel of Figure \ref{fig:pebbles_surf_size} shows the gas and pebble surface densities as a  function of orbital distance as the disk evolves. The bottom panel of Figure \ref{fig:pebbles_surf_size} shows the Stokes number and sizes of pebbles in our disk with nominal pebble fluxes and sizes at different disk ages.

To account for the sublimation of the water component of pebbles crossing the ice line we assume a reduction of the pebble mass fluxes to half. This is consistent with the assumed composition of seeds forming inside and outside the snow line. In our disk model, the water ice line moves inwards as the disk dissipates, and reaches 1~AU at 2~Myr.

A planet only accretes a fraction $f_{\rm acc}$ of the flux of pebbles $\dot{M}_{\rm peb}$ drifting past its orbit:
\begin{equation}
 f_{\rm acc} = \frac{\dot{M}_{\rm core}}{\dot{M}_{\rm peb}} \ .
\end{equation}
The pebble flux arriving at interior planets is thus reduced by this fraction $f_{\rm acc}$, reducing the accretion rates onto the interior planets. 


A sufficiently massive planet opens a partial gap and creates an inversion in the radial pressure gradient of the disk, halting the inward drift of pebbles \citep{paardekooperetal06, morbidellinesvorny12,lambrechtsetal14}. The mass at which this takes place is called the pebble isolation mass. The pebble isolation mass in itself is a function of the local properties of the protoplanetary disks, namely the viscosity, aspect ratio, and radial pressure gradient, and of the Stokes number of the particles, which can diffuse through the partial gap of the planet \citep{bitschetal18b,weberetal18}. Here, we follow the exact description of \citet{bitschetal18b}, who give the pebble isolation mass including diffusion as
\begin{equation}
\label{eq:MisowD}
  M_{\rm iso} = 25 f_{\rm fit} {\rm M}_{\rm E} + \frac{\Pi_{\rm crit}}{\lambda} {\rm M}_{\rm E} \ ,
\end{equation}
with $\lambda \approx 0.00476 / f_{\rm fit}$, $\Pi_{\rm crit} = \frac{\alpha}{2\tau_{\rm f}}$, and
\begin{equation}
\label{eq:ffit}
 f_{\rm fit} = \left[\frac{H/r}{0.05}\right]^3 \left[ 0.34 \left(\frac{\log(\alpha_3)}{\log(\alpha)}\right)^4 + 0.66 \right] \left[1-\frac{\frac{\partial\ln P}{\partial\ln r } +2.5}{6} \right] \ ,
\end{equation}
where $\alpha_3 = 0.001$.

After planets reach the pebble isolation mass, they can start to accrete gas from the disk.  That process is modeled in our companion paper by \citet{bitschetal18c}. Here we assume that the contraction of the gaseous envelope \citep[][]{pisoyoudin14} is sufficiently slow\footnote{A sufficiently high envelope opacity  may prevent fast contraction \citep{pisoyoudin14}.} that our seeds would not transition into runaway gas accretion and stay at super-Earth mass. Three-dimensional hydrodynamical simulations show  that only planets more massive than about 15~${\rm M_{\oplus}}$ and in disks with opacities lower than $\sim$0.01~$cm^2/g$ can transition to runaway gas accretion~\citep{lambrechtslega17}. In this paper, we neglect the effects of gas accretion in all simulations but we flag a growing planet as a giant planet core when it reaches a mass larger than 15~${\rm M_{\oplus}}$ during the gas disk phase.

\section{The role of the pebble flux}\label{rolepebbleflux}

Our simulations were conducted in two phases. First, we ran simulations considering a wide range of pebble fluxes (${S_{{\rm peb}}=0.2}$, 0.4, 1, 2.5, 5, and 10; see Table \ref{tab:model}).  We used the outcome of these simulations to inspect which pebble fluxes could  lead to the formation of planets in the super-Earth/Neptune mass range  -- with masses smaller than $\sim$15 ${\rm M_{\oplus}}$ -- during the gas disk phase. The second phase of long-term integrations beyond the gas disk phase is discussed in Section 5.  We remind the reader that our goal is to model the formation of hot super-Earth systems; the formation of rocky terrestrial planets and rocky super-Earths as well as giant planets are modeled in companion papers of this trilogy \citep{lambrechtsetal18,bitschetal18c}. In Paper I we presented our findings that an integrated pebble flux of ${\rm 114~M_{\oplus}}$ leads to the formation of classical terrestrial planets while simulations with integrated pebble fluxes of ${\rm 190~M_{\oplus}}$ (or more) form super-Earth systems. As discussed before, one should not expect that by considering the same integrated pebble flux as Paper I our simulations will produce  the same types of planets (super-Earths or terrestrial planets). This is because our planets may grow outside the snow line by accreting larger pebbles. We discuss this issue again below.

We performed ten simulations for each value of the pebble flux scaling ${S_{{\rm peb}}}$. We did not model the long-term dynamical evolution of these systems. We stop our simulations at the end of the gas disk phase, namely at 5~Myr. Due to the large number  of simulations to be conducted in this section and in order to save central processing unit (CPU) time, we also set the disk inner edge in these simulations to about ${\rm r_{\rm in}\simeq0.3~AU}$. The location of the disk inner edge in simulations modeling the formation of hot super-Earth system have been typically assumed to be  around 0.1~AU \citep{cossouetal14,izidoroetal17,ogiharaetal18,lambrechtsetal18}. We likewise set the disk inner edge to  ${\rm r_{in}\simeq0.1~AU}$ in our simulations of  Section \ref{sec:super-Earth}. However, here we set  ${\rm r_{in}\simeq0.3~AU}$ and use a larger integration time-step to conduct this large batch of simulations without degrading the quality of our results. In this section, we infer the final compositions of the 
planets by tracking the source of the accreted material in terms of icy and silicate pebbles. 

\subsection{Model I}
\begin{figure*}
\centering
\includegraphics[scale=.4]{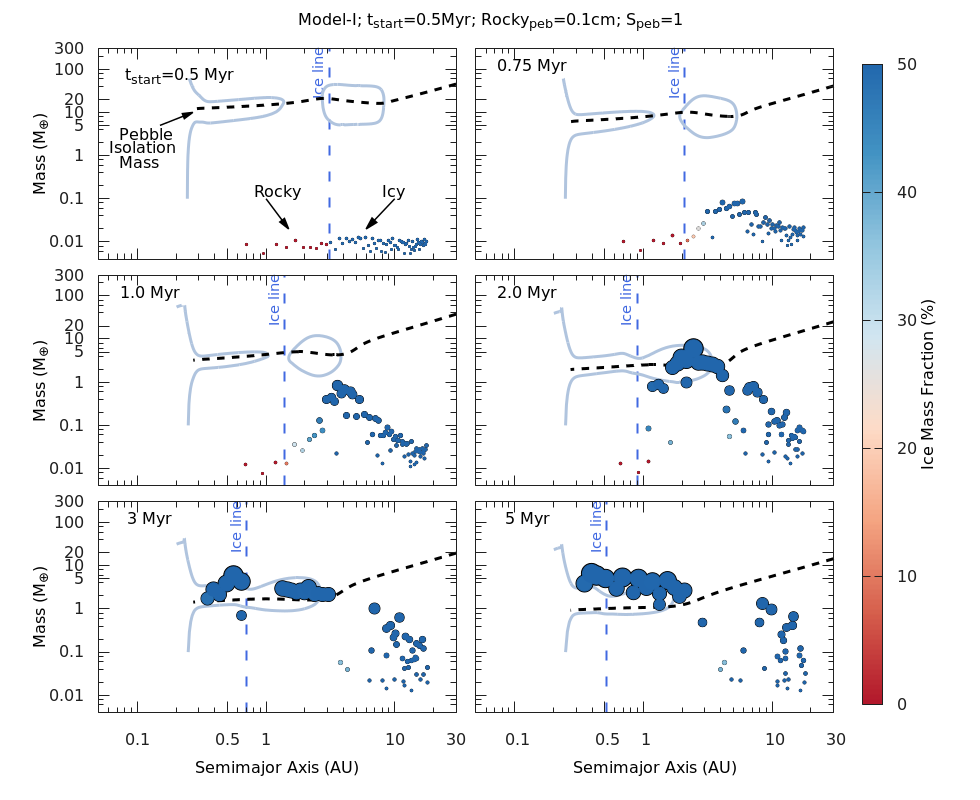}
\caption{Evolution of an example simulation from model I.  The panels represent snapshots of the growth of protoplanetary embryos in a simulation with ${\rm t_{start}=0.5~Myr}$,  ${S_{{\rm peb}}= 1,}$ and  a size of the rocky pebbles equal to ${\rm Rocky_{{\rm peb}}=0.1~cm}$. The vertical and horizontal axes represent mass and semi-major axis, respectively. The location of the disk snow line (blue vertical dashed line) and the pebble isolation mass (black dashed line) are also shown for reference as the disk evolves. Planetary embryos growing inside the snow line accrete silicate pebbles while those outside the snow line accrete icy pebbles. The gray solid lines delimit the region of outward migration. The disk inner edge is shown at ${\rm \sim0.3~AU}$, where planets are trapped as well. The color of each dot gives the ice mass fraction. The size of each dot scales as $m^{1/3}$, where ${m}$ is the planetary mass. The exact time evolution is shown in Fig.~\ref{fig:dynamics_S1}.}
    \label{fig:snapshots}
\end{figure*}

Figure \ref{fig:snapshots} shows the growth and dynamical evolution of protoplanetary embryos in one simulation of Model I (see Table \ref{tab:model} for the definition of our models) with ${\rm t_{start}=0.5~Myr}$ and ${S_{{\rm peb}}=1}$ during the gas disk phase. In this simulation, planetary embryos grow by pebble accretion more quickly outside the snow line than inside because pebbles are typically larger in the cold regions of the disk  in our model. At 1~Myr, the largest planetary embryo outside the snow line is about ${\rm 1~M_{\oplus}}$. At 2~Myr, the mass of the largest planetary embryo is about ${\rm \sim7~M_{\oplus}}$. As the disk evolves, the pebble isolation mass (black dashed line) decreases across the entire disk because the disk cools down and gets thinner \citep{bitschetal15}. The gray curves give the boundary of the (a,M) region where migration is outwards.  We note that the pebble isolation mass is within the range of the region of outward migration in some parts of the disk.

\begin{figure*}
\centering
\includegraphics[scale=.9]{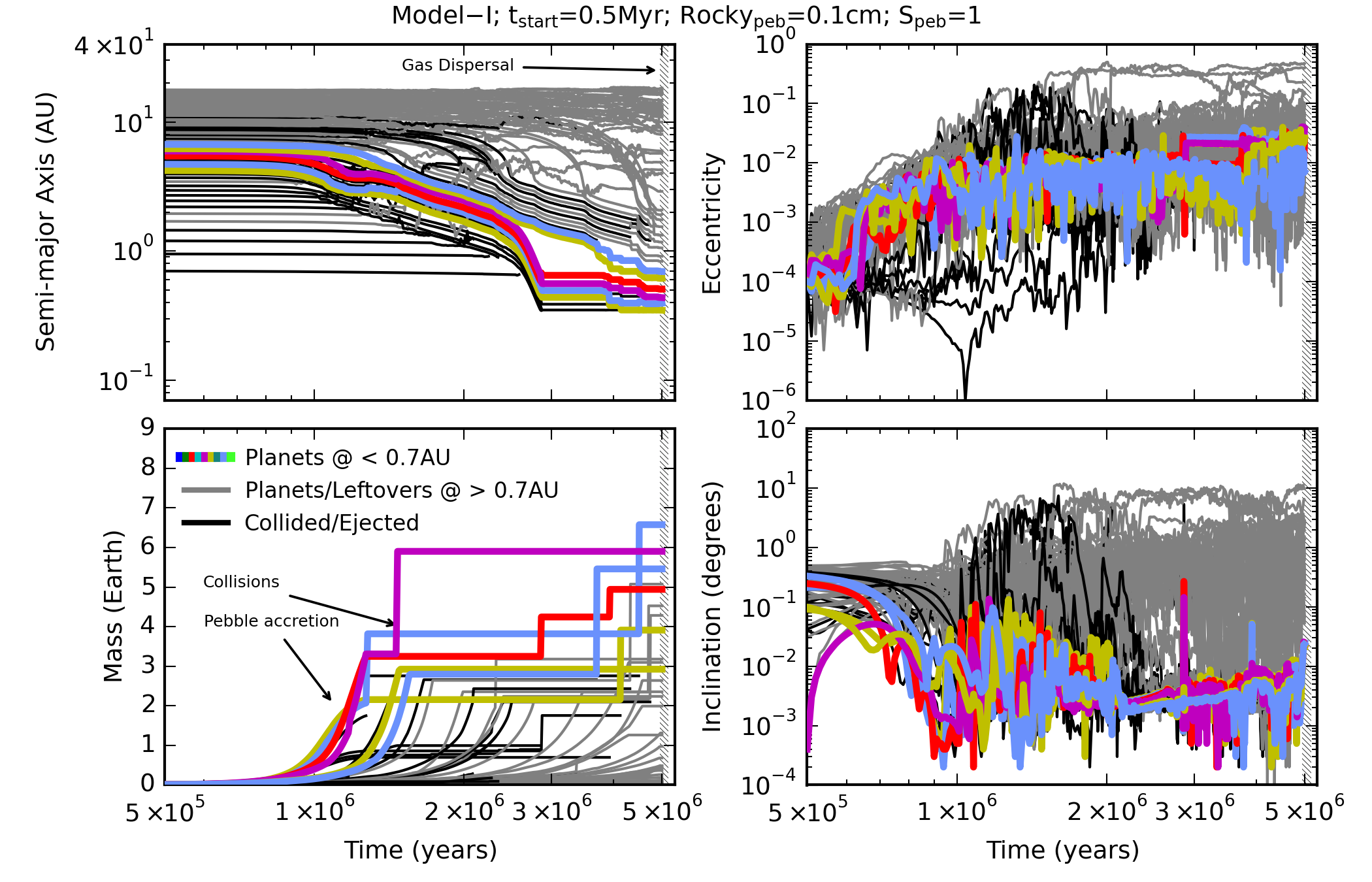}
\caption{Dynamical evolution of protoplanetary embryos in a simulation of Model I where ${\rm t_{start}= 0.5~Myr}$,  ${S_{{\rm peb}}= 1}$ and the size of the rocky pebbles is ${\rm Rocky_{{\rm peb}}=0.1~cm}$ (same simulation as in Fig. \ref{fig:snapshots}). The top left-hand panel shows the evolution of the semi-major axes. The top right-hand panel shows the evolution of the eccentricities. The bottom left-hand and right-hand panels show mass growth and the orbital inclinations, respectively. The dashed vertical line shows the end of the gas disk dispersal (corresponding to $\dot{M}\sim 10^{-9}M_{\odot}/yr$ in Eq. \ref{eq:2}). Colored lines show the final planets inside 0.7~AU. The gray line shows final planets and leftover protoplanetary embryos with orbits outside 0.7~AU. Leftover embryos are those with masses smaller than ${\rm \sim0.1~M_{\oplus}}$. Finally, the black line shows collided or ejected objects over the course of the simulation.}
    \label{fig:dynamics_S1}
\end{figure*}

Some planetary embryos in the simulation from Fig. \ref{fig:snapshots} grew massive enough to enter the outward migration region, yet they did not always migrate outwards. This is because the mutual gravitational interaction with other growing planetary embryos acts to excite eccentricities and to reduce the contribution of the co-orbital torque responsible for driving outward migration~\citep[see][]{bitschkley10,cossouetal13}. Outer nearby embryos may also act as a dynamical barrier for embryos  in the outward-migration region. As the disk evolves  further, the  outward migration region  quickly shrinks.
Typically, the outermost embryo in the outward migration zone is eventually caught by the inward-moving outer edge of the outward-migration region. We do not see a significant level of outward migration in these simulations (see Figure \ref{fig:dynamics_S1}). Instead, planetary embryos inside the outward migration region typically migrate very slowly inwards in a chain of mean-motion resonances. In this configuration, the outermost embryo sits at the edge of the outward migration region (see  panel corresponding to 2~Myr in Figure \ref{fig:dynamics_S1}).

Dynamical instabilities and collisions take place during the migration phase. As embryos in the outer disk grow fast and migrate inwards, they rapidly join the chain of embryos trapped at the outer edge of the outward migration region. This strong convergent migration tends to destabilize the resonant chain and promote additional collisions between embryos and further growth beyond the pebble isolation mass \citep[][]{wimarssonetal20}.  As the outward-migration region shrinks further, the more massive planetary embryos eventually get out of the region, becoming free to quickly migrate inwards.

At 3~Myr, several embryos have reached the inner edge of the disk to form a long resonant chain in mutual first-order mean motion resonances. At this time, some planets within 0.5~AU lie inside the outward-migration region. These planets have been pushed inwards by planets quickly migrating inwards with masses that fall above the outward-migration region. Figure \ref{fig:snapshots} shows that at the end of the gas disk phase, planetary embryos in the resonant chain at the inner edge of the disk have masses lower than ${\rm \sim 10~M_{\oplus}}$. 

The final compositions of all planets in the simulation from Fig.~\ref{fig:snapshots} are dominated by ice-rich material. Although this simulation starts with small planetary embryos inside the snow line (see snapshot corresponding to ${\rm t=0.5~Myr}$ in Figure \ref{fig:snapshots}) the growth of planetary embryos beyond the snow line is much more efficient. Planetary embryos growing beyond the snow line quickly reach pebble isolation mass blocking the pebble flux to inner regions, and consequently starving the innermost planetary embryos, in particular the rocky ones. Moreover, as larger planetary embryos migrate inwards they either collide with or scatter away small rocky planetary embryos.  Consequently, the final system of close-in planets has a water-rich composition.

Figure \ref{fig:dynamics_S1} shows the mass and orbital evolution of the simulation from Fig.~ \ref{fig:snapshots}. In Fig. \ref{fig:dynamics_S1}, the evolutions of the planets with final orbital semi-major axes within  ${\rm 0.7~AU}$  are shown in color. We highlight the innermost objects of the system because we are interested in the formation of close-in super-Earths. The Kepler sample is almost complete for transiting planets larger than Earth and orbital periods smaller than 200 days. Thus, we compare our results with Kepler observations taking into account only  planets with orbital periods shorter than about 200 days. As our simulations consider a solar-mass central star, this corresponds to planets with an orbital radius smaller than about 0.7~AU. In Figure \ref{fig:dynamics_S1}, embryos that underwent collisions or were ejected from the system are shown in bold black. The gray color is used to show planetary objects on final orbits outside 0.7~AU and also leftover planetary embryos. 

Figure \ref{fig:dynamics_S1} shows how the orbital eccentricities of embryos on average increase  from the start of the simulation to about 1~Myr as they grow by pebble accretion and consequently their mutual gravitational interaction becomes stronger. The gas tidal effects damp the orbital eccentricity and inclination and counter-balance the effects of mutual gravitational stirring. Orbits of larger protoplanetary embryos are more efficiently damped by the gas.  Figure \ref{fig:dynamics_S1} shows that only after ${\rm 1~Myr}$ planetary embryos have reached masses large enough to enter in a regime where type-I migration is reasonably fast. Migration leads to resonant shepherding and scattering events among planetary embryos. As consequence of close encounters, leftover planetary embryos were typically scattered by the largest migrating bodies when the latter moved towards the disk inner edge. Scattered bodies tend to reach dynamically excited orbits which may not be efficiently damped during the gas lifetime to allow residual growth by pebble accretion and migration \citep{levisonetal15b}. At the end of the gas disk phase, this simulation produced six planets with masses larger than ${\rm  2~M_{\oplus}}$ inside 0.7~AU. As shown in the panel illustrating mass evolution, the first phase of growth of these final planets is characterized by pebble accretion. However, collisions with other embryos, which  also take place during  early times (e.g., ${\rm \sim1-1.5~Myr}$), become more common when the planets approach the inner edge of the disk. Of course, this is a consequence of convergent migration  but also an effect of short dynamical timescales in the inner parts of the disk. At 3~Myr, multiple planetary embryos have reached the inner edge of the disk forming a long resonant chain.
\begin{figure*}
\centering
\includegraphics[scale=.9]{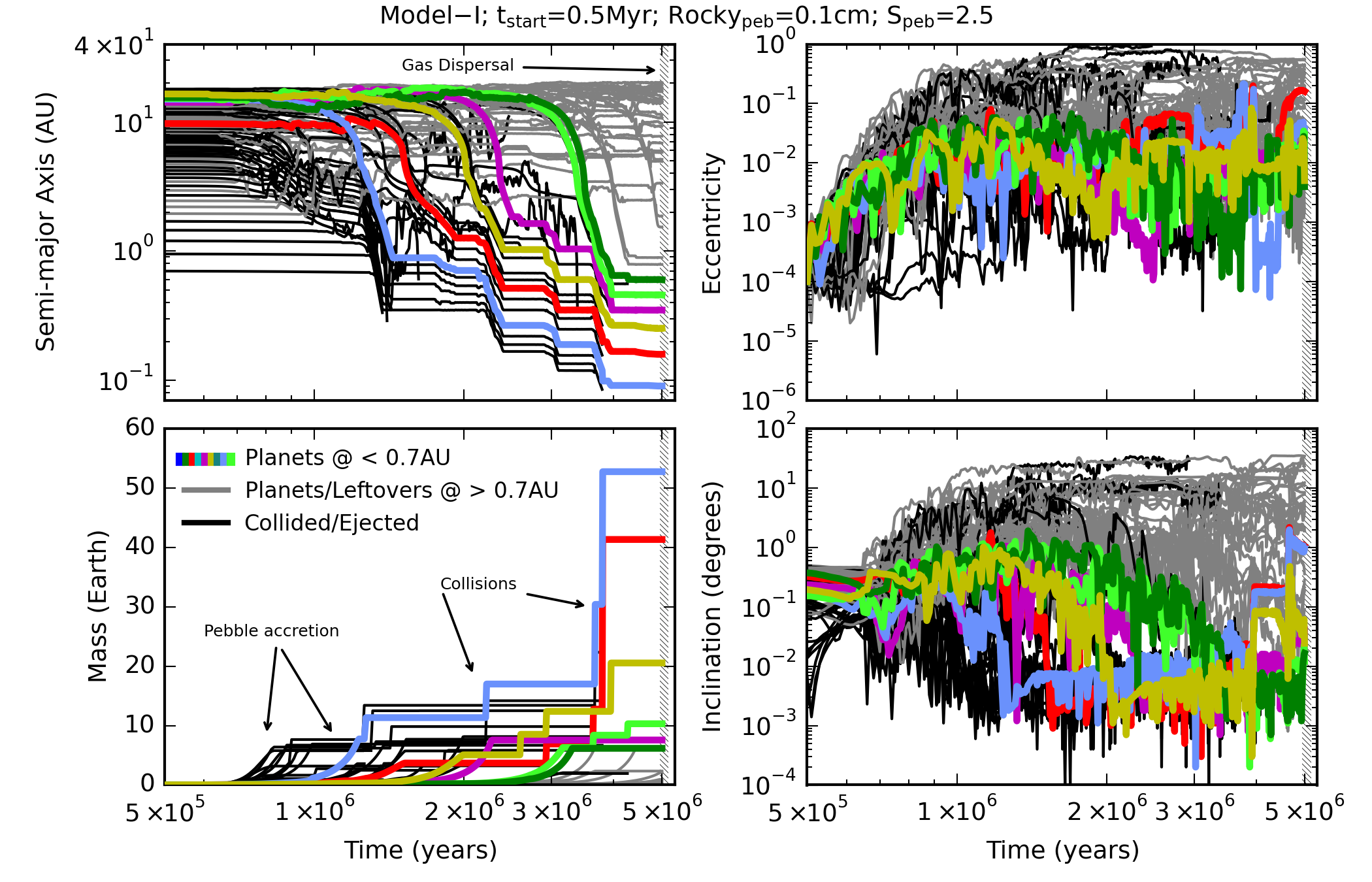}
\caption{Dynamical evolution of a simulation similar to the one shown in Fig. \ref{fig:dynamics_S1}, but using ${S_{{\rm peb}}= 2.5}$. The larger pebble flux leads to faster growth and thus larger planets via pebble accretion. Even before the disk dissipates, planetary embryos collide and form massive bodies.}
    \label{fig:dynamics_S2.5}
\end{figure*}


Figure \ref{fig:dynamics_S2.5} shows the evolution of a simulation like the one in Figure \ref{fig:dynamics_S1} but for a case where the pebble flux is 2.5 times higher. Overall, the dynamical behavior of protoplanetary embryos in this simulation is similar to that shown in Figure \ref{fig:dynamics_S1}. However, as expected, a higher pebble flux promotes a much faster growth of protoplanetary embryos by pebble accretion. In this simulation, during the first ${\rm \sim1.5~Myr}$ protoplanetary embryos have already reached masses of greater than ${\rm \sim6~M_{\oplus}}$ by pure accretion of pebbles. We note that, broadly speaking, this corresponds to the final masses of planetary objects at the end of the gas disk phase in the simulation of Figure \ref{fig:dynamics_S1}.  As these larger planetary objects migrate inwards, they collide with each other and grow even further. Most of these collisions happen inside the disk inner cavity as this region becomes overcrowded because of the successive arrival of planetary embryos migrating from more distant regions of the disk. At the end of the gas disk phase, the most massive planetary embryo in this simulation is ${\rm \sim50~M_{\oplus}}$. Thus, the final masses of planetary objects in the simulations of Figures \ref{fig:dynamics_S1} and \ref{fig:dynamics_S2.5} are drastically different. 

In our simulations we neglect gas accretion onto protoplanetary embryos. The three most massive final planets produced in the simulation of Figure \ref{fig:dynamics_S2.5}  -- with masses larger than ${\rm \sim20~M_{\oplus}}$ -- represent very good candidates for accretion of massive gas atmospheres to become gas giants. We do not model the formation of gas giant planets in this paper, but we dedicated a companion paper  \citep{bitschetal18c} to address this issue more carefully. In this work, whenever a planetary embryo becomes more massive than ${\rm \sim15~M_{\oplus}}$ during the gas disk phase we flag it as a giant planet core rather than a super-Earth. We use the simulations of this section only to identify the pebble fluxes that lead to the formation of super-Earths rather than giant planet cores. Once we know which pebble fluxes produce planets of ${\rm \leq15~M_{\oplus}}$ during the gas disk phase, in Section \ref{sec:super-Earth} we  use this information to conduct a larger set of simulations  dedicated to modeling the formation and long-term dynamical evolution of super-Earths. However, we note that our assumed threshold mass of ${\rm \sim15~M_{\oplus}}$ for transition to runaway gas accretion (our flag of giant planet core) depends on dust opacity in the envelope, which is uncertain \citep{pollacketal96b,ormeletal15,leechiang15,lambrechtslega17}.


\begin{figure*}
\centering
\includegraphics[scale=.15]{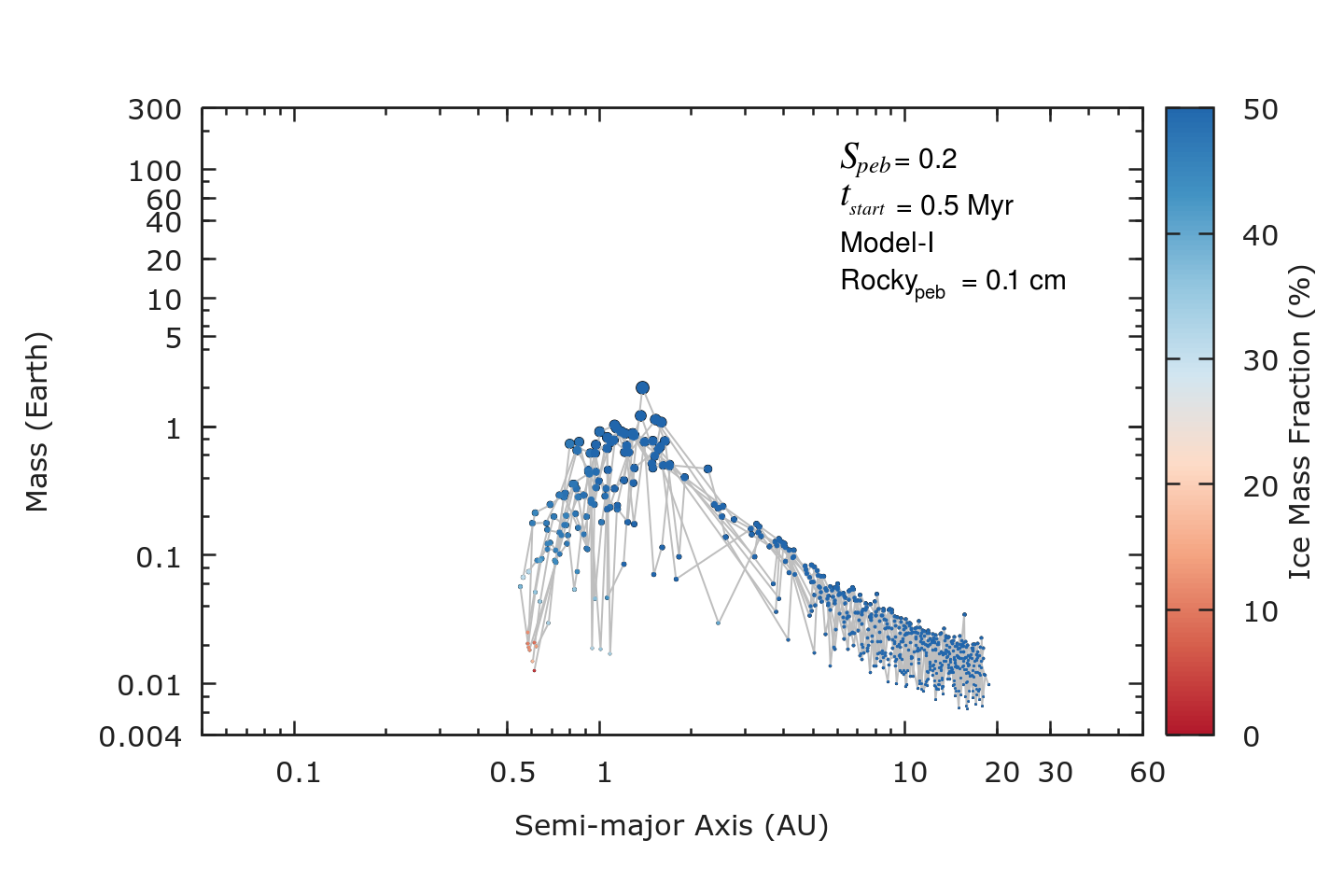}
\includegraphics[scale=.15]{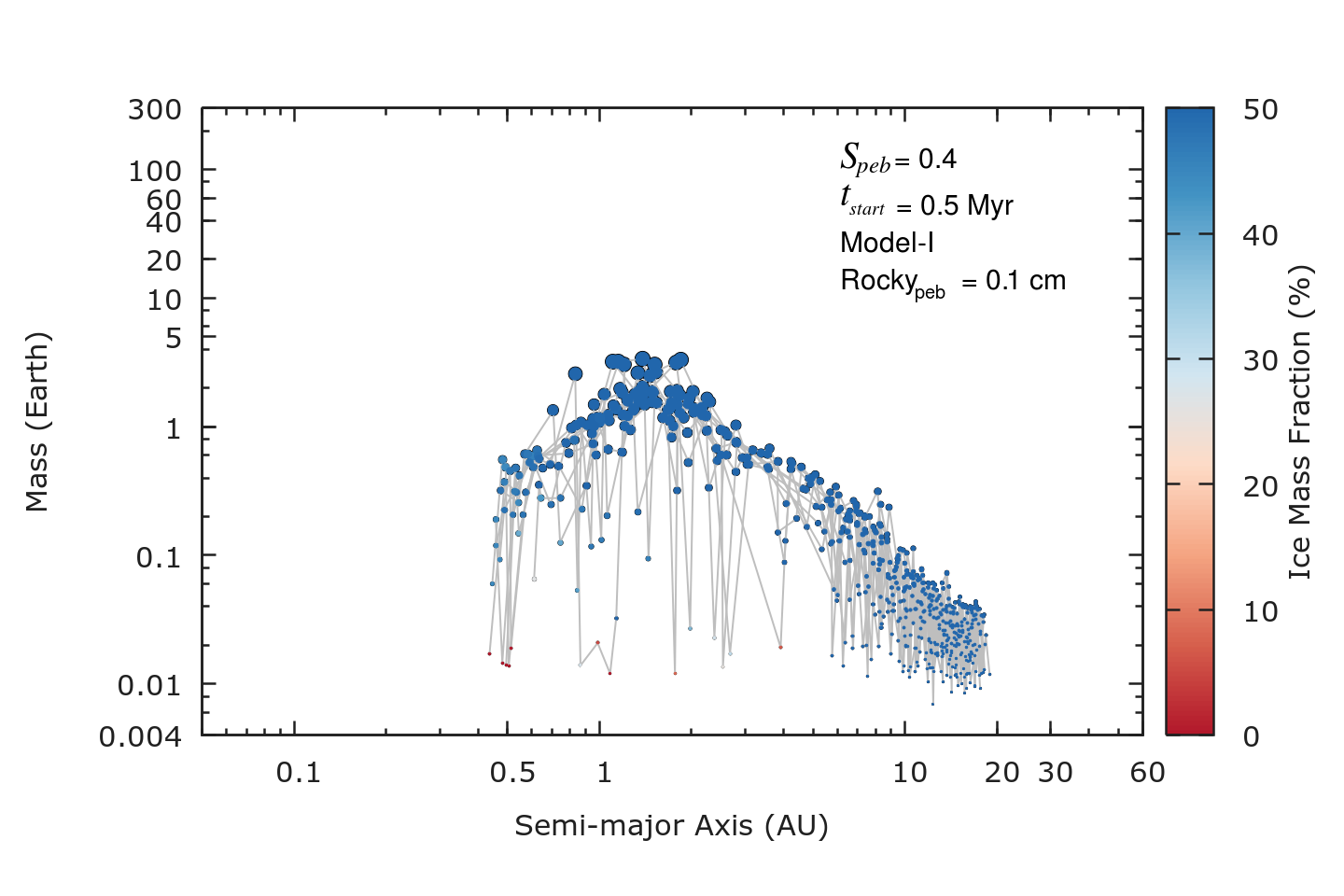}
\includegraphics[scale=.15]{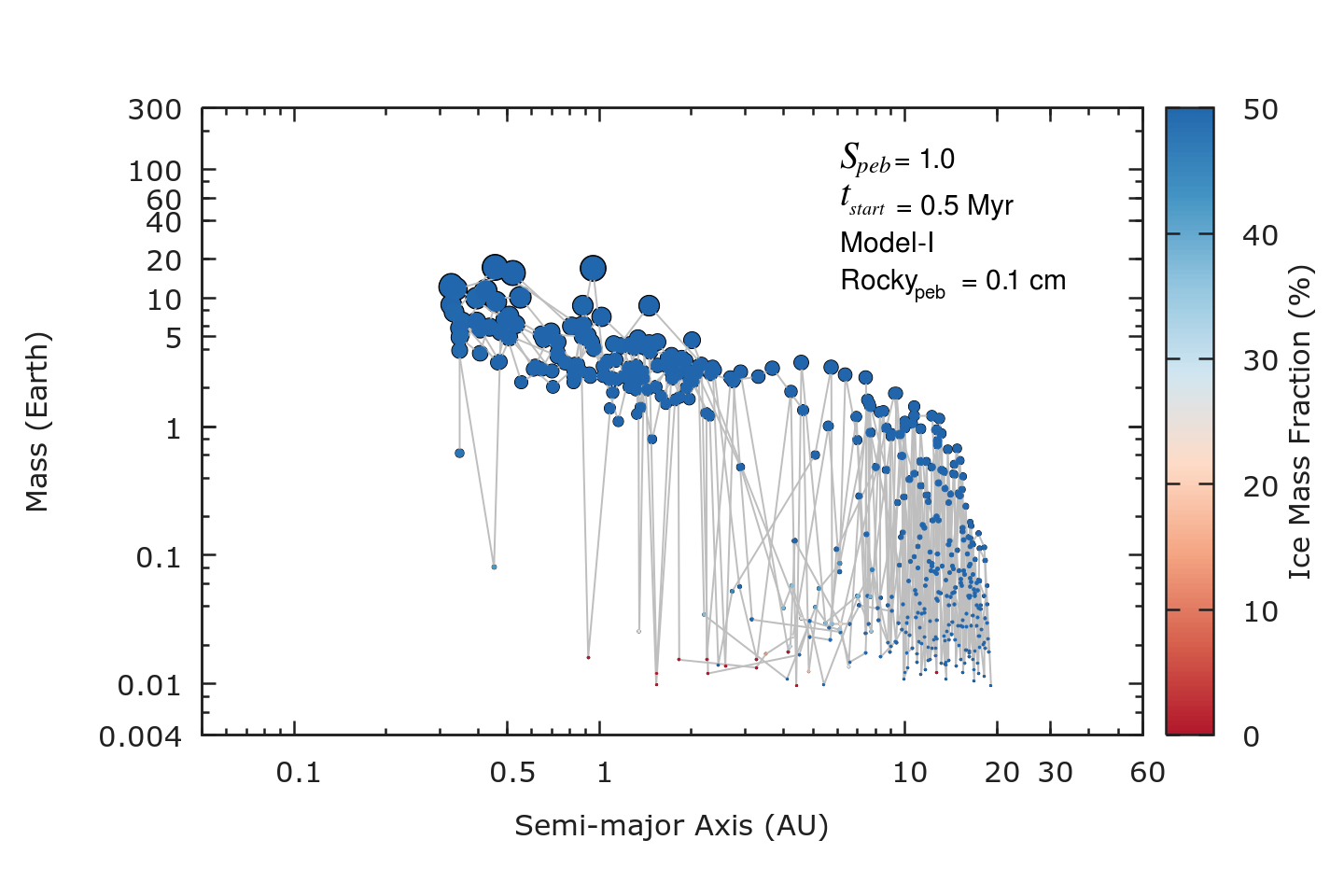}
\includegraphics[scale=.15]{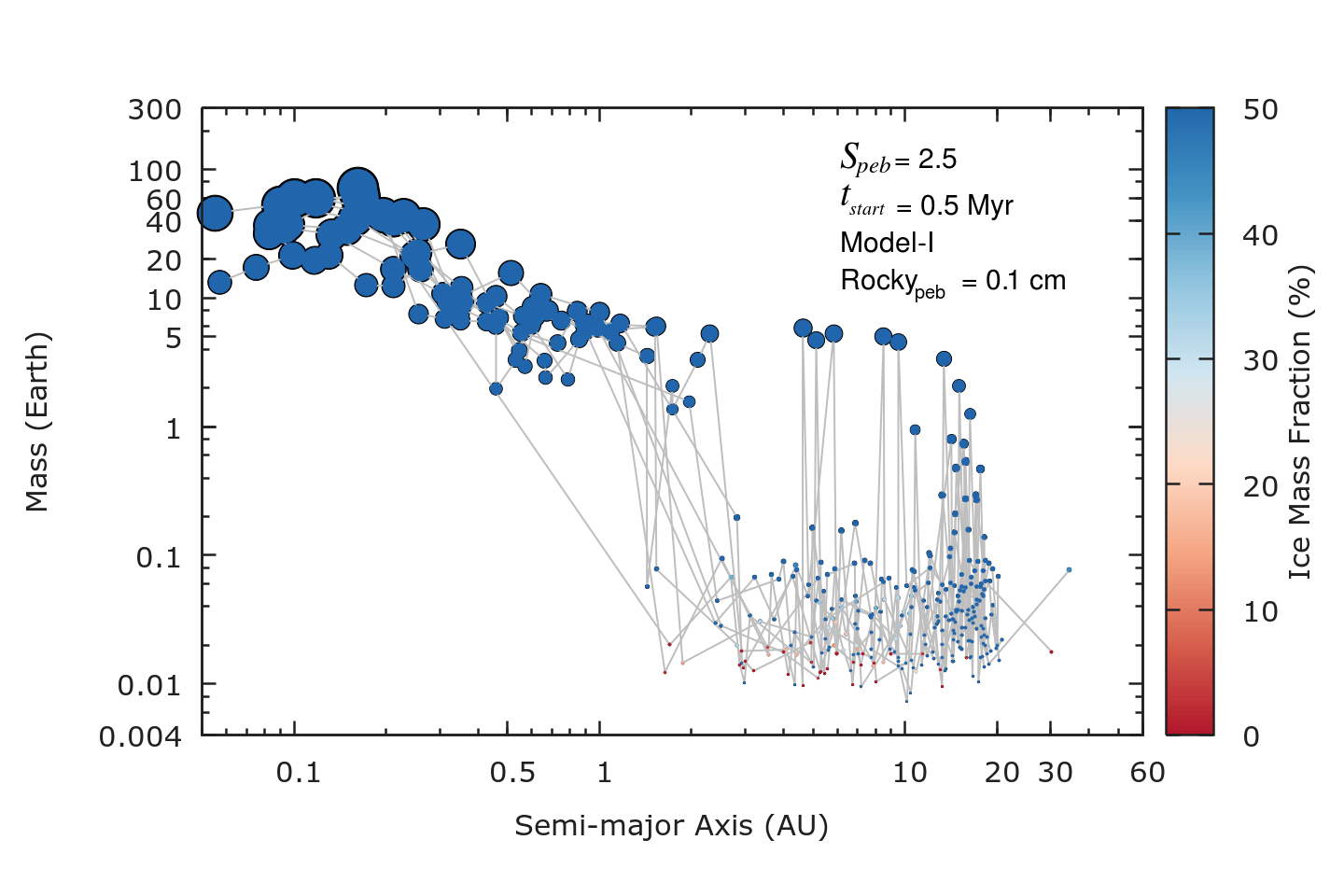}
\caption{Final masses of protoplanetary embryos in simulations of  Model I where ${\rm t_{start}=0.5~Myr}$ and the size of the rocky pebbles is ${\rm Rocky_{{\rm peb}}=0.1~cm}$ for different pebble fluxes ${S_{{\rm peb}}}$. Each panel shows the outcome of ten different simulations with slightly different initial conditions. Each final planetary object is represented by a colored dot where the color represents its final ice mass fraction. Planetary objects belonging to the same simulation are connected by lines. The integrated pebble fluxes are  ${\rm \sim 39~M_{\oplus}}$, ${\rm \sim 78~M_{\oplus}}$, ${\rm \sim 194~M_{\oplus}}$, and ${\rm \sim 485~M_{\oplus}}$ for  ${S_{{\rm peb}}=0.2,~0.4,~1,~{\rm and}~2.5}$, respectively. The efficiency of pebble accretion (fraction of integrated pebble flux used to build planets) is about 17\%, 25\%, 32\%, and 30\% for  $S_{\rm peb}=0.2$, 0.4, 1, and~2.5, respectively. }
    \label{fig:panels_fluxes_model_1}
\end{figure*}

%

Figure \ref{fig:panels_fluxes_model_1} shows the final masses of planetary embryos in all simulations of Model I with scaling pebble fluxes  varying from ${S_{{\rm peb}}=0.2}$ to 2.5. Each panel of Figure \ref{fig:panels_fluxes_model_1} shows the outcome of ten simulations in a diagram depicting semi-major axis versus mass. Figure \ref{fig:panels_fluxes_model_1} shows a clear trend: the final planet masses are higher for higher pebble fluxes (larger ${S_{{\rm peb}}}$). Planetary embryos growing in low-pebble-flux environments do not grow massive enough to migrate to the disk inner edge. We recall that in these particular simulations, planetary embryos are initially distributed from 0.7 to 20~AU. For ${S_{{\rm peb}}=0.2}$ and ${S_{{\rm peb}}=0.4}$ the amount of radial migration of planetary embryos is typically modest. In these two cases, most planetary embryos remain at sub-Earth mass and beyond 0.5~AU (see also Paper I for simulations showing the long-term dynamical evolution of a similar population of rocky embryos). Earth-mass or more-massive planets at 1-2 AU produced for ${S_{{\rm peb}}=0.2}$ and 0.4 (for Model I)  are planets that started forming farther out and migrated down to their final position. We compare our results with those of Paper I in the following section.
Before discussing the details of these results we also recall that the integrated pebble flux is the flux of pebbles during the entire course of the simulation. A simulation with ${S_{{\rm peb}}=0.2}$ and ${\rm t_{start}=0.5~Myr}$, for example, features an integrated pebble flux of ${\rm \sim 0.2\times194~M_{\oplus}=39~M_{\oplus}}$. A simulation with ${S_{{\rm peb}}=1}$ and ${\rm t_{start}=0.5~Myr}$ results in an integrated pebble flux of ${\rm \sim 1\times194~M_{\oplus}}$ (see Figure \ref{fig:flux}). In both cases,  only a fraction of the integrated pebble flux is used to build planets. In Figure \ref{fig:panels_fluxes_model_1}  these numbers are 17\%, 25\%, 32\%, and 30\% for  $S_{\rm peb}=0.2$, 0.4, 1, and~2.5, respectively. 

Figure \ref{fig:panels_fluxes_model_1} also shows that our nominal pebble flux ${S_{{\rm peb}}=1}$ results in the formation of planets that do indeed migrate to the inner edge of the disk  (assumed at ${\rm \sim0.3~AU}$ in these simulations). Planets inside ${\rm 0.7~AU}$ in simulations with  ${S_{{\rm peb}}=1}$ have masses lower than ${\rm \sim10-15~M_{\oplus}}$ (see also Figure \ref{fig:dynamics_S1}). A further factor of 2.5 increase in the pebble flux results in the formation of multiple planets  with masses as large as ${\rm \sim40-50~M_{\oplus}}$ (see also  Fig. \ref{fig:dynamics_S2.5}). We note that planets do not necessarily stay beyond the disk inner edge. Planets migrating to the inner edge of the disk pile up in long resonant chains. In some cases, the innermost planets  anchored at the inner edge of the disk are pushed inside the disk cavity \citep{cossouetal14,izidoroetal17,brasseretal18,carreraetal19}. Some planets are pushed to distances as close as 0.06~AU from the star.

Our results also show that a  simple difference of a factor of approximately two in pebble flux (from ${S_{{\rm peb}}=1}$ to 2.5) is enough to bifurcate  the evolution of our planetary systems from one leading to a system of super-Earth mass planets to one hosting several massive protoplanetary cores which are very likely to become gas giants. For example, several planets forming in our simulations with ${S_{{\rm peb}}=2.5}$ are more massive than the Solar System ice giants by a factor of a few.  Although not shown in Figure \ref{fig:panels_fluxes_model_1}, for completeness, we also inspected the results of simulations considering higher scaling pebble fluxes of ${S_{{\rm peb}}=5}$ and 10. In these simulations, the final planets reaching the inner edge of the disk are as massive as ${\rm 100~M_{\oplus}}$ due to convergent migration towards the disk inner edge and successive collisions. \citet{bitschetal18c} present the results of simulations with higher pebble fluxes where gas accretion onto planetary cores is consistently modeled.

\subsection{Model
II}

\begin{figure*}
\centering
\includegraphics[scale=.15]{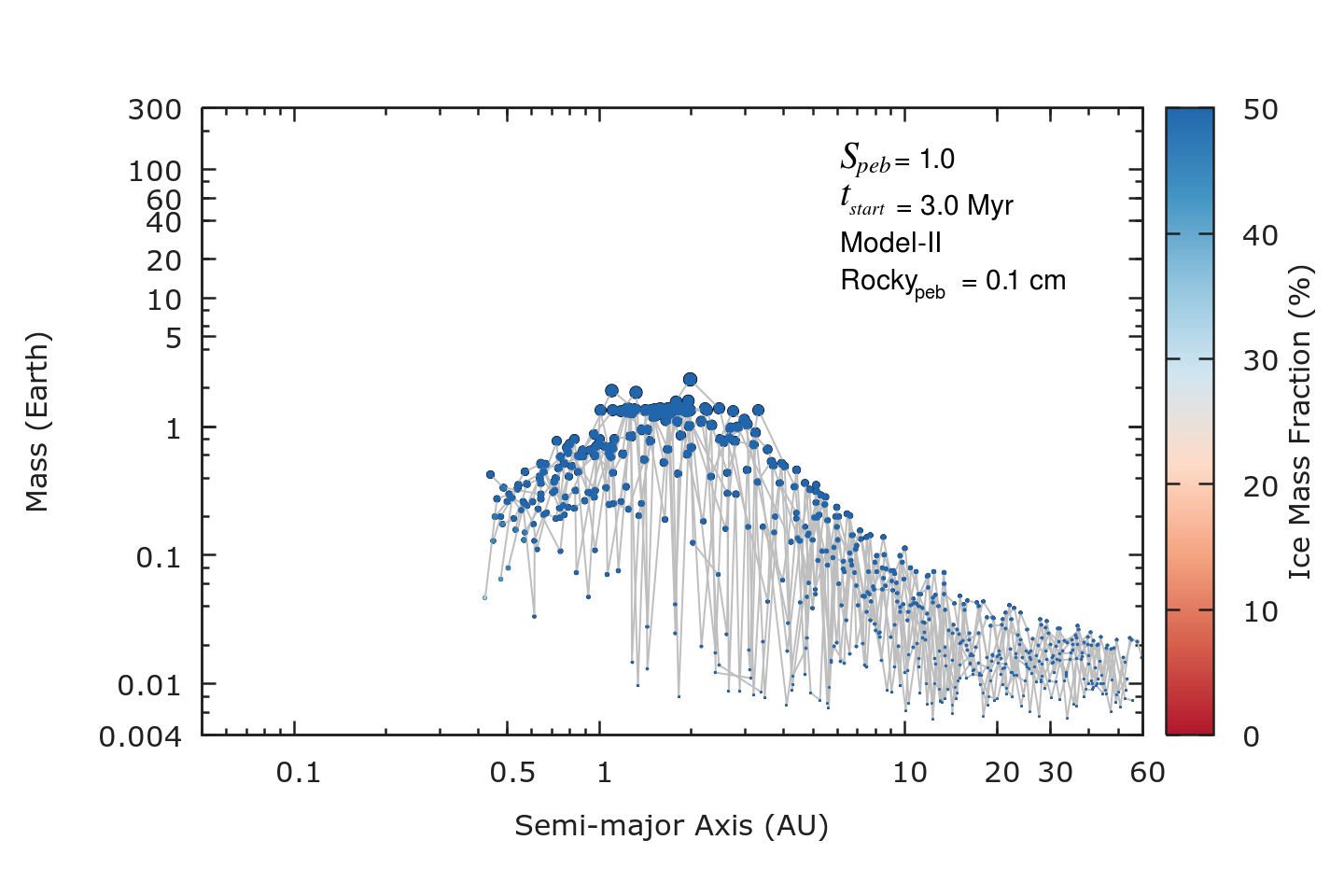}
\includegraphics[scale=.15]{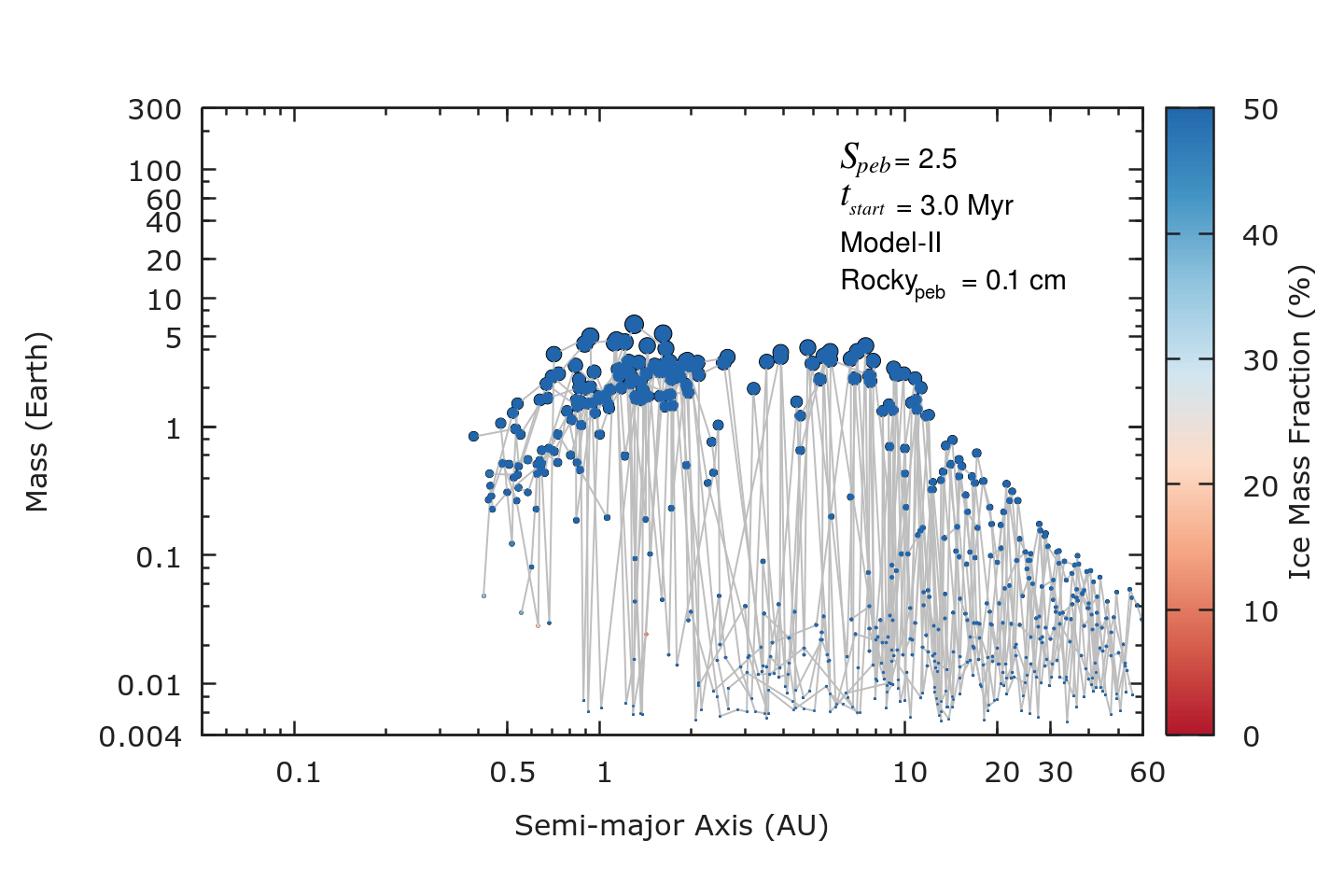}
\includegraphics[scale=.15]{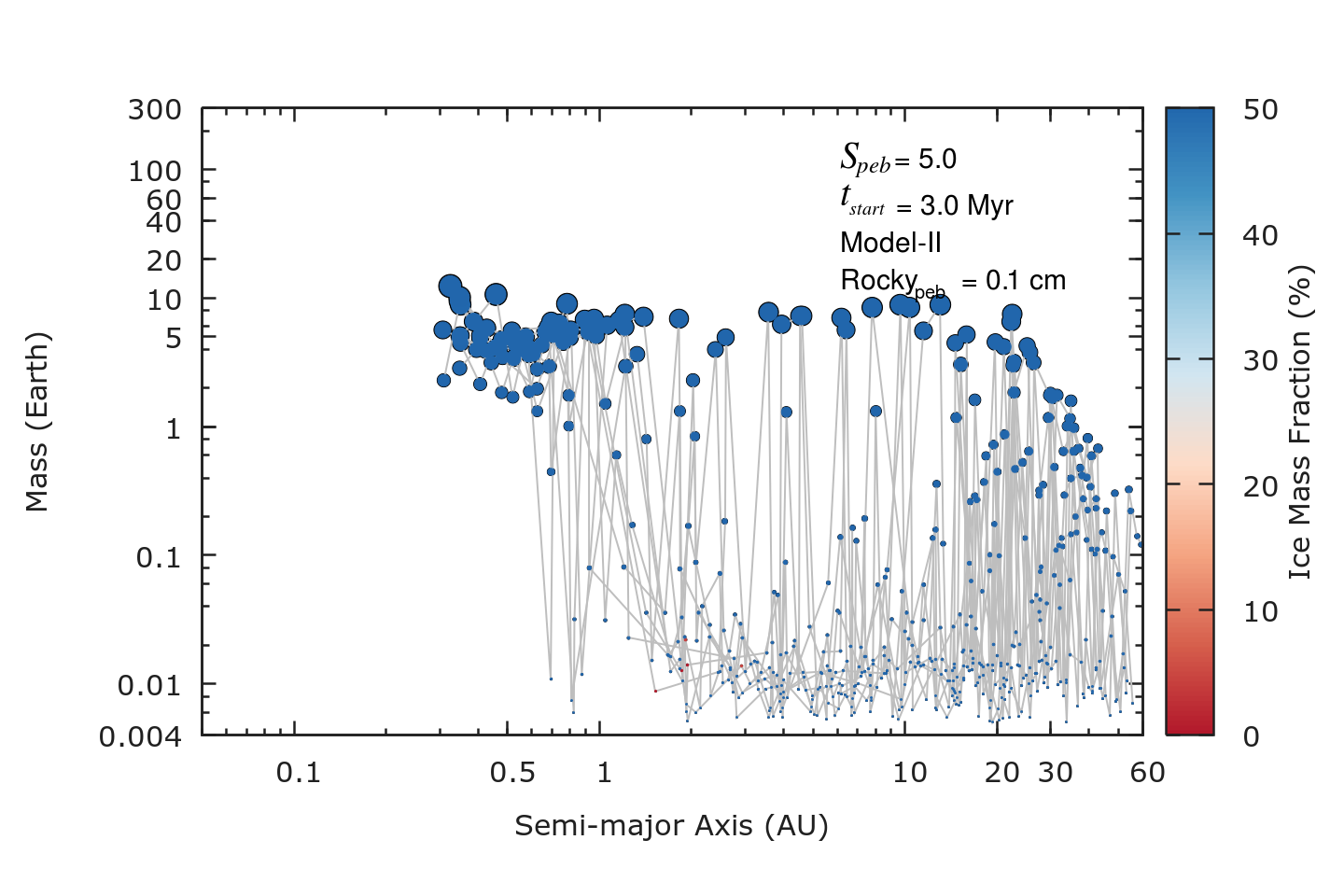}
\includegraphics[scale=.15]{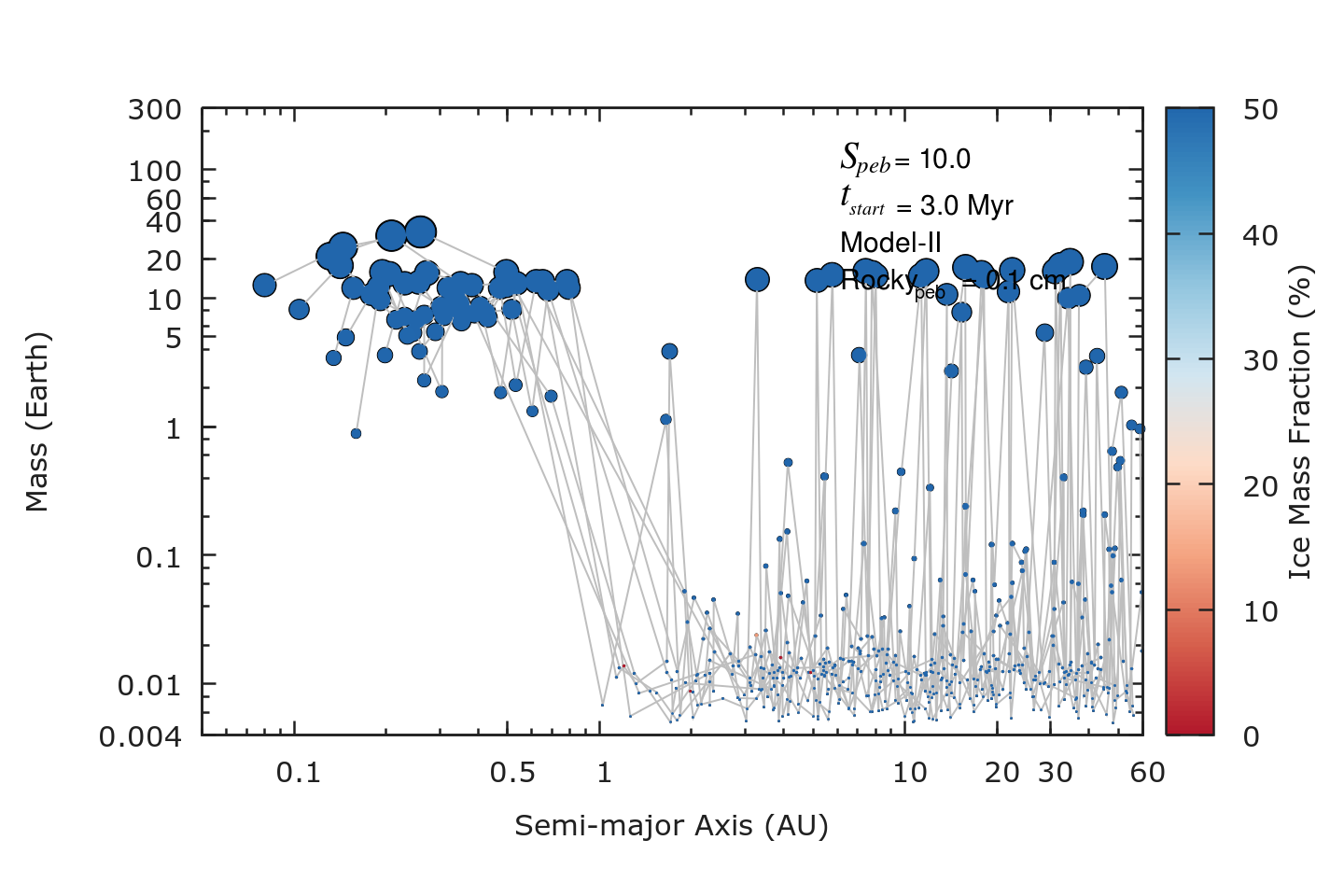}
\caption{Same as Figure \ref{fig:panels_fluxes_model_1} but using  Model II where ${\rm t_{start}=3.0~Myr}$ and the size of the rocky pebbles is ${\rm Rocky_{{\rm peb}}}$=0.1~cm for different pebble fluxes ${S_{{\rm peb}}}$, as annotated in each panel.  As these simulations start at ${\rm t_{start}=3.0~Myr}$, the integrated pebble fluxes are  ${\rm \sim 30~M_{\oplus}}$, ${\rm \sim 75~M_{\oplus}}$, ${\rm \sim 150~M_{\oplus}}$, and ${\rm \sim 300~M_{\oplus}}$ for  ${S_{{\rm peb}}=1,~2.5,~5,~{\rm and}~10}$, respectively. The efficiency of pebble accretion is about 49\%, 43\%, 39\%, and 33\% for  ${S_{{\rm peb}}=1,~2.5,~5,~{\rm and}~10}$, respectively.}
    \label{fig:panels_fluxes_model_2}
\end{figure*}

Figure \ref{fig:panels_fluxes_model_2} shows the distribution of planets produced in simulations of  Model
II, which is similar to Model I but the seeds are assumed to appear only at ${\rm t_{start}=3.0~Myr}$, i.e., towards the end of the  lifetime of the disk (5 Myr). For the nominal pebble flux scaling ${S_{{\rm peb}}=1}$, the integrated drifted pebble flux for a disk with ${\rm t_{start}= 0.5~Myr}$ is about ${\rm \sim194~M_{\oplus}}$. In a disk with ${\rm t_{start}= 3.0~Myr}$ and ${S_{{\rm peb}}=1}$ the total integrated pebble flux is about ${\rm \sim 30~M_{\oplus}}$ (see Figure \ref{fig:flux}). This is the total reservoir of mass available for planetary growth.

We note that the results from Models I and II  shown in Figures \ref{fig:panels_fluxes_model_1} and 7, respectively, are not dramatically different: with ${S_{{\rm peb}}=0.2}$ and ${\rm t_{start}= 0.5~Myr}$ versus ${S_{{\rm peb}}=1}$ and ${\rm t_{start}= 3.0~Myr}$. For example, the largest planets in the top left-hand panel of Figure \ref{fig:panels_fluxes_model_2} are around 1~AU and  have masses of between 1 and ${\rm 2~M_{\oplus}}$. Planets around 1~AU in the top left-hand panel of Figure \ref{fig:panels_fluxes_model_1}  have also masses in this range.  The reason for this is that the total drifted pebble flux is about the same in these two setups (${\rm \sim 39~M_{\oplus}}$ for Model-I  with ${S_{{\rm peb}}=0.2}$ and ${\rm \sim 30~M_{\oplus}}$ for Model
II with ${S_{{\rm peb}}=1.0}$; note that they have different ${\rm t_{start}}$).  However, even though the integrated pebble fluxes are relatively similar,  the Stokes numbers of the pebbles (see Eq. \ref{stokesnumber}) are different because of the different pebble and gas surface densities at the planetary location, which can lead to different growth. This effect becomes more clear in the outer regions of the disk when comparing, for example, the top right panel of Figure \ref{fig:panels_fluxes_model_1} (${S_{{\rm peb}}=0.4}$) with the top right  panel of Fig. \ref{fig:panels_fluxes_model_2} (${S_{{\rm peb}}=2.5}$). We note that these two cases have also comparable integrated pebble fluxes (${\rm \sim 75~M_{\oplus}}$). In the top right-hand panel of Figure \ref{fig:panels_fluxes_model_1}, the typical final mass of planetary embryos around 10~AU is sub-Mars mass while those in top right-hand panel of Figure \ref{fig:panels_fluxes_model_2} have masses of about ${\rm \sim 2~M_{\oplus}}$, at least a factor of 20 larger.
 

Figure \ref{fig:panels_fluxes_model_2} shows that increasing pebble flux from ${S_{{\rm peb}}=2.5}$ to 5 is enough to promote the delivery of planetary embryos to the inner edge of the disk (see bottom left-hand panel of Figure \ref{fig:panels_fluxes_model_2}). The final planets anchored at the inner edge of the disk in this case have typical masses of a few Earth masses to Neptune mass. A further increase in the pebble flux with ${S_{{\rm peb}}=10}$ (see bottom left-hand panel of Figure \ref{fig:panels_fluxes_model_2}) promotes the formation of planetary embryos with masses greater than ${\rm \sim20~M_{\oplus}}$. This reinforces our previous finding of Figure \ref{fig:panels_fluxes_model_1} where a simple increase in the pebble flux by a factor of two is enough to bifurcate the growth pattern of planetary embryos from one leading to super-Earths or hot-Neptunes to another producing massive planetary cores which would be very likely to become gas giants \citep{lambrechtslega17}. 

\subsection{Model
III}

In Model III we test an extreme scenario in which planetary seeds are assumed to have only formed in the innermost rocky regions of the disk (see Table \ref{tab:model}). Our goal is not only to inspect the final masses of the planets as a function of the pebble flux but also to study how the initial distribution of planetary embryos may impact the final planet compositions. We note that all our simulations of Models I and
II failed to grow and deliver rocky planetary embryos to the disk inner edge. We naively expect this to be more easily achieved if the initial distribution of planetary embryos is restricted to the region inside the snow line. 
\begin{figure*}
\centering
\includegraphics[scale=.15]{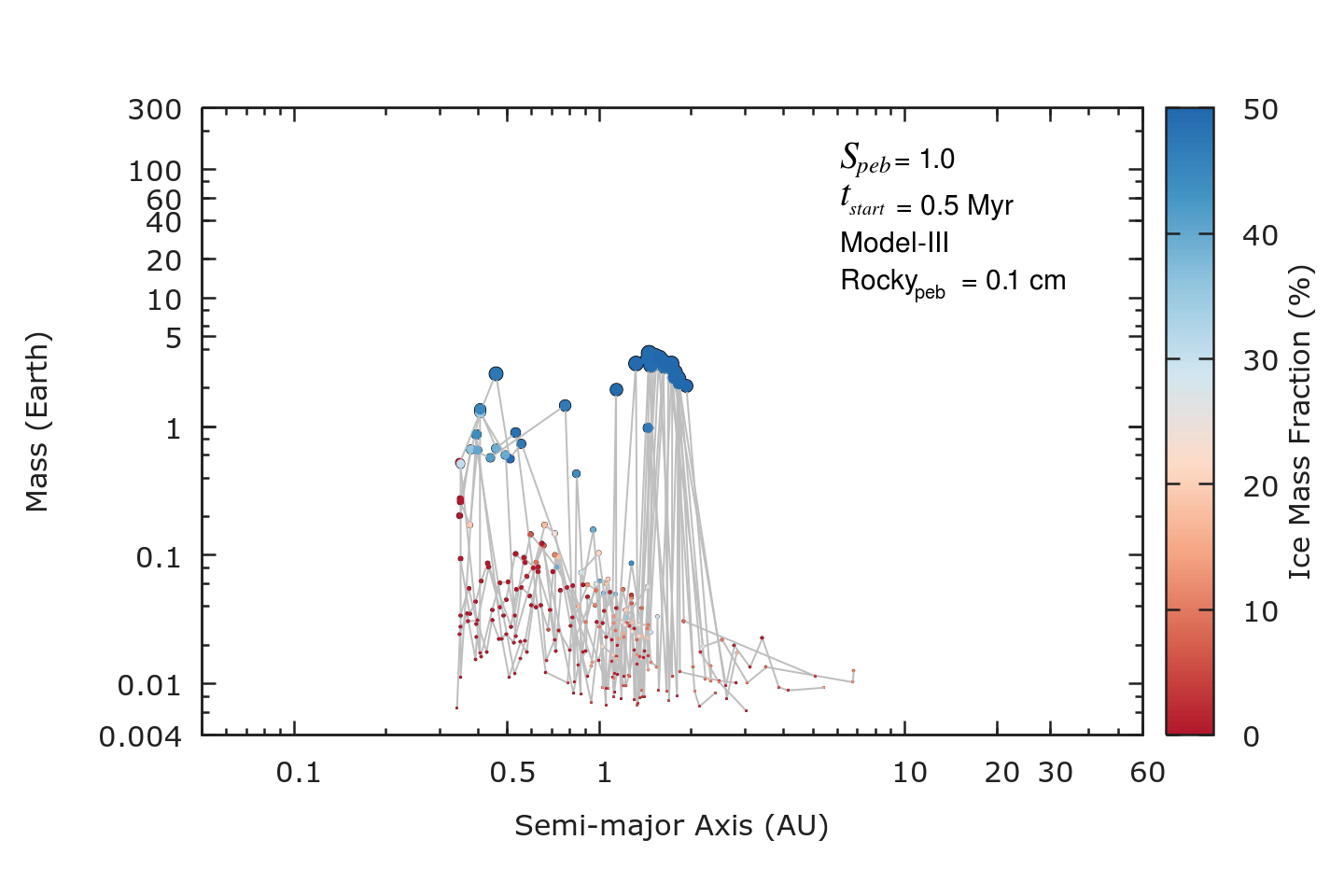}
\includegraphics[scale=.15]{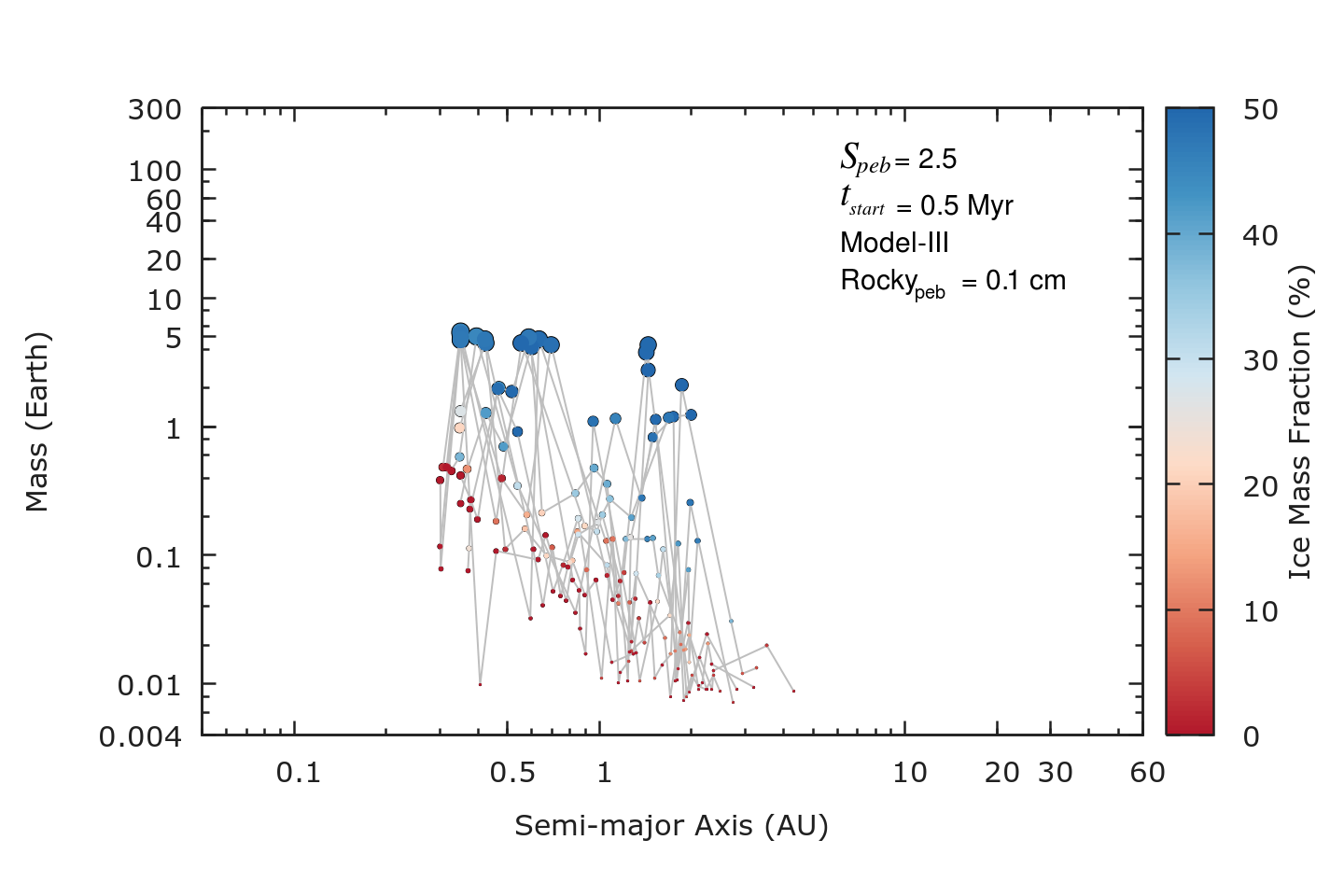}
\includegraphics[scale=.15]{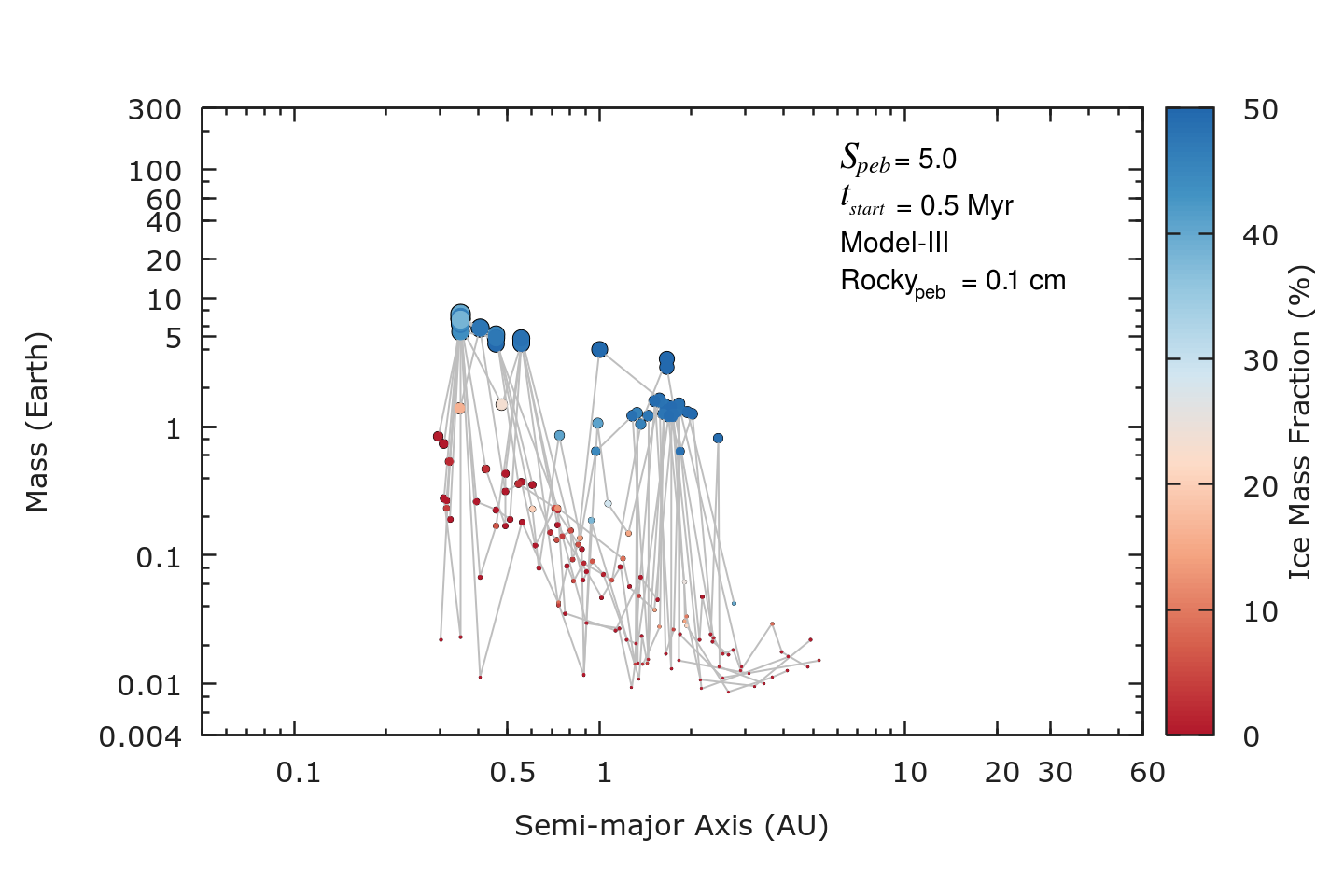}
\includegraphics[scale=.15]{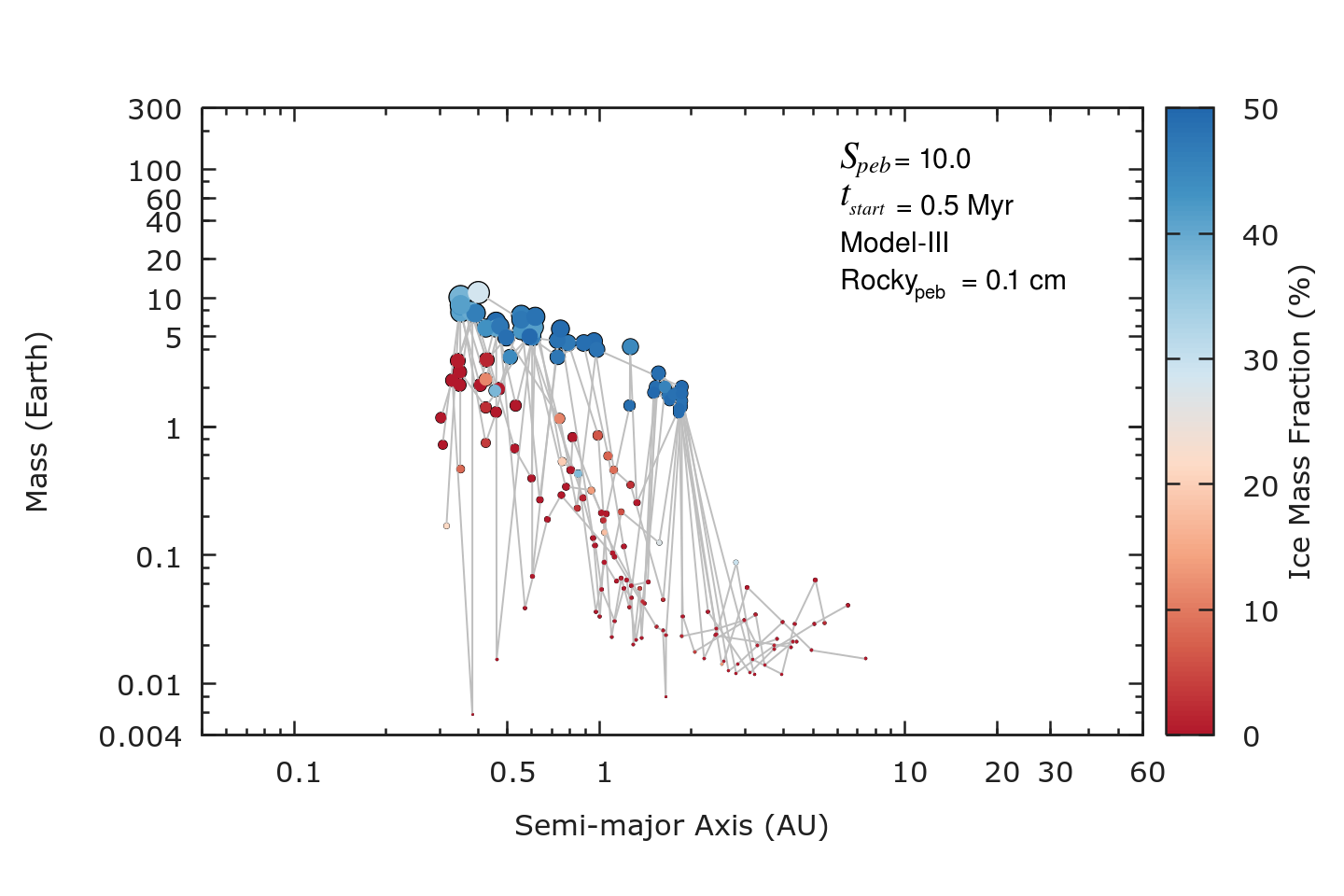}
\caption{Same as Figure \ref{fig:panels_fluxes_model_2} except that we now use Model
III, where the embryos are initially spread from 0.3 to 2.0 AU. The integrated pebble fluxes are  ${\rm \sim 194~M_{\oplus}}$, ${\rm \sim 485~M_{\oplus}}$, ${\rm \sim 970~M_{\oplus}}$, and ${\rm \sim 1940~M_{\oplus}}$ for  ${S_{{\rm peb}}=1,~2.5,~5,~{\rm and}~10}$, respectively. The efficiency of pebble accretion is about 4.3\%, 2.1\%, 1.6\%, and 1.4\% for  ${S_{{\rm peb}}=1,~2.5,~5,~{\rm and}~10}$, respectively. Clearly,  the most massive bodies in each simulation have a large water fraction.}
    \label{fig:panels_fluxes_model_3_mm}
\end{figure*}

\begin{figure*}
\centering
\includegraphics[scale=.15]{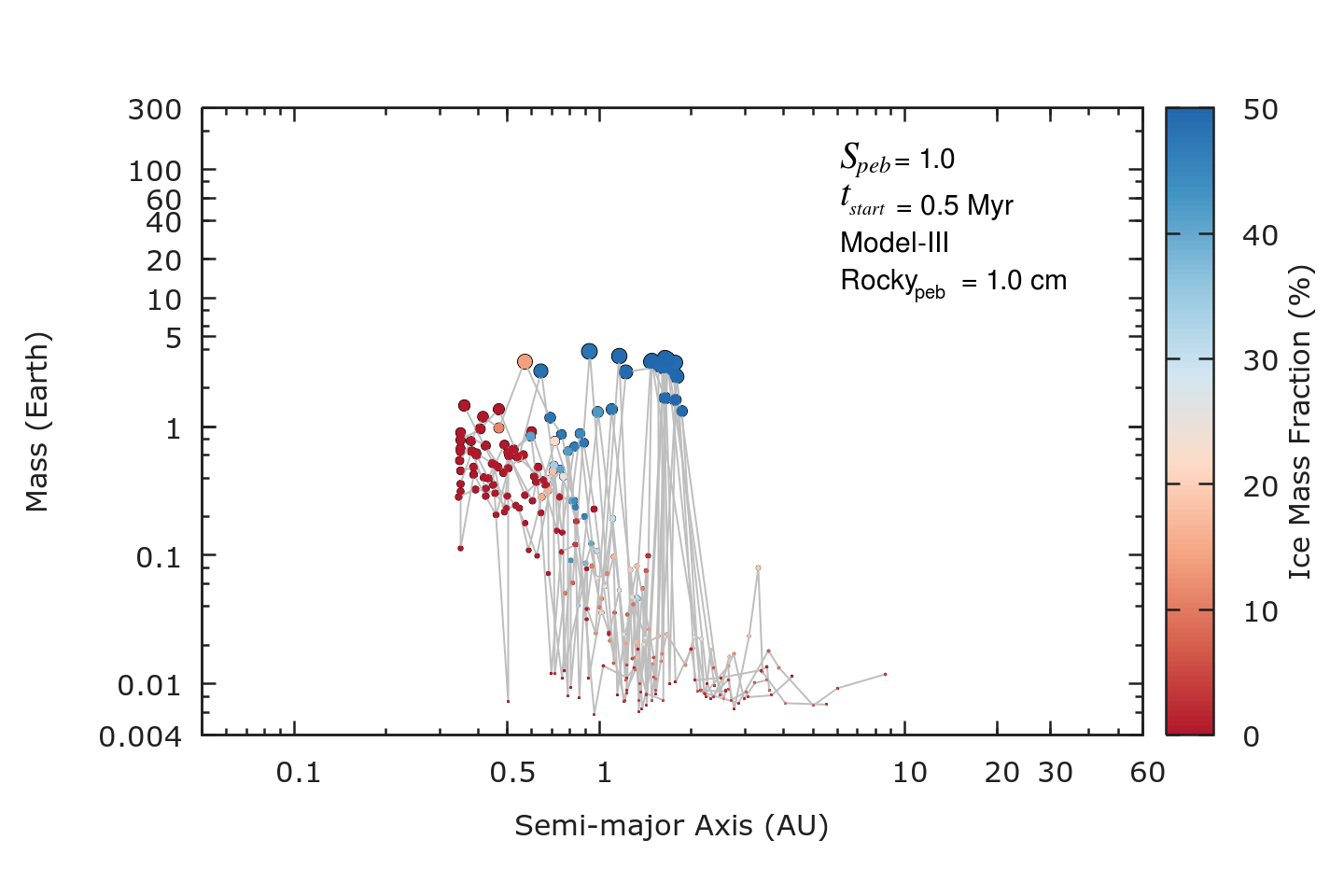}
\includegraphics[scale=.15]{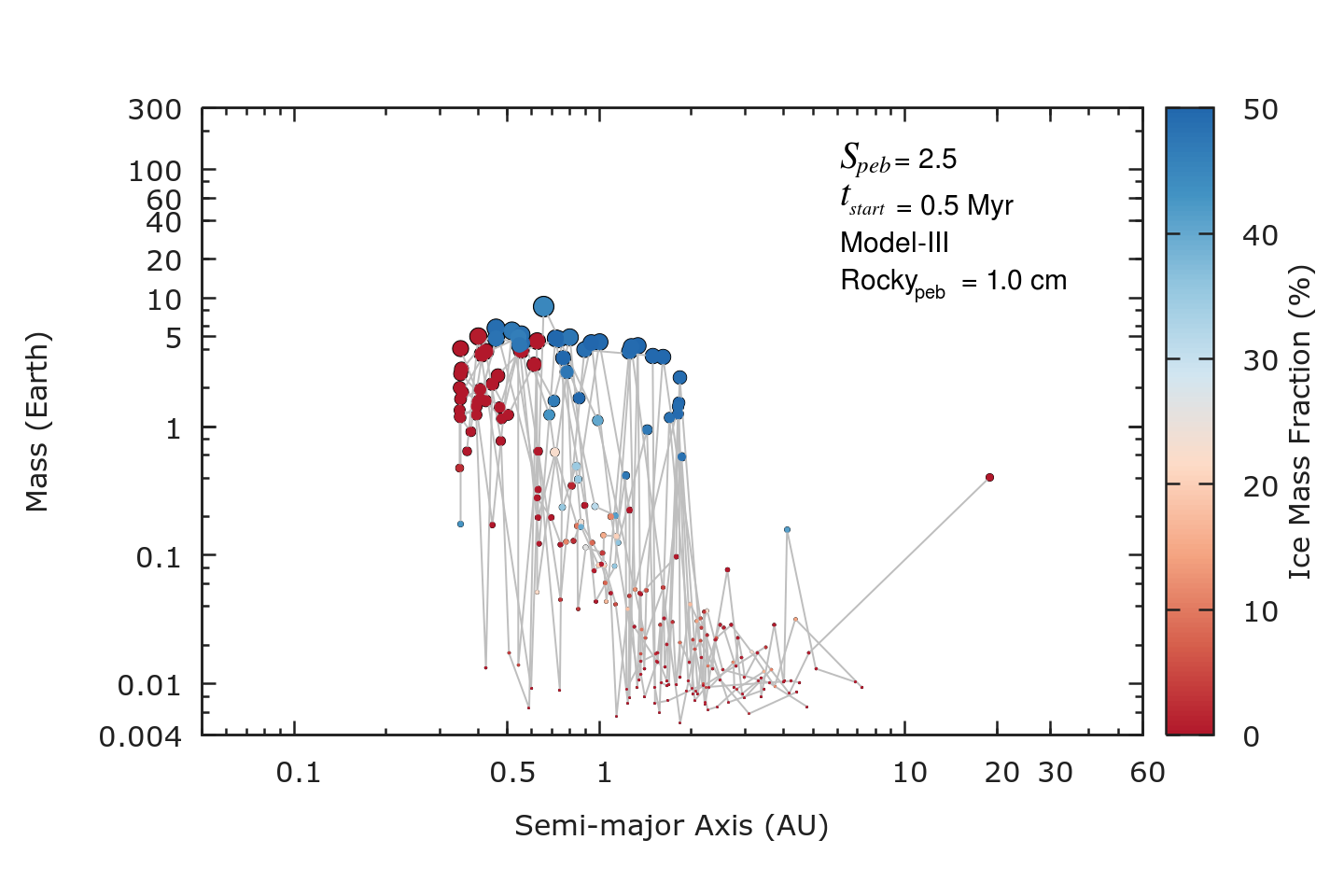}
\includegraphics[scale=.15]{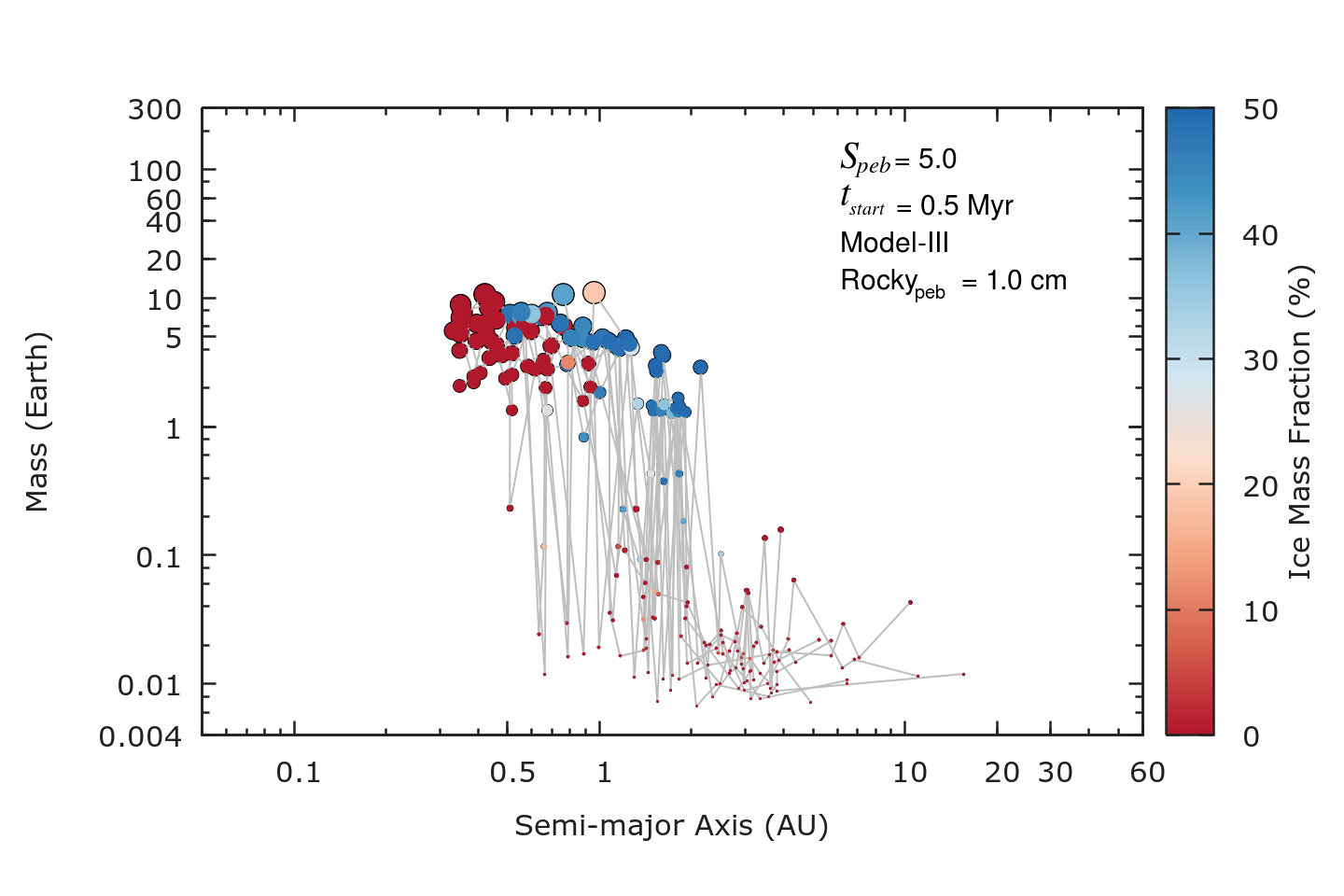}
\includegraphics[scale=.15]{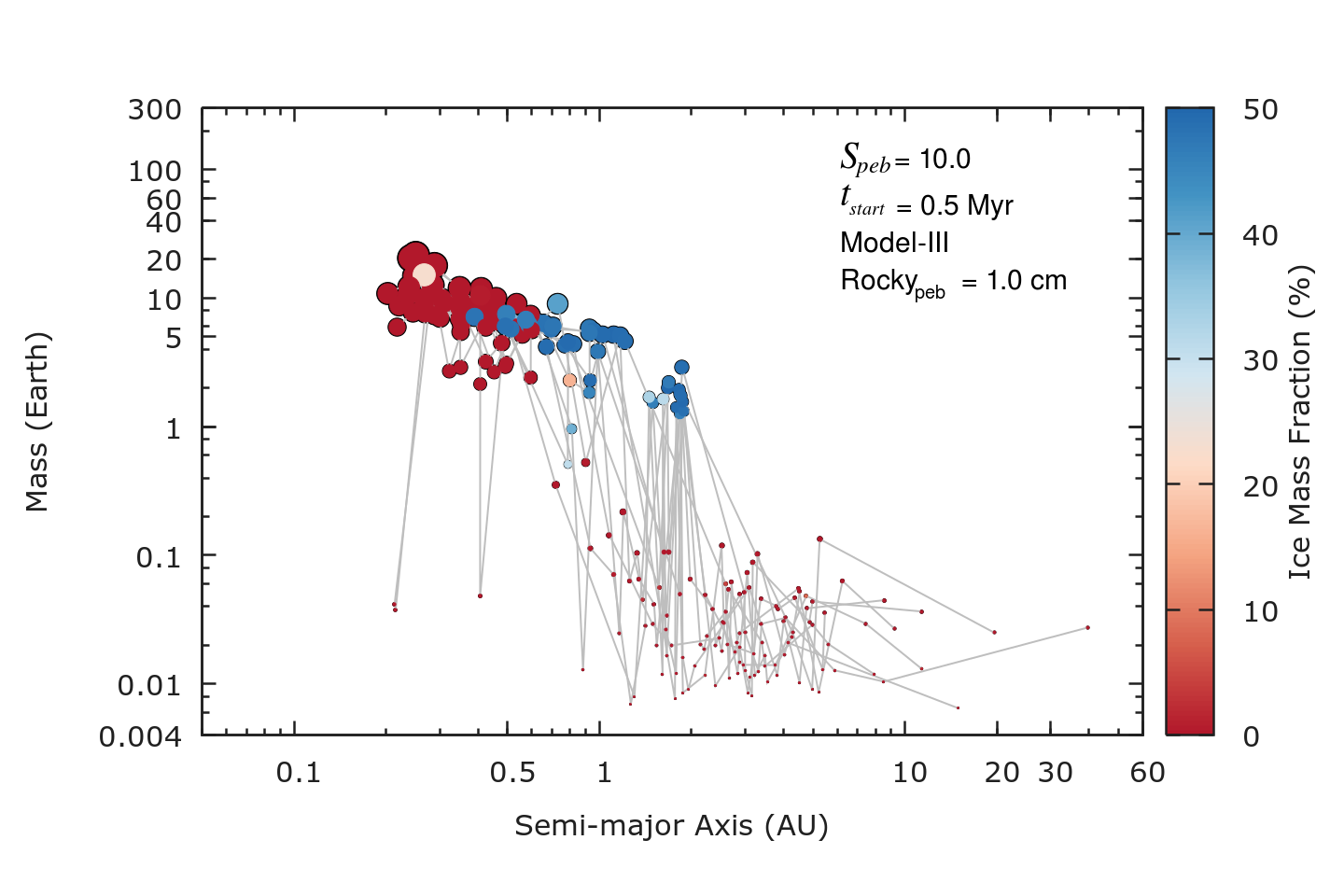}
\caption{Same as Figure \ref{fig:panels_fluxes_model_3_mm}, except that we now use rocky pebbles of ${\rm Rocky_{{\rm peb}}=1~cm}$. This leads to enhanced growth of the inner embryos in contrast to Fig. \ref{fig:panels_fluxes_model_3_mm}. The efficiency of pebble accretion is about 5.6\%, 4.8\%, 4.1\%, and 3.4\% for  ${S_{{\rm peb}}=1,~2.5,~5,~{\rm and}~10}$, respectively.}
    \label{fig:panels_fluxes_model_3_cm}
\end{figure*}

Figures \ref{fig:panels_fluxes_model_3_mm} and \ref{fig:panels_fluxes_model_3_cm} show the planets produced by simulations of Model
III with pebbles of  different sizes inside the snow line. In Figure \ref{fig:panels_fluxes_model_3_mm}  the size of the silicate pebble is ${\rm Rocky_{{\rm peb}}=0.1~cm}$ while in Figure \ref{fig:panels_fluxes_model_3_mm} this is ${\rm Rocky_{{\rm peb}}=1~cm}$. In both cases ${\rm t_{start}= 0.5~Myr}$. 



In simulations of Figure \ref{fig:panels_fluxes_model_3_mm}, small  planetary embryos are initially distributed between $ \sim0.3$ and 2~AU.  The disk snow line is at about  3~AU at 0.5~Myr. Consequently, all planetary embryos grow at first  by the accretion of silicate pebbles. However, as the disk evolves, the snow line moves inwards eventually sweeping  the outermost seeds from outside in. Thus, the outermost seeds (at about 2~AU) eventually switch to accreting larger icy pebbles until they reach pebble isolation mass, and thus block the flux of pebbles from the outer disk, starving the inner disk. We note that the final ice mass fraction of a planetary seed initially inside the snow line is primarily regulated by the total mass in icy dust particles available in the outer disk at the time when the snow line crosses the seed's orbit \citep{idaetal19}. Nevertheless, the pebble flux filtering by  outer seeds  (if they exist) also plays a crucial role. Figure \ref{fig:panels_fluxes_model_3_mm} shows that protoplanetary seeds swept up by the snow line may reach masses large enough  to move by  type-I migration to the inner edge of the disk before the gas is dispersed. However, at the end of the gas disk phase the very final shape of the outward migration region is such that it favors ${\rm \sim2~M_{\oplus}}$ protoplanetary embryos becoming stranded between 1 and 2~AU (see e.g., top left-hand panel of Fig. \ref{fig:panels_fluxes_model_3_mm} and the shape of the migration zone in Fig. \ref{fig:snapshots}).


Figure \ref{fig:panels_fluxes_model_3_mm} shows that simulations with ${S_{{\rm peb}}=1}$ failed to form multiple planets with masses larger than a few ${\rm M_{\oplus}}$ of any composition anchored at the inner edge of the disk. Increased  pebble fluxes (${S_{{\rm peb}} =2.5}$, 5, and 10)  successfully promote the formation and delivery of icy planetary embryos with masses of ${\rm\sim5-10~M_{\oplus}}$ to the inner edge of the disk at 0.3~AU. However, only simulations with  ${S_{{\rm peb}}=10}$ (botton right-hand panel of Fig. \ref{fig:panels_fluxes_model_3_mm})  produce final rocky protoplanetary embryos with masses larger than ${\rm\sim1~M_{\oplus}}$, i.e., in the super-Earth mass range. Higher pebble fluxes also tend to produce a  larger number of closer-in icy planetary embryos compared to lower pebble fluxes because initially more distant seeds (swept by the snow line) can grow faster and to larger masses, consequently leaving the outward migration region and migrating inwards.

Figure \ref{fig:panels_fluxes_model_3_cm} shows the planets produced in our Model III simulations with larger silicate pebbles (${\rm Rocky_{{\rm peb}}=1~cm}$) for different pebble fluxes. As expected, comparing the results of  Figure \ref{fig:panels_fluxes_model_3_mm} and Figure \ref{fig:panels_fluxes_model_3_cm}  it is clear that the growth of rocky planetary embryos is far more efficient when  ${\rm Rocky_{{\rm peb}}=1~cm}$  for all pebble fluxes. Interestingly, only simulations with  ${S_{{\rm peb}}\gtrsim2.5}$ produced concomitantly and systematically  rocky and icy super-Earths of similar mass. Low pebble fluxes tend to favor the formation of large icy super-Earths as opposed to rocky ones. Simulations with ${S_{{\rm peb}}=10}$ produce at least a few planetary embryos as large as ${\rm \sim20~M_{\oplus}}$. Results presented in  Figs. \ref{fig:panels_fluxes_model_3_mm} and
\ref{fig:panels_fluxes_model_3_cm} clearly show that avoiding the formation of icy super-Earths is a difficult task even in the scenario where seeds only form well inside the snow line.  Figures \ref{fig:panels_fluxes_model_3_mm}  and \ref{fig:panels_fluxes_model_3_cm} show that Model-III  only produces systems dominated by rocky  super-Earths if seeds starting inside 1~AU  grow  to Earth-mass or larger sufficiently quickly, allowing them to grow above pebble-isolation mass. In order to ensure fast growth of these seeds we invoked the existence of cm-sized silicate pebbles in the inner disk (${\rm Rocky_{{\rm peb}}=1~cm}$). However, faster growth could be equally achieved with our nominal mm-sized silicate pebbles if we reduce the level of turbulent vertical stirring of mm-sized silicate pebbles.  We performed two simple simulations to confirm this claim. We find that the growth of a ${\rm 0.01~M_{\oplus}}$ planetary seed  at 1~AU growing in a 0.5~Myr-old disk where  ${S_{{\rm peb}}=5}$ and  ${\rm Rocky_{{\rm peb}}=1~cm}$ is very similar to that of an identical seed in a simulation with ${S_{{\rm peb}}=5}$, ${\rm Rocky_{{\rm peb}} = 0.1~cm,}$ and a disk where the scale height of the pebble layer is smaller by a factor of approximately three. This scale height corresponds to a disk where $\alpha_{set}  \simeq  \alpha/10$ (see Section \ref{subsec:accretion}). We discuss this issue again in Section \ref{sec:observations}.

The main difference between the results of Model III and Models I and II is in the efficiency of pebble accretion; in other words, the fraction of the integrated pebble flux converted into planets. In Models I and II -- depending on ${S_{{\rm peb}}}$ and ${t_{{\rm start}}}$ -- from 17\% to 49\% of the pebble flux is converted into planets while in Model III this quantity varies between 1.4\% and 5.6\%. This represents a significant reduction. We now list a few possible reasons for this difference but probably not all. First, in Model III (both Figures \ref{fig:panels_fluxes_model_3_mm} and \ref{fig:panels_fluxes_model_3_cm}), seeds starts only inside the snow line where pebbles are typically smaller than beyond the snow line (even when ${\rm Rocky_{{\rm peb}}=1.0~cm}$) and, consequently, the efficiency of pebble accretion is reduced. Second, in simulations of Model-III with ${\rm Rocky_{{\rm peb}}=0.1~cm}$ the outermost seed is typically swept by the moving snow line and grows really fast to pebble isolation mass suppressing planetary growth in the inner regions. Finally, the pebble isolation masses in the inner regions of the disk are small thus planets stop growing at lower masses.

Although the parameters ${S_{{\rm peb}}}=5~{\rm and}~10$ result in a large mass in pebbles available for planetary growth (if ${t_{{\rm start}}}=0.5~{\rm Myr}$ they imply ${\rm \sim 970~M_{\oplus}}$  and ${\rm \sim 1940~M_{\oplus}}$, respectively), we stress that the actual amount of mass in pebbles required to grow our planetary systems is always much smaller.  Most of the planets in our simulation are already fully formed after a relatively short time, especially the ones migrating to the inner system. Thus, to form most of these planets it is only required to have a sufficiently  large pebble flux in the beginning of our simulations and  not necessarily during the whole gas disk lifetime.

The pebble accretion efficiency in our simulations varies between 1 and 50\%. We note that in simulations with single planets, the reported pebble accretion efficiencies are typically lower, only a few percent per planet \citep[e.g.,][]{linetal18,liuormel18}. Our simulations start with many protoplanetary seeds, so pebble accretion becomes far more efficient simply because multiple planets can accrete pebbles.

\section{Comparison with the results of Paper I}\label{sec:sec4}


Paper I defined super-Earths as planets that reached a mass large enough for significant migration during the gas disk lifetime, and showed that rocky super-Earths tend to be clustered close to the inner edge of the disk, particularly at the end of the gas-disk phase (some spreading occurring during instabilities after gas removal; see section 4 below). The rocky planets at ${\rm \sim1~AU}$ are typically terrestrial planets, that is, planets that grew mostly after gas removal from a disk of planetary embryos too small to migrate (less than 3 Mars masses). The results presented here in Figs. \ref{fig:panels_fluxes_model_1}-\ref{fig:panels_fluxes_model_3_cm} show that super-Earths (in the sense of Paper I) can also be at 1 AU or even significantly beyond this region. Their existence is also suggested by micro-lensing observations  \citep{beaulieuetal06,murakietal11,bennettetal07,kubasetal12,sumietal10,sumietal16,koshimotoetal17}. However, in our model these planets are never rocky. They either start their formation beyond the snow line and migrate inwards towards their final position, or are swept by the snow line (in the case of model III), so that they accrete most of their mass from icy pebbles and experience a temporary phase of outward migration. In this paper, rock-dominated planets are seen only in model III and their distribution mirrors the results of Paper I: only small rocky embryos remain near 1 AU (e.g., Fig. \ref{fig:panels_fluxes_model_3_mm} upper-left panel) while rocky super-Earths migrate towards the inner edge of the disk (here placed farther out than in Paper I for the reasons explained in section 2).

The sensitive dependency of the final planet mass on the pebble flux discussed in paper I is recovered here. However, a quantitative comparison with Paper I shows some apparent differences. For instance, in Paper I a total pebble flux of 200 Earth masses (5/3$\times$nominal) leads to the formation of a super-Earth close to the disk's inner edge. Here, a pebble flux of 194 Earth masses (Fig. \ref{fig:panels_fluxes_model_3_mm}) leads only to small (typically submartian) rocky planetary embryos. However, the pebble flux reported in Paper I is the flux within the snow line. The flux reported here is beyond the snow line, and should be divided by two inside of the snow line because of ice sublimation. Moreover, pebble size, pebble flux, pebble scale height, and the gas disk model differ from the model presented in Paper I. Even when the pebble fluxes are truly equivalent,  the final masses of planets in our simulations and those in Paper I will probably differ. In Paper I, the disk snow line location is fixed throughout the entire  lifetime of the disk. Thus, because  in our simulations the disk snow line moves inside 1 AU   at late stages,  the pebble sizes around 1-2~AU may become much larger than the fixed pebble size considered in Paper I, resulting in different growth modes. Moreover, in our simulations, a significant fraction of the pebble flux is filtered by the growing planets in the outer disk. Hence, the rocky pebble flux seen by the embryos in the inner disk in the simulation of Fig. \ref{fig:panels_fluxes_model_3_mm} (upper-left panel), for example, corresponds to a subnominal flux in Paper I. There is therefore no inconsistency on the final masses of rocky bodies.   


The main difference between this paper and Paper I is that we simulate the concurrent growth of rocky and ice-rich super-Earths ---as well as dynamical perturbations between them--- whereas the existence of icy planets was neglected in Paper I. Moreover, we show that, if seeds are present throughout the outer disk, the growth of rocky planets is precluded because almost all of the pebble flux is intercepted or blocked by the outer, icy embryos. Only if the seed distribution ends near the snow line (model -III) can the growth of rocky planets be significant. If the rocky pebbles are smaller than the icy pebbles, the ice-rich planets and the smaller, rocky-dominated planets are radially mixed (Fig. \ref{fig:panels_fluxes_model_3_mm}). If the rocky pebbles are as big as icy pebbles, rocky-dominated planets are typically the innermost ones and ice-rich planets are those farther out (Fig. \ref{fig:panels_fluxes_model_3_cm}). Although both cases can be consistent with the observations of extra-solar planets (see Sections \ref{sec:super-Earth} and \ref{sec:observations} below), they are inconsistent with the Solar System, where rocky planets are much smaller than the ice-rich planets (Uranus, Neptune, and the cores of Jupiter and Saturn) but the two types of planets are not radially mixed. A discussion of the Solar System case is deferred to Section \ref{sec:solarsystem}.  Paper I can therefore be seen as a subcase where some process, for example the formation of giant planets (see also Paper III), impedes the migration of ice-rich planets into the inner system.

\section{Formation of close-in super-Earths}\label{sec:super-Earth}

We now aim to model the formation of super-Earth systems like those observed.  As shown in the previous section, different combinations of parameters (${t_{{\rm start}}}$,  ${S_{{\rm peb}}}$, and ${\rm Rocky_{{\rm peb}}}$) produce dramatically different planetary systems. To conduct our new simulations we have purposely selected pebble fluxes that can successfully lead to the formation of hot super-Earth systems rather than systems of terrestrial planets \citep[see ][]{lambrechtsetal18} or gas giants \citep{bitschetal18c}. We are interested in setups where at the time of the gas disk dispersal most planets anchored at the disk inner edge have masses of between ${\rm \sim1~ M_{\oplus}}$ and ${\rm \sim15~M_{\oplus}}$. This makes them reasonably consistent with the expected masses for most close-in super-Earths observed by Kepler \citep[e.g.,][]{wolfgangetal16}.

To systematically evaluate the performance of the migration model we performed 250 simulations considering four different selected scenarios. These are shown in Table \ref{tab:2}, which also presents the integrated pebble flux of each setup, which depends on ${S_{{\rm peb}}}$ and ${t_{{\rm  start}}}$. 


To remain consistent with previous simulations of the formation of close-in super-Earth systems \citep[e.g.,][]{izidoroetal17} we set the disk inner edge at $\rm r_{in}$=0.1~AU rather than at $\sim$0.3~AU.  This also matches the location of the actual inner edge of the disk according to \cite{mulders18}.  This is essentially the only difference between the setup of the simulations presented in this section and those presented above. However, to have better statistics when analyzing the long-term dynamical evolution of planetary systems, for each scenario of Table \ref{tab:2} we performed 50 simulations with slightly different initial conditions.  As before, each model is represented by an initial distribution of planetary embryos as described in Table \ref{tab:model}. After the gas disk dispersal,  simulations are continued for an additional  $\sim$50~Myr. Some particularly interesting cases were integrated up to 300~Myr.
\begin{table}
\caption{Selected setups to model the formation and long-term dynamical evolution of  close-in super-Earths based on the results of Section \ref{rolepebbleflux}. From left to right, the columns are name of the model, disk age at the start of the simulation (${t_{{\rm start}}}$),  the scaled pebble flux (${S_{{\rm peb}}}$), and the integrated pebble flux (${\rm I_{{\rm peb}}}$). In our nominal simulations, ${\rm Rocky_{{\rm peb}}=0.1~cm}$ but we also performed simulations with ${\rm Rocky_{{\rm peb}}=1~cm}$.}
   \centering 
\begin{tabular}{@{}lccc@{}}
  \hline
Scenarios   &  ${t_{{\rm start}}}$               & ${S_{{\rm peb}}}$ &     ${\rm I_{{\rm peb}}~(M_{\oplus})}$    \\
              &                                     &            & \\
  \hline\hline
Model-I         & 0.5 Myr   &  1  & ${\rm \sim194}$ \\
\hline
Model-I         & 3.0 Myr   &  5  & ${\rm \sim 150}$\\
\hline
Model-II        & 3.0 Myr    &   5  & ${\rm \sim 150}$ \\
\hline
Model-III       &  0.5 Myr     &  10 & ${\rm \sim 1940}$\\

\hline
\label{tab:2}
\end{tabular}
\end{table}

\subsection{Long-term dynamical evolution}

\begin{figure*}
\centering
\includegraphics[scale=.9]{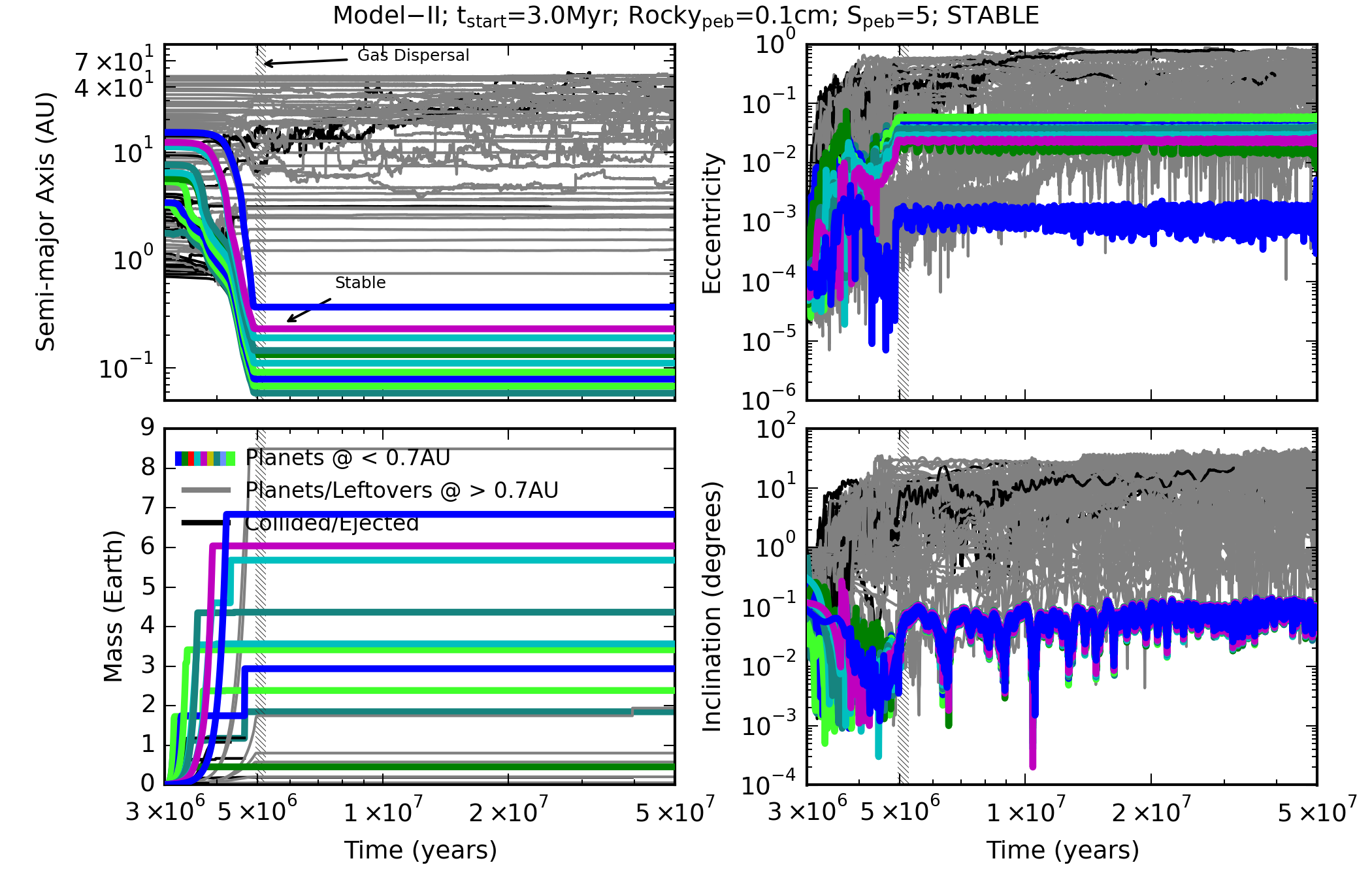}
\caption{Growth and long-term dynamical evolution of protoplanetary embryos in a simulation of Model II where ${\rm t_{start}= 3.0~Myr}$,  ${S_{{\rm peb}}=5,}$ and the size of the rocky pebbles is ${\rm Rocky_{{\rm peb}}=0.1~cm}$. The top left panel shows the evolution of semi-major axes. The top right panel shows the evolution of the eccentricities. The bottom left and right panels show mass growth and the orbital inclinations, respectively. Colored lines show the final planets orbiting inside 0.7~AU. The gray lines show final planets and leftover protoplanetary embryos with orbits outside 0.7~AU. Leftover embryos are those with masses smaller than ${\rm \sim0.1~M_{\oplus}}$. Finally, the black lines show collided or ejected objects over the course of the simulation.  The dashed vertical line shows the instant of the gas disk dispersal. After gas disk dispersal the resonant chains of super-Earths anchored at the inner edge of the disk remain dynamically stable during the total integration time of 50~Myr.}
    \label{fig:dynamics_stable}
\end{figure*}

Figure \ref{fig:dynamics_stable} shows the evolution of a simulation of Model II where ${\rm t_{start}= 3.0~Myr}$,  ${S_{{\rm peb}}= 5}$, and ${\rm Rocky_{{\rm peb}}=0.1~cm}$. The dynamical evolution during the gas disk phase of the growing planetary embryos  is qualitatively similar to those in Figures \ref{fig:dynamics_S1} and \ref{fig:dynamics_S2.5}. Essentially, planetary embryos grow and migrate inwards establishing a long chain of planets mutually captured in first-order mean motion resonances. At the end of the gas disk phase, planetary embryos anchored at the inner edge of the disk have masses lower than  ${\rm \sim7~M_{\oplus}}$.  After gas dispersal, the simulation is continued for an additional 45~Myr in a gas-free scenario but the resonant chain remains dynamically stable. The right-hand panels of Figure \ref{fig:dynamics_stable} show the orbital eccentricities and inclinations of all planetary embryos in this simulation. As shown, planetary embryos inside 0.7~AU at the end of the simulation  have orbits with very low eccentricities and inclinations.
\begin{figure*}
\centering
\includegraphics[scale=.9]{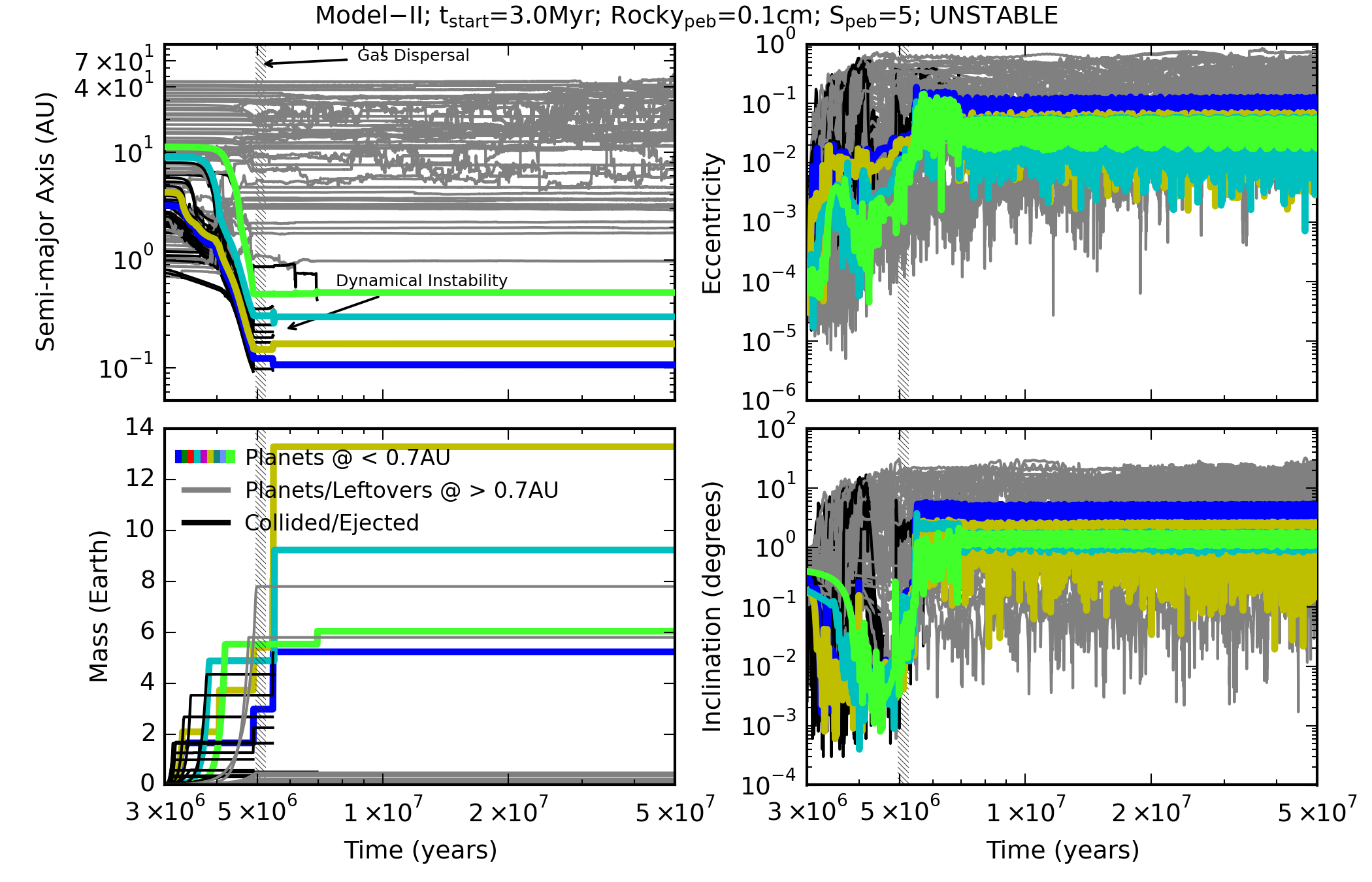}
\caption{Same as Figure \ref{fig:dynamics_stable}, except that we show here a  case where the system undergoes a dynamical instability.}
    \label{fig:dynamics_unstable}
\end{figure*}

Figure \ref{fig:dynamics_unstable} shows the evolution of another simulation of  Model II in which $t_{{\rm start}}={\rm 3.0~Myr}$,  $S_{{\rm peb}}={\rm 1}$, and ${\rm Rocky_{{\rm peb}}}$=0.1~cm. Again, planetary embryos grow and migrate to the disk inner  edge forming a long resonant chain. The long-term dynamical evolution of planets in this simulation is in great contrast with that of Figure \ref{fig:dynamics_stable}. After the gas disk phase  (see vertical gray line in the figure), the resonant chain of planets anchored at the inner edge becomes dynamically unstable at about 6~Myr. The dynamical instability promotes collisions and further planetary growth. The instability phase leads to a final planetary system which is dynamically more excited. Planets' orbital eccentricities reach values of a few percent while orbital inclinations increase by a few degrees.

The dynamical evolutions presented in Figures \ref{fig:dynamics_stable}  and \ref{fig:dynamics_unstable} are representative of all our simulations. Some planetary systems present a phase of dynamical instability after gas dispersal but some do not. Following \citet{izidoroetal17}, we refer to these two classes of planetary systems as ``{stable}'' and ``{unstable}''. The fraction of  stable and unstable systems varies in our simulations. We stress that once a system becomes dynamically unstable, the duration of the instability phase is typically  short and the system rapidly evolves to a less compact but typically long-term stable configuration.

\begin{figure*}
\centering
\includegraphics[scale=.5]{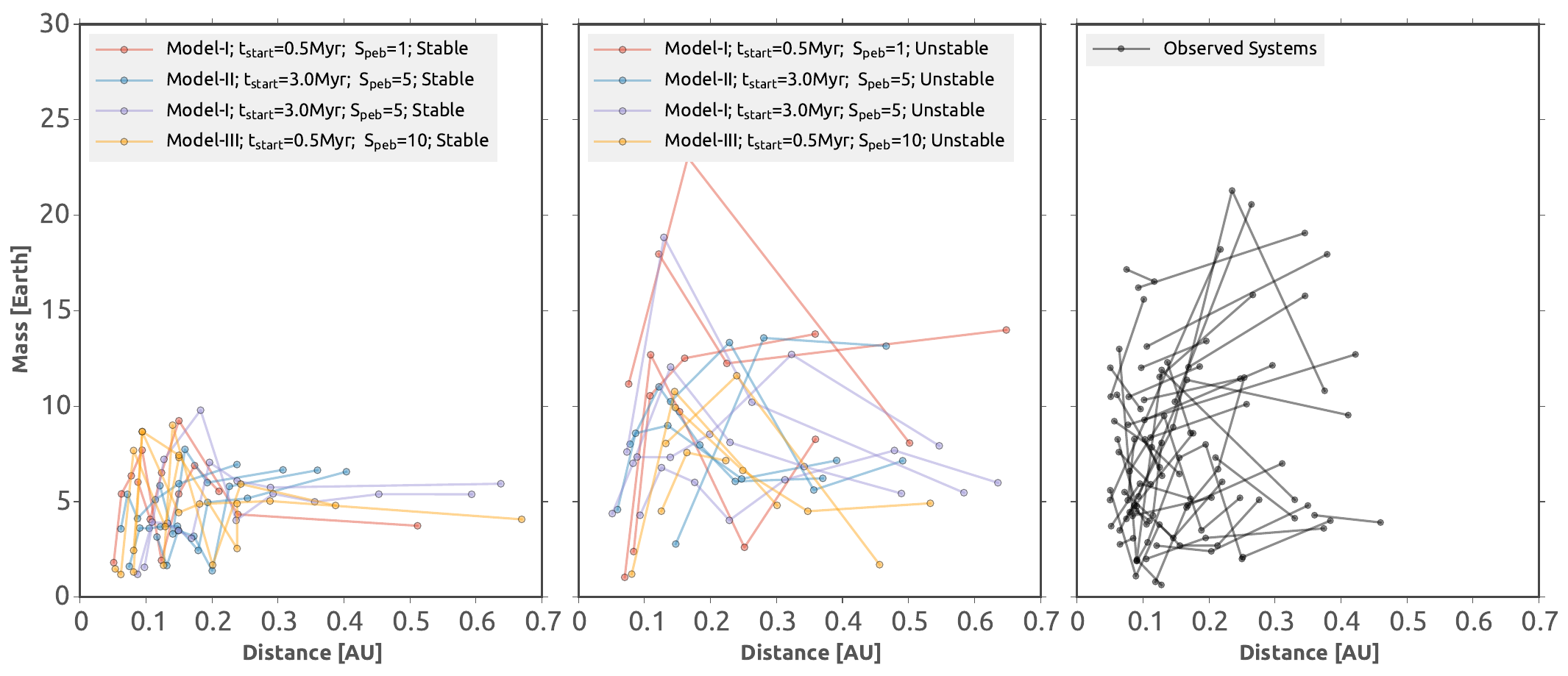}
\caption{Planetary systems produced in different simulations in a diagram showing semi-major axis versus mass after 50~Myr of integration. The left panel shows  stable planetary systems. The middle panel shows the final architecture of selected dynamically  unstable systems, and the right panel shows close-in observed planetary systems where all planet masses (or $msin(i)$ in the case of radial velocity detections) are estimated to be lower than ${\rm 22~M_{\oplus}}$. Only planets within 0.7~AU are shown even though Earth-mass planets may exist in these systems farther out. In these simulations, ${\rm Rocky_{{\rm peb}}}$=0.1~cm.}
    \label{fig:masses}
\end{figure*}

Figure \ref{fig:masses} shows selected  stable (left panel) and unstable (middle panel) planetary systems produced in our different models. The right-hand panel shows selected observed planetary systems. Planets in  stable systems have masses lower than $\sim$10${\rm M_{\oplus}}$. There is no clear radial mass ranking in planetary systems produced in our simulations (in contrast, see Figure 3 of \cite{ogiharaetal15a}). In the following section we take a closer look at the orbital architecture of these systems. Finally, we note from comparing the left and middle panels of Figure \ref{fig:masses}  that  unstable systems are relatively more spread and have fewer but larger planets. The results of Figure \ref{fig:masses} are  qualitatively similar to those in \citet{izidoroetal17} where the  formation of close-in super-Earth systems is modeled assuming ad hoc initial distributions of Earth-mass planetary embryos in the outer parts of the disk. 

\subsection{Fraction of  unstable systems and timing of the instability}

\begin{figure}
\centering
\includegraphics[scale=.41]{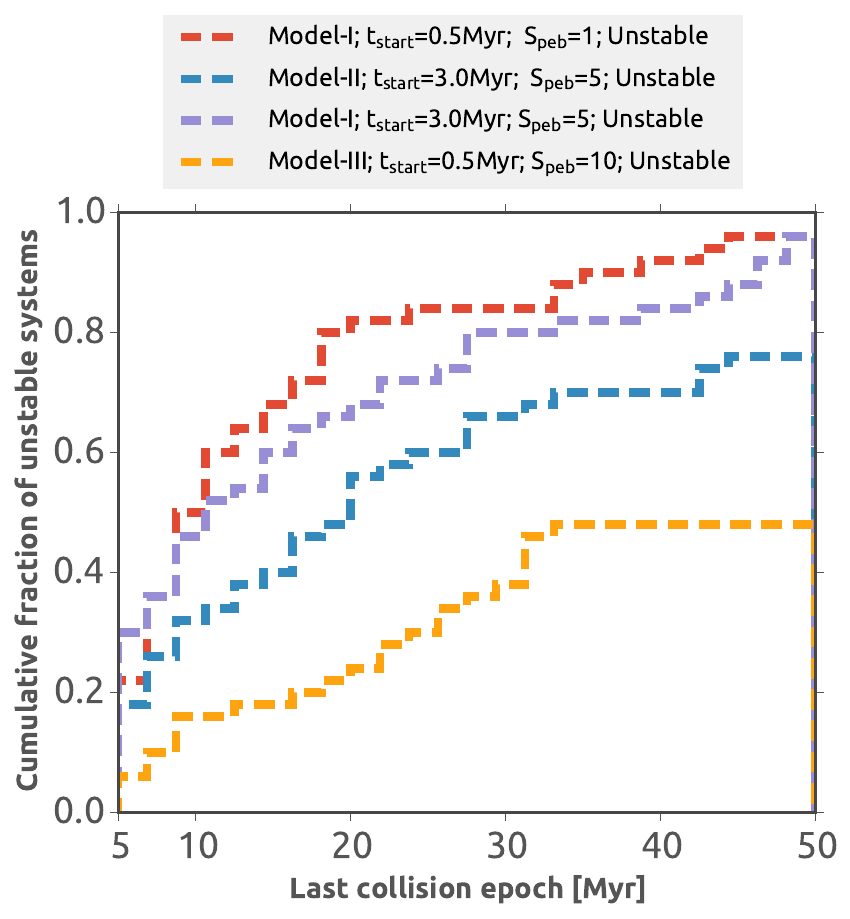}
\caption{Cumulative distributions of the last collision epoch in all our unstable systems also with ${\rm Rocky_{{\rm peb}}}$=0.1~cm. The distributions only account for collisions happening after the gas disk dispersal.}
    \label{fig:lastcollision}
\end{figure}

Figure \ref{fig:lastcollision} presents a cumulative distribution of the epoch of the last collision after gas disk dispersal for different sets of simulations. Dynamical instabilities start to occur as soon as the gas goes away (at 5~Myr). In all our simulations, the fraction of dynamically {\it unstable} systems is always smaller than 1. However, some scenarios of Table \ref{tab:2} produce a much higher rate of instabilities than others. For example, red and purple dashed lines representing  Model I show that more than 90\% of planetary systems anchored at the inner edge of the disk become dynamically unstable after gas dispersal. On the other hand,  the fractions of unstable systems in simulations of Models II and III drop to 70\% and 45\%, respectively.

In the simulations of  \citet{izidoroetal17}, only $\sim$50\% of planetary systems became  dynamically unstable after gas dispersal. However, \citet{izidoroetal17} also showed that, in order to match the period ratio distribution of observations, more than 75\% of the planetary systems are required to become dynamically unstable. Why so many systems become dynamically unstable after the gas dispersal remained unsolved in \citet{izidoroetal17}. Figure \ref{fig:lastcollision} shows that in some of our systems this fraction  may be as high as $\sim 95\%$. 
A discussion about why, in some of our models, a higher fraction of the systems become dynamically unstable  compared to that of \citet{izidoroetal17} is presented in Section \ref{sec:whyusntable}. In later sections we also evaluate how these simulations match other observational constraints. 

\subsection{Why do some systems become dynamically unstable?}\label{sec:whyusntable}
A critical question remains as to what determines whether or not a given close-in planetary system will become dynamically unstable. To answer this question, we analyze the orbital architecture of our protoplanetary systems at the end of the gas disk dispersal and prior to the timing of the instability, at 5 Myr. Some planetary systems started the instability phase at the very end of the disk dispersal (e.g., at $\sim$4.9~Myr) and these systems were discarded from this analysis. Of course, because the rate of dynamically unstable systems is very high in some of our models we end up with only a few stable planetary systems. We do not include  in our analysis cases that could suffer dramatically from small number statistics (e.g., stable systems of Model I).
\begin{figure*}
\centering
\includegraphics[scale=.41]{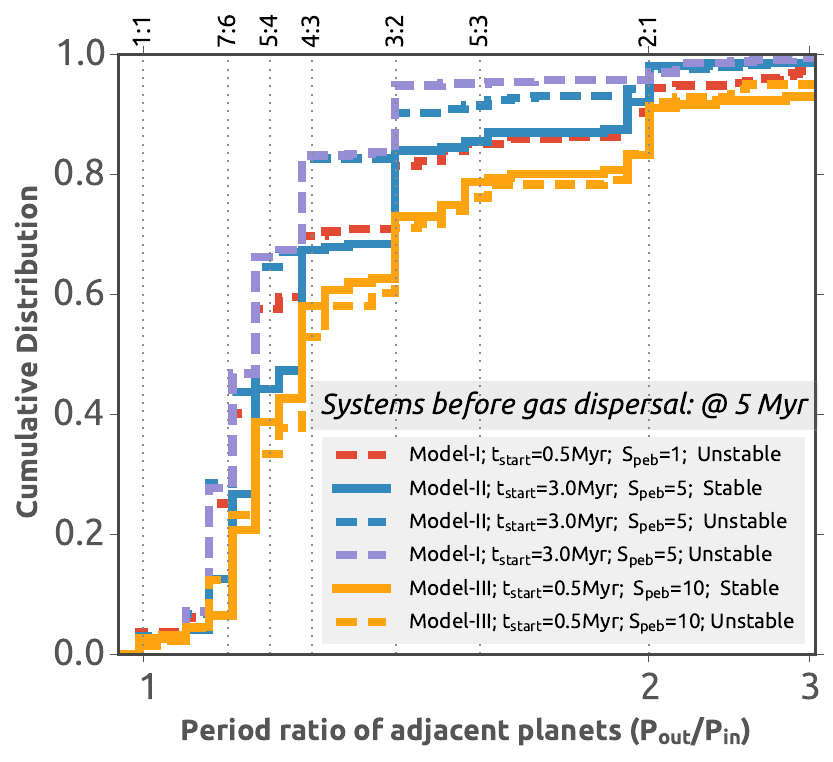}
\includegraphics[scale=.41]{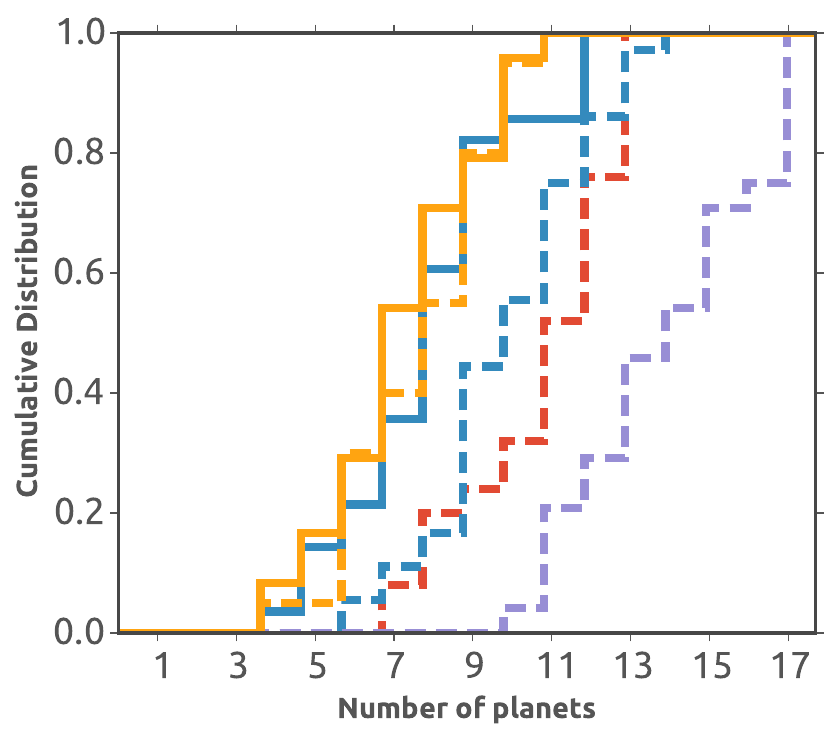}
\includegraphics[scale=.41]{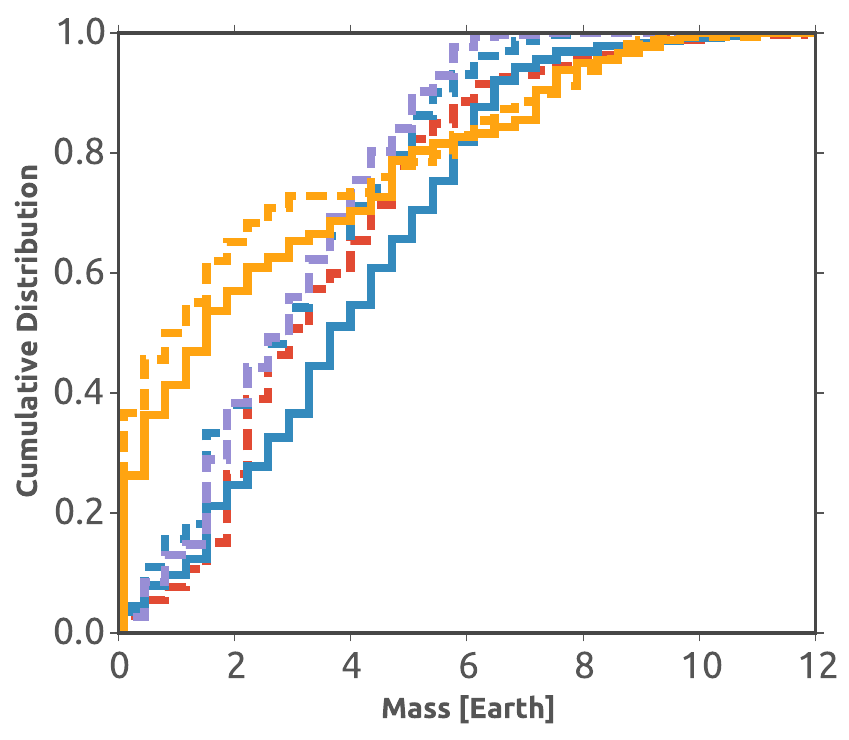}
\caption{Dynamical architecture of stable and unstable planetary systems anchored at the disk inner edge at the end of the gas disk dispersal (before the onset of dynamical instabilities).   From left to right, the panels show the period ratio distribution, the distribution of the number of planets, and planet mass distributions. These distributions were calculated considering planets inside 1~AU with masses larger than $0.1{\rm M_{\oplus}}$.}
    \label{fig:whyunstable}
\end{figure*}

The left-hand panel of Figure \ref{fig:whyunstable} shows the period ratio of unstable and {\it stable} systems at 5~Myr (before the instability). The higher rate of dynamical instability was observed for simulations of Model I (purple and red dashed lines in Figure \ref{fig:lastcollision}). The purple dashed line of Figure \ref{fig:whyunstable} shows that  Model I with ${\rm t_{start}= 3.0~Myr}$ and ${S_{{\rm peb}} = 5}$ produces the most compact planetary systems at the end of the gas disk phase. The middle panel of Figure \ref{fig:whyunstable} shows that  these same planetary systems correspond to those with the larger number of planets anchored at the disk inner edge.


The lowest fraction of dynamically unstable systems belongs to Model III with ${\rm t_{start}= 0.5~Myr}$,  ${S_{{\rm peb}} = 10}$, and ${\rm Rocky_{{\rm peb}}=0.1~cm}$, in which fewer than half of the planetary systems become dynamically unstable after gas dispersal (yellow dashed line in Figure \ref{fig:lastcollision}). The period ratio distribution of planet pairs for this set of simulations shows that they are the least dynamically compact systems and also the least crowded ones, although stable and unstable systems within Model III do not show significant differences. Thus, the fate of a planetary system in becoming dynamically unstable or not after gas dispersal depends on the number of planets in the chain, compactness, and planet masses. Indeed, \cite{matsumotoetal12} showed numerically that, for a given resonant chain, the less massive the planets are, the more numerous they need to be to become dynamically unstable or, equivalently, for a given mass, there is a maximal number N of planets that can be stable. \citet{pichierimorbidelli20} performed a detailed analytical analysis of the dynamics of resonant chains and found a theoretical justification for this observation.

\section{Matching observations}\label{sec:observations}

We now evaluate how well our simulations match observations. In Section \ref{ref:observationalconstraints} we first lay out five observational constraints related to the observed super-Earth population, mainly based on observations with the NASA Kepler space telescope~\citep{Boruckietal10}.  Our simulations  already match two constraints by consistently forming super-Earth systems that have similar masses to the real ones.  Next we discuss three constraints in detail and perform synthetic observations to quantitatively compare our simulations with observations.  We first explore the period ratio distribution (Section \ref{sec:periodratiodisttribution}) and then the Kepler dichotomy (Section \ref{sec:keplerdichotomy}). Finally, we discuss the rocky versus icy nature of super-Earths (Section \ref{sec:rockyicy}) and perform several additional sets of simulations to explore the conditions required to form rocky super-Earths.

\subsection{Observational constraints}\label{ref:observationalconstraints}

To be considered successful, any super-Earth formation model must match the available observational constraints.  Yet all constraints are not equally important.  We now briefly discuss five constraints that we, quite subjectively, order by relative strength.

\begin{itemize}
\item {\bf The large abundance of super-Earths.}  At least 30\%, and perhaps up to 90\%, of main sequence stars host close-in super-Earths~\citep{mayoretal11,howardetal12,fressinetal13,petiguraetal13,zhuetal18,mulders18,muldersetal18}.  Any model must explain why such planets form so readily, and are found on a wide range of stellar metallicity, with more massive super-Earths being at most slightly more abundant around high-metalicity stars~\citep[e.g.,][]{buchhaveetal12,petiguraetal18,kutrawu20}.
\item {\bf The super-Earth period ratio distribution.} In multiple-planet systems, the relative spacing of planetary orbits is a measure of the dynamical state of the system. This is measured via the period ratio of adjacent planets, which has been well characterized by the Kepler mission~\citep[modulo selection effects such as missing certain planets][]{lissaueretal11a,fabryckyetal14}.
\item {\bf Approximate masses and sizes of super-Earths.} While super-Earths are typically between 1 and 4 Earth radii, determining their masses has required great investment in radial velocity~\citep[e.g.,][]{marcyetal14} and transit timing variation~\citep[e.g.,][]{lithwicketal12,mazehetal13} measurements. This has led to the derivation of mass--radius relationships for close-in small planets~\citep{weissetal13,weissmarcy14,wolfgangetal16}, which suggests that most super-Earths (and mini-Neptunes) are a few to ten Earth-masses. This is a mass at which migration is very efficient.
\item {\bf The super-Earth multiplicity distribution, or `Kepler dichotomy'.}  While a large fraction of stars are found to host super-Earths, most systems seem to host only a single super-Earth whereas a small fraction of the systems host many planets.  This has been referred to as the Kepler dichotomy~\citep{johansenetal12,ballardetal16}. The debate continues as to whether or not systems with a single super-Earth are truly single, and as to whether or not these systems have different origins from systems with multiple super-Earths. In our previous paper we proposed that most single super-Earths are not single and that the dichotomy is simply a consequence of the relatively broad distribution of mutual inclinations between super-Earths which reduces the probability of multiple-transiting systems~\citep{izidoroetal17}. From that perspective, the dichotomy reflects two kinds of planetary systems in nature: those that underwent dynamical instabilities  and inclination excitation after gas dispersal and those that avoided this. However, we note that the rate of false positive single super-Earth systems is likely to be higher than for multiple systems, potentially affecting this constraint at a quantitative level. 
\item {\bf Compositional trends among super-Earths.}  Atmospheric photoevaporation models suggest that many super-Earths are likely to be purely rocky. On very close-in orbits, planets are unable to retain thick envelopes in the face of strong UV-driven photoevaporation~\citep[e.g.,][]{hubbardetal07,owenwu13}. \cite{fultonetal17} showed that there is a dip in the size distribution of close-in planets between roughly 1.5 and 2$\rearth$. Interior modeling of planets with inferred masses and radii suggests that the photo-evaporation valley favors mainly a rocky composition of super-Earth cores over icy cores~\citep{owenwu17,jinmordasinietal18,vaneylenetal18}. However, at present, there is no consensus as to whether or not all super-Earths are rocky. A recent analysis of super-Earth compositions by \cite{zengetal19} suggests that super-Earths with sizes larger than 2-3$\rearth$ are more likely to be water-rich worlds.
\end{itemize}

While each of these constraints is important, we consider the compositional constraints on close-in super-Earths to be the weakest.  This is also because (a) the division between rocky and ice-rich is poorly defined in terms of the actual water mass fraction; and (b) there are uncertainties in both measured planetary radii and masses and in the equation of state of planetary constituents at high pressure \citep[e.g.,][]{valenciaetal07,rogersseager10,swiftetal12, lopezfortney14,duffyetal15,zengetal16,dornetal17,millsmazeh17,bergeretal18,wicksetal2018,smithetal2018}. Nonetheless, we discuss the composition of super-Earths in Section \ref{sec:rockyicy}.

\subsection{The period ratio distribution}\label{sec:periodratiodisttribution}

In this section we first examine the period ratio distribution from our simulated systems and then perform synthetic observations of those systems to statistically compare them with Kepler data.

\subsubsection{Dynamical architecture of unstable and stable simulated systems}\label{architecture}

We first divide our simulations within each scenario of Table \ref{tab:2} into groups of stable and unstable planetary systems. As above, unstable systems are those that have undergone dynamical instabilities after the gas dispersal and stable systems are those that did not present instabilities from the end of the gas disk phase (5 Myr) to the end of our simulations (typically at 50 Myr, but in some cases at 300 Myr). In order to compare our results with the sample of planet candidates from the Kepler mission we only consider in our analysis planets with semi-major axes smaller than 0.7~AU (${\rm P\lesssim200~days}$) and with masses larger than ${\rm 1~M_{\oplus}}$ at the end of the simulation (50~Myr). These cut-offs are applied because the Kepler sample is almost complete for transiting  planets larger than Earth and orbital periods smaller than 200 days \citep{petiguraetal13,silburtetal15}. In our analysis, we used observational data downloaded from NASA Exoplanet archive. We selected planets with sizes between 1 and ${\rm 4~R_{\oplus}}$, stars with effective temperature between 3660 and 7600~K and, finally, we removed potential false positives, that is, planet candidates in the dataset with {\it score parameter} smaller than 0.5. In this section, we compare the results of our simulations directly with the Kepler sample. However, as observational data are expected to suffer from observational bias, in the following section we perform a more systematic analysis where we attempt to quantify the effects of observational data when comparing our synthetic planetary systems with Kepler observations.
\begin{figure*}
\centering
\includegraphics[scale=.4]{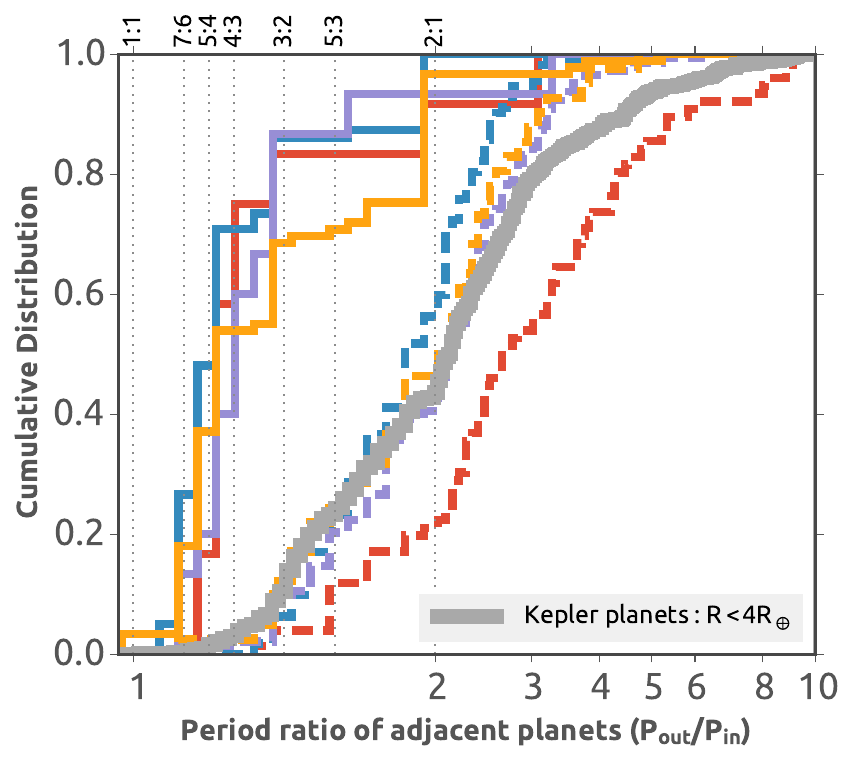}
\includegraphics[scale=.4]{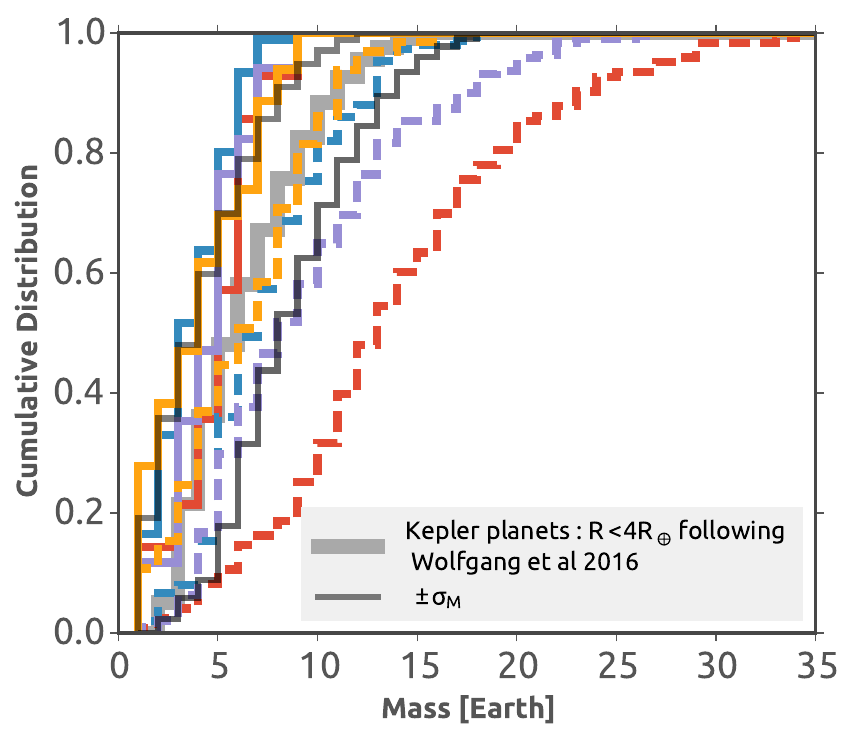}
\includegraphics[scale=.4]{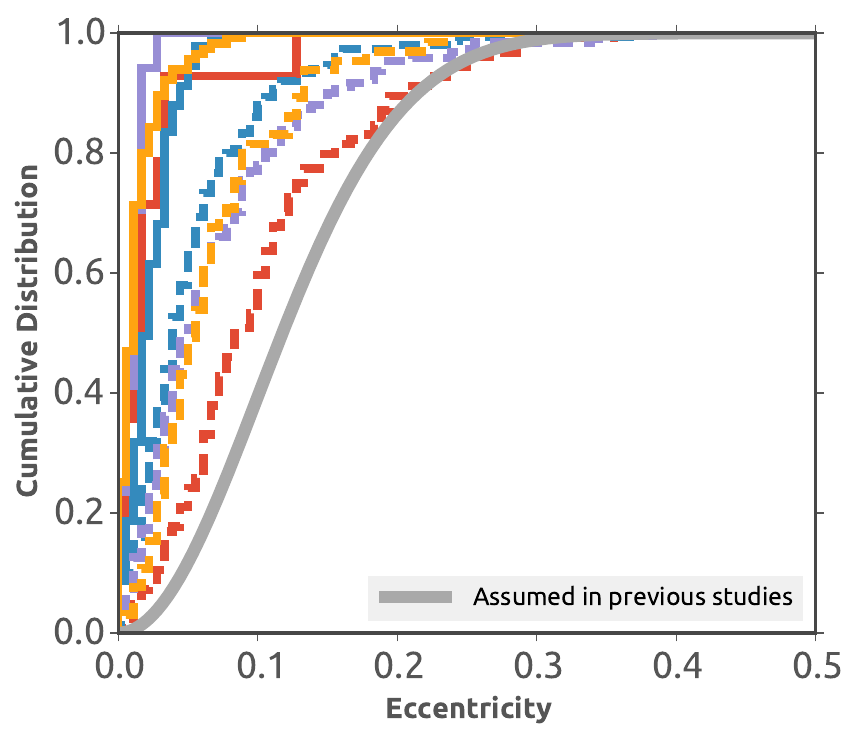}

\includegraphics[scale=.4]{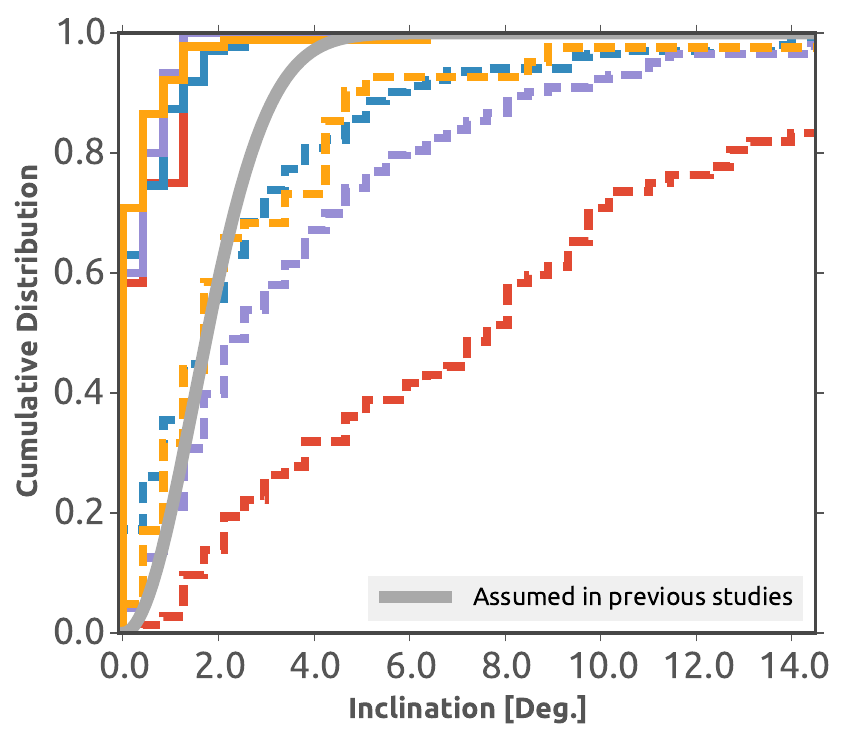}
\includegraphics[scale=.4]{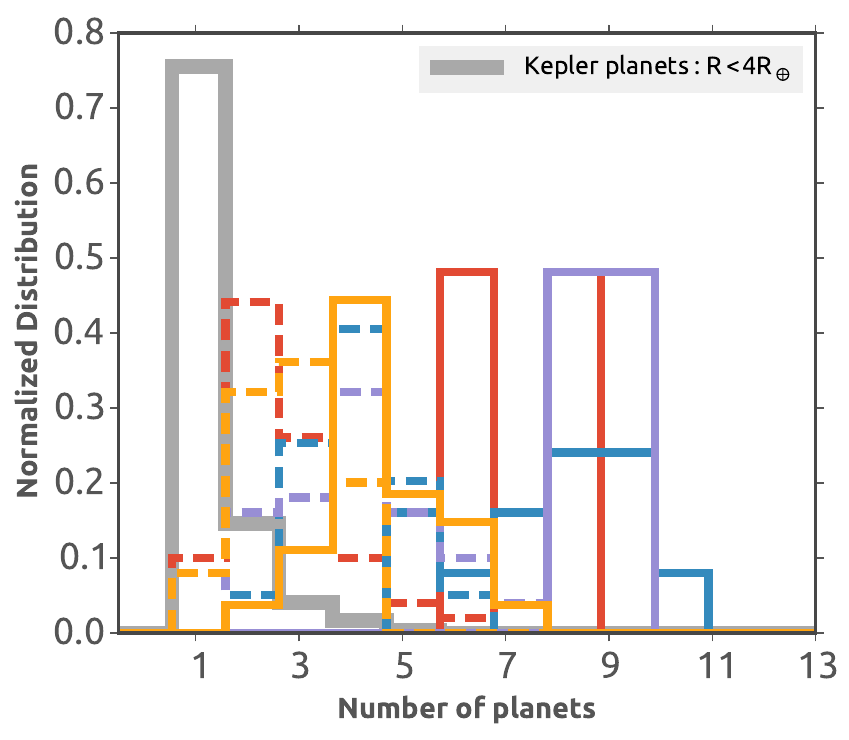}
\includegraphics[scale=.4]{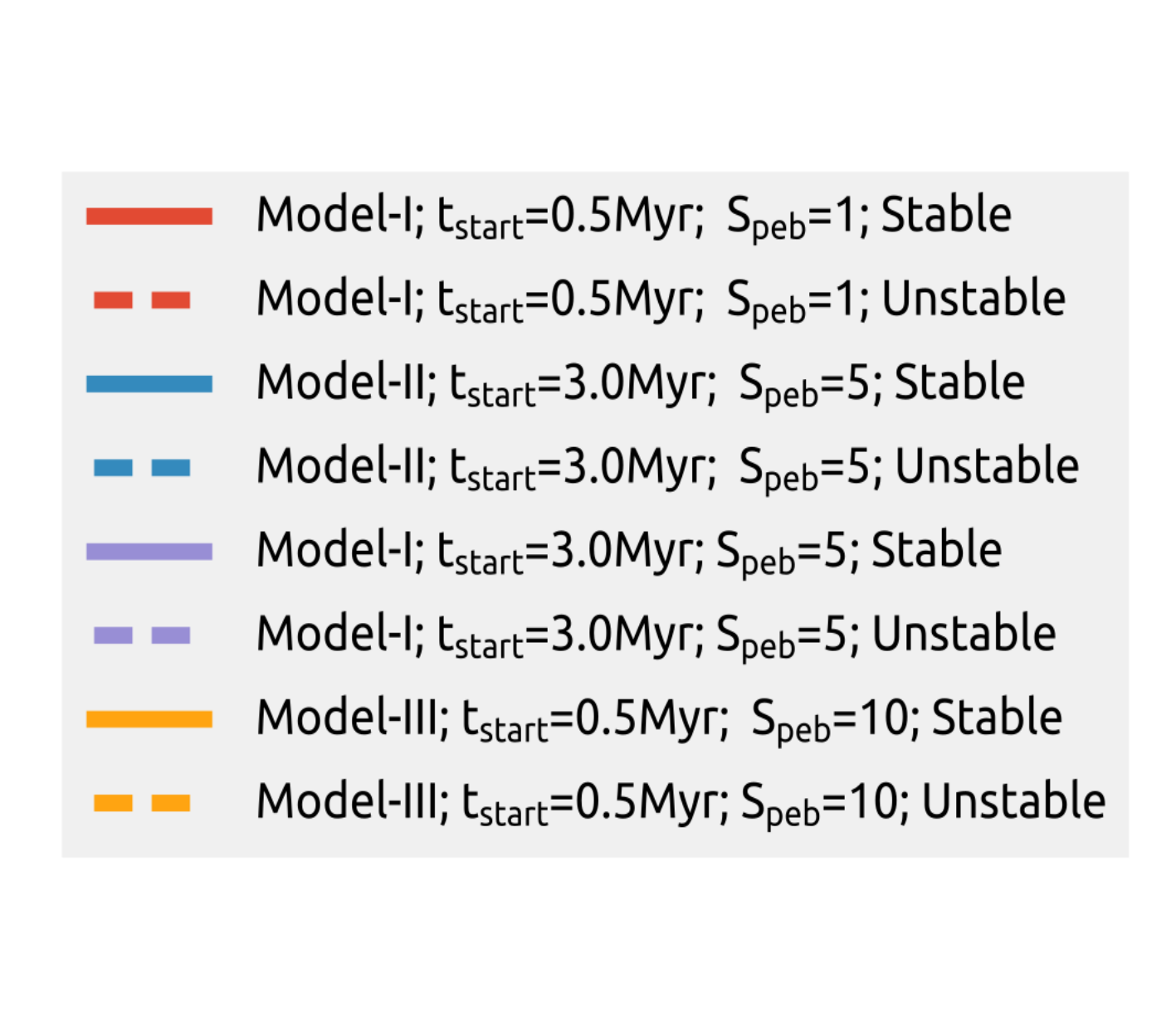}
\caption{Dynamical architecture of stable and unstable planetary systems anchored at the disk inner edge at the end of the simulations (50~Myr).  From left to right, the top panels show the cumulative distributions of period ratio, mass, and orbital eccentricity. From left to right, the bottom panels show the distributions of orbital inclination and number of planets. Only planets with semi-major axes smaller than 0.7~AU and masses larger than ${\rm 1~M_{\oplus}}$ are considered. In all considered models ${\rm Rocky_{{\rm peb}}}$=0.1~cm.  The Kepler planet mass distribution inferred from the mass--radius relationship of \citet{wolfgangetal16} is shown as a thick gray line. The  uncertainty in the mass distribution is shown by the thin gray lines (${\rm M \pm \sigma_M}$).}
    \label{fig:statistics}
\end{figure*}

Figure \ref{fig:statistics} shows the cumulative distributions of (a) the period ratio of adjacent planet pairs, (b) planet masses, (c) number of planets, (d) orbital eccentricities, and (e) orbital inclinations of all four different setups of Table \ref{tab:2}. Planets or planet pairs belonging to stable systems are shown with solid lines and unstable ones with dashed lines. The period ratio distribution of adjacent planet pairs shows that stable systems are dynamically very compact at the end of the gas disk phase. Most adjacent planet pairs belonging to dynamically stable systems exhibit period ratios smaller than two. stable adjacent planet pairs are typically locked in first-order mean motion resonances.  Systems becoming dynamically unstable at the very end or after the gas dispersal break resonances  and become dynamically much less compact. Planet pairs in unstable systems are typically not locked in first-order mean motion resonances. These results are consistent with those of \cite{izidoroetal17}.

The gray line in the top left-hand panel of Figure \ref{fig:statistics} shows the period ratio distribution of adjacent Kepler planets (${\rm R < 4~R_{\oplus}}$). While the period ratio distribution of stable systems is drastically different from the Kepler sample for all our models, the period ratio distributions of adjacent planets in unstable systems in the various models span from slightly more  compact than observed to even more spread out  (see e.g., the purple and red dashed lines in the top-left panel of Figure \ref{fig:statistics}).

Figure \ref{fig:statistics} (middle-top panel) shows that stable systems have lower mass planets than unstable ones, which is expected. Dynamical instabilities promote a late phase of accretion. Indeed, at the inner edge of the disk, dynamical instabilities rarely result in ejection of planetary bodies from the system, but instead result in accretion.  Typical stable systems have planets with masses  lower than ${\rm \sim10~M_{\oplus}}$ and a median value of ${\rm \sim3-4~M_{\oplus}}$ for all scenarios of Table \ref{tab:2}. However, the mass distributions of unstable systems are dramatically different.

The blue dashed line representing unstable systems of Model II with ${\rm t_{start}=3.0~Myr}$,  ${S_{{\rm peb}}= 5,}$ and ${\rm Rocky_{{\rm peb}}=0.1~cm}$ produced final planets with a median mass of ${\rm \sim7.5~M_{\oplus}}$. The red dashed line representing unstable systems of Model I with ${\rm t_{start}= 0.5~Myr}$,  ${S_{{\rm peb}}= 1,}$ and ${\rm Rocky_{{\rm peb}}=0.1~cm}$ shows a median planet mass of about ${\rm \sim15~M_{\oplus}}$, a factor of two larger. This result  becomes easier to understand by revisiting Figure \ref{fig:whyunstable}. Although the mass distributions of Figure \ref{fig:statistics} are similar for stable systems, the differences among the mass distributions of unstable systems  are remarkable because the typical number of planets in the system in each of these models is different (see middle panel of Figure \ref{fig:whyunstable}). Thus, even if planets produced in different scenarios have similar masses before gas dispersal, very compact systems with a larger number of planets in the resonant chain generally produce more massive planets after instabilities.

It is not straightforward to compare the masses of simulated planets with those in the Kepler sample. Kepler observations typically provide planet radii rather than  masses. The masses of Kepler planets have been estimated in several  studies, for example by fitting empirical mass--radius relations from well-characterized planets or by exploring probabilistic aspects of the mass--radius relation \citep{lissaueretal11b,fangmargot12,weissmarcy14,wolfgangetal16}.  Here we use the  probabilistic mass--radius relation of   \citet{wolfgangetal16} which yields ${\rm M = 2.6 \left(R/R_\oplus\right)^{1.3}}$ with a standard deviation ${\rm \sigma_M = \sqrt{4.41 + 1.5\left(R/R_\oplus -1\right)}}$. Because we impose a cutoff of ${\rm R<4~R_{\oplus}}$ in the Kepler sample, the maximum mass of Kepler planets inferred from Wolfgang's mass--radius relation is ${\rm \sim 18.9~M_{\oplus}}$. With this relation, the median planet mass in our Kepler sample is ${\rm6.4~M_{\oplus}}$. The gray lines in the top-middle panel of Figure \ref{fig:statistics} show the expected mass (${\rm M}$; thick line) and mass dispersion (${\rm M \pm \sigma_M}$; thin lines) distributions of  Kepler planets.


The planet-mass distribution of unstable systems of model I (${\rm t_{start}= 0.5~Myr}$, ${S_{{\rm peb}}= 1,}$ and ${\rm Rocky_{{\rm peb}}=0.1~cm}$; red dashed line in Figure \ref{fig:statistics}) shows that these planets are too massive compared to the expected masses of Kepler planets. About 20\% of planets in these systems have masses larger than ${\rm 18.9~M_{\oplus}}$.  Planet masses in unstable systems of model I with ${\rm t_{start}= 3.0~Myr}$ and  ${S_{{\rm peb}}= 5}$ are  marginally consistent with those expected for  Kepler planets. Masses of planets in unstable systems of Models II and III are typically  lower than ${\rm18.9~M_{\oplus}}$, in good agreement with the expected masses of  Kepler planets. Figure \ref{fig:statistics}  also shows a clear trend. Planet pairs in Model I  with ${\rm t_{start}=0.5~Myr}$ typically have  larger period ratios than planet pairs composed of lower mass planets (Model I  with ${\rm t_{start}= 3.0~Myr}$). This result is consistent with those of \citet{izidoroetal17}.

The orbital eccentricities and inclinations of planets in stable and unstable systems are also dramatically different. Planets in stable systems have low eccentricity and obits with low orbital inclination, while planets in unstable systems are dynamically excited. Statistical analyses have inferred the orbital eccentricity and inclination distributions of Kepler planets  \citep{lissaueretal11a,kaneetal12,tremainedong12,figueiraetal12,fabryckyetal14,plavchanetal14,ballardjohnson14,vaneylenalbrechts15,xieetal16,vaneylenetal18}. Following  \citet{izidoroetal17}, these distributions are represented by Rayleigh distributions with $\sigma_e=0.1$ (e.g., \cite{moorheadetal11}) and $\sigma_i=1.5^\circ$ (e.g., \cite{fangmargot12}) for eccentricity and inclination, respectively. These distributions are represented by the gray lines in the top right-hand and bottom left-hand  panels of Figure \ref{fig:statistics}. The median orbital eccentricities of planets in unstable systems range from 0.05 to 0.1. unstable planetary systems produced in Model I are clearly the most excited ones. This result is easy to understand  by inspecting the corresponding mass distributions, which are the most skewed towards massive planets. The red dashed line representing  Model I in Fig. 15 also corresponds to the most massive planets.  Higher mass planets have more violent dynamical instabilities and consequently their final orbital architectures are more excited \citep{puwu15,izidoroetal17}. Finally, the bottom middle panel of Figure \ref{fig:statistics}  shows the planet multiplicity distributions in our systems and in observations. As already discussed,  stable systems have a larger number of planets  than unstable systems. None of our systems match the high number of Kepler systems with a single transiting planet but we revisit this issue below when we account for the observational effects in our simulations.

\subsubsection{Synthetic observations  and a quantitative comparison with Kepler data}

Figure \ref{fig:statistics} shows that, for some of our models, the  period ratio distributions of unstable systems constitute a reasonable match to observations by themselves. Although encouraging, a more effective comparison requires an attempt to quantify the effects of observational bias. We start by first comparing the period distribution of planets in our simulations with the debiased observed period distribution from~\cite{fressinetal13}. Later in this section, we describe simulations of observations of our planetary systems to compare with other observed distributions (e.g., period ratio and planet multiplicity) of the  Kepler sample.

\begin{figure}
\centering
\includegraphics[scale=.5]{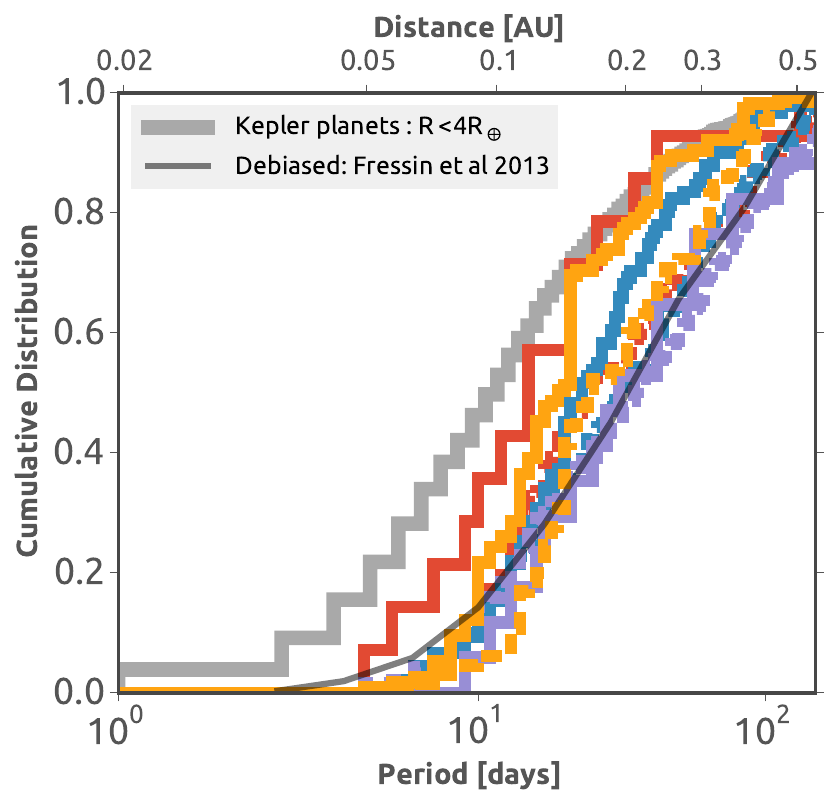}
\caption{Orbital period distribution of stable and unstable planetary systems anchored at the disk inner edge at the end of the simulations (50~Myr).  The color and style of each line representing our simulations correspond to those shown in the legend of Figure \ref{fig:statistics}. The thick gray line shows the period distribution of Kepler planets before correction for transit probabilities and detection biases. The thin gray line shows the debiased period distribution following \cite{fressinetal13}. The top x-axis shows the corresponding orbital distance calculated assuming a solar-mass star.}
    \label{fig:period}
\end{figure}

Figure \ref{fig:period} shows the period distribution of planetary systems in each of our models (see legend of Figure \ref{fig:statistics} for the description of each model). The Kepler period distribution (thick gray line) and the distribution corrected (thin gray line) for transit probability and detection biases~\citep{fressinetal13} are also shown. While the Kepler sample shows a distribution that is  shifted towards low orbital periods, the period distribution of our simulations match the debiased  observations fairly well. Our Model  (purple lines) provides a good match to the debiased sample, in particular for P$\gtrapprox$10~days (which corresponds to orbital distances greater than $\sim$0.09~AU for a solar mass star). It is  important to keep in mind that the inner edge of our disk is at about $\sim$0.1~AU (P$\approx$11~days), so it is natural to expect that planet occurrence in our simulation will be lower inside 0.1~AU. However, the innermost planets may still be pushed inside the disk inner cavity by the outermost migrating planets, as one can see for Model I (solid red line; see also \cite{carreraetal19}). Finally, the location of the disk inner edge is a free parameter in our model and we do not include the effects of tides in this work. A disk inner edge closer to the star and stellar tides may have a significant impact on the occurrence of planets at very short orbital periods~\citep{leechiang17,carreraetal19}.

To compare our model with the observed distribution of period ratios between adjacent planets (and planet multiplicity distribution), we need to apply observational biases to our model. To understand why biases may be important, imagine for instance a model system of three planets with adjacent period ratios P2/P1 and P3/P2. If, for some reason, the probability of observing the middle planet is small compared that of observing each of   the other two, then the observed period ratio of what appears to be adjacent planets would be P3/P1.  Unlike the period distribution, it is very hard to debias the observed data on period ratios, and therefore
applying the observational biases to the model (planetary systems produced in our simulations) is a preferable approach.

The transit probability of a planet in a circular orbit is ${ R_{\star}/a}$, where ${ R_{\star}}$ and ${ a}$ are the stellar radius and planet semi-major axis, respectively. More importantly, transits are only detectable if the planet's orbital plane is sufficiently near to the line of sight between the observer and the star. Following  \citet{izidoroetal17}  we simulated transit observations of the planetary systems produced in our N-body simulations. Each planetary system coming from our N-body simulations  is observed from a large number of lines of sight evenly spaced by 0.1 degree from angles spanning from 30 to -30 degrees in relation  to the arbitrary plane i=0. Azimuthal viewing angles are evenly spaced by 1 degree and span from 0 to 360 degrees.

For a given line of sight, if at least one planet is detected\footnote{In our synthetic observation model, a detection is characterized by a planet transiting in front of the star. We do not attempt to model the signal-to-noise ratio of the light curve, for example, and so our biases are simply due to the geometry of the system relative to  the view direction.}   we store the orbital details of each detect planet and from the detected planet(s) we create a new observed planetary system.
The top left-hand panel of Figure \ref{fig:simulated_observations} shows the period ratio distribution of stable and unstable detected systems of Model I with ${\rm t_{start}= 0.5~Myr}$. In the whole of our analysis, we only consider  adjacent planet pairs with ${\rm P_{out}/P_{in}} < 10$. The synthetic observed distributions (black dashed/solid lines) are more compact than the respective original ones. This is because if two adjacent planets are widely spaced radially we are more likely to observe only one of them (typically the inner one) and therefore miss the information on their wide separation. Instead, planets close to each other are more likely to be observed (i.e., both of them) and therefore the datum on their narrow separation is unlikely to be missed.


This effect is more pronounced for unstable systems where adjacent planet pairs are typically not resonant but mutually inclined. Only planet pairs with sufficiently small mutual orbital inclinations are detected in unstable systems. Planet pairs with small mutual orbital inclinations are generally the more compact planet pairs as well (see Figure \ref{fig:statistics}). Thus, the period ratio distribution of detected unstable systems tends  to be significantly more compact than the real one.  We anticipate that this result is qualitatively different from that found in \citet{izidoroetal17}, where synthetic observations of unstable systems produced an even greater spread of planetary systems for the following reason.


In our simplistic synthetic simulations of transit observations, a single planetary system produced in our N-body simulations is observed from several different lines of sight.  Thus, synthetic observations of a single N-body system result in several observed systems, which can contribute to the observed period ratio distribution with thousands of identical planet pairs observed from different lines of sight. One of the challenges in performing this analysis is to decide how to weight such contribution when calculating the period ratio distribution. The total number of retrieved planet pairs  after combining observations from all lines of sight may vary drastically from one N-body system to another. For example, a planetary system with planets in almost coplanar orbits  is probably successfully observed for several viewing angles aligned with the planets' orbital plane, thus producing a very large number of planet pairs through the survey simulator.  On the other hand, a planetary system with planets on inclined orbits is probably more rarely  observed, producing a smaller number of planet pairs. \citet{izidoroetal17} computed the period ratio distribution of observed planet pairs through the survey simulator, weighting the cumulative distribution by N-body system. In other words, even if a given N-body system was observed $n$ times by the survey simulator, its characteristics were counted in the resulting distribution only once. In this work, we decide to normalize the distributions by  ``detected system'' rather than by N-body system. Thus, the characteristics of an N-body system detected $n$ times are counted $n$ times in the resulting distributions. We believe our new approach is more appropriate than that used in \citet{izidoroetal17}.

The difference between the observed and real distributions is more marginal for all other models (see middle and bottom left-hand panels in Fig. \ref{fig:simulated_observations}) because the orbital separation between adjacent planets  and mutual inclinations are significantly smaller than in the case of Model I with ${\rm t_{start}= 0.5~Myr}$.


\paragraph{Matching Kepler observations with a mix of stable and unstable systems}\mbox{}\\

Figure \ref{fig:simulated_observations}  shows that the detected period ratio distributions of our unstable systems can in some case match the Kepler distribution reasonably well. Nevertheless, here we attempt to statistically match  Kepler observations by mixing a  fraction of stable and unstable systems. As discussed in \citet{izidoroetal17}, though most Kepler planet pairs are not resonant, a few known planetary systems do have planets locked in long resonant chains. This is the case of Kepler-223 \citep{millsetal16} and TRAPPIST-1 \citep{gillonetal17,lugeretal17}, for instance. Detected unstable planet pairs (from simulations)  alone do not account for these long resonant chain planetary systems. In order to truly match observations, we therefore need to invoke a mixture of stable and unstable systems \citep{izidoroetal17}. We recall that  we have from our N-body simulations the ratio of stable and unstable systems  in each of our models (see Figure \ref{fig:lastcollision}). Now, to compare the outcome of our simulated detections with the Kepler data we take that ratio of real stable and real unstable systems to be a free parameter whose value is determined by fitting the observations. To perform this fit, we proceed as follows.

For each assumed fraction of detected stable planet pairs we randomly select from our extensive database of  stable and unstable simulated observations a number of stable and unstable planet pairs that, when added together, yields a number of planet pairs equivalent to that in the Kepler sample. For each assumed fraction of detected stable pairs we repeat this procedure 1000 times calculating the KS p-value of each subsample and Kepler observation. The p-value associated with a given fraction of detected stable planet pairs is the  mean p-value computed from these 1000 subsamples. As we are fundamentally interested in assessing the real fraction of stable systems, we transform the fraction of detected stable planet pairs into an estimate of the real fraction of stable systems. We do this by dividing the fraction of detected stable planet pairs by the ratio between the mean number of planet pairs detected in stable systems and the mean number of planet pairs detected in unstable systems. This transformation is important because synthetic observations of stable systems tend to retrieve a much larger number of planets pairs than those of unstable systems.

The right-hand panels of Figure \ref{fig:simulated_observations}  show the p-values for samples mixing different real fractions of stable  and unstable  systems. We consider that whenever the p-value of the KS-test is higher than 5\% the model distribution and the observed distribution  are in good statistical agreement, that is, the assumed fraction of stable systems is acceptable.

The top-right panel  of Figure \ref{fig:simulated_observations} shows that our simulated observations match the Kepler sample if less than ${\rm \sim1-5\%}$ of systems remain stable. This also holds true for simulated observations of Model I with ${\rm t_{start}= 3.0~Myr}$ and Model III (right-second-from-top and bottom panels of Figure \ref{fig:simulated_observations}). However, our KS-tests show that the period ratio distribution of observed planets in Model II do not match observations for any real fraction of stable systems because its unstable systems  are not sufficiently spread.

\citet{izidoroetal17} obtained a larger upper bound to the fraction of possible stable systems ($\lesssim 25\%$) than the one obtained here (${\rm \lesssim 5-10\%}$) because the p-values calculated in \citet{izidoroetal17} take  a very reduced effective sample size. Moreover, we note that. in their analysis. \citet{izidoroetal17} only considered  planet pairs with ${\rm P_{out}/P_{in} < 3}$ and with masses lower than ${\rm 18.9~M_{\oplus}}$. We have relaxed these cutoffs  in our analysis because our simulations produced lower mass planets than those of \citet{izidoroetal17} (compare Figure \ref{fig:statistics} with Figure 12 of  \citet{izidoroetal17}). 
More importantly, we recall that the construction of the observed distributions in this work is different from that of \cite{izidoroetal17}. In the new method, stable systems  ---which are very coplanar and consequently commonly observed--- have a huge weight in the distribution. Therefore,  only a few of them can be accommodated when combining stable and unstable systems to match observations. 


Although the results of our study suggest that $>$95\% of the systems have to become unstable after gas dispersal, only the simulations of Model I achieve this result (Figure \ref{fig:lastcollision}). These simulations match the period ratio distribution of observations naturally, with their intrinsic fraction of unstable and stable systems. Therefore, when we consider the mixing ratio of unstable and stable systems as a free parameter we are just inferring the range of mixing ratios that would still be consistent with observations. This is different for Model II and Model III.  Both Model II and Model III give a significantly smaller fraction of unstable systems, and so their  period ratio distribution produced via simulated observations does not naturally match Kepler observations. This implies that for these models, some additional mechanism is needed to trigger instabilities and rise the fraction of unstable systems (e.g., that discussed in \citet{spaldingbatygin16}).

The high fraction of unstable systems produced in Model I is  nevertheless a significant improvement relative to \cite{izidoroetal17}, where only 50\% of the systems happened to become unstable after gas removal. On the other hand, we reiterate the fact that not all our models are equally successful in matching other observational constraints. For instance, model I with  ${\rm t_{start}= 0.5~Myr}$ and  ${S_{{\rm peb}}= 1}$ and model II with ${\rm t_{start}= 3.0~Myr}$ and  ${S_{{\rm peb}}= 5}$  are the least favored models overall. The former model produces planets that are systematically too massive (top-middle panel of Figure \ref{fig:statistics}) while the latter produces systems that are 
too dynamically compact, and do not match the Kepler period ratio distribution for any real fraction of stable systems. Therefore, our favored  models are model I with ${\rm t_{start}= 3.0~Myr}$ and  ${S_{{\rm peb}}= 5}$ and model III with ${\rm t_{start}= 0.5~Myr}$ and  ${S_{{\rm peb}}= 10}$.

Each of our favored models has its own caveats. The masses of planets  in model I with ${\rm t_{start}= 3.0~Myr}$ and  ${S_{{\rm peb}}= 5}$ are dangerously high compared to the estimated masses of Kepler planets, but our masses should be taken as upper limits because we do not model fragmentation and volatile loss in giant impacts \citep{marcusetal10,stewartleinhardt12}. Also, perhaps we are simply missing some ingredient to make the fraction of unstable systems in model III higher. A possibility is that our simulations, covering just 50~Myr, are too short. On longer timescales, more systems would become dynamically unstable. Another possibility is that we are missing some physics, such as for example the interaction of the planets with planetesimals remaining in the system \citep{chatterjeeford15}, tidal interactions with the star \citep{bolmontmathis16}, or even spin-orbit misalignments effects \citep{spaldingbatygin16}. \citet{adamsetal08} suggested that turbulence in the disk may prevent deep locking in resonance, increasing the post-gas instability fraction. However, \citet{izidoroetal17} found no difference between systems produced in simulations with or without turbulence. Thus, we did not attempt turbulent simulations here.

Next, we evaluate how these simulations match other observational constraints.




\subsection{The Kepler dichotomy}\label{sec:keplerdichotomy}

The Kepler sample has a much larger number of planetary systems exhibiting single transiting planets than  multi-transiting ones \citep{lissaueretal11a,fabryckyetal14}. Almost 80\% of the Kepler candidates are in single transiting systems. This has been referred to as the Kepler dichotomy \citep{johansenetal12,fangmargot12,ballardjohnson14,moriartyballard16}. Different scenarios have been proposed as an attempt to explain this dichotomy. On one hand, it has been suggested that this remarkable excess of single transiting planets is indeed real, that is, single transiting planets are truly singles \citep[e.g.,][]{johansenetal12}. On the other hand, it has been suggested that the dichotomy is only apparent and that the excess of single transiting planets arises from observing higher multiplicity systems where only one planet transits \cite[e.g.,][]{izidoroetal17}.


Synthetic transit observations of a mix of stable and unstable systems produced in \citet{izidoroetal17} suggest that the Kepler dichotomy arises from observing multi-planet systems  with planets in mutually inclined orbits. Systems with low mutual inclination (mostly  stable systems) contribute mostly to observed high-N planet systems and large mutual inclinations of unstable systems  naturally produce a peak in the observed planet-multiplicity distribution at ${\rm N_{det}=1}$. \citet{izidoroetal17} matched the Kepler dichotomy assuming that Kepler planets comprise  about $\lesssim10\%$ of stable and $\gtrsim$90\% of unstable systems. If this is correct, the Kepler dichotomy consists of a dichotomy in the inclination distribution rather than in planet multiplicity. Therefore, here, we also investigate how our simulations match the Kepler dichotomy and how our results compare to \citet{izidoroetal17}.
\begin{figure*}
\centering
\includegraphics[scale=.45]{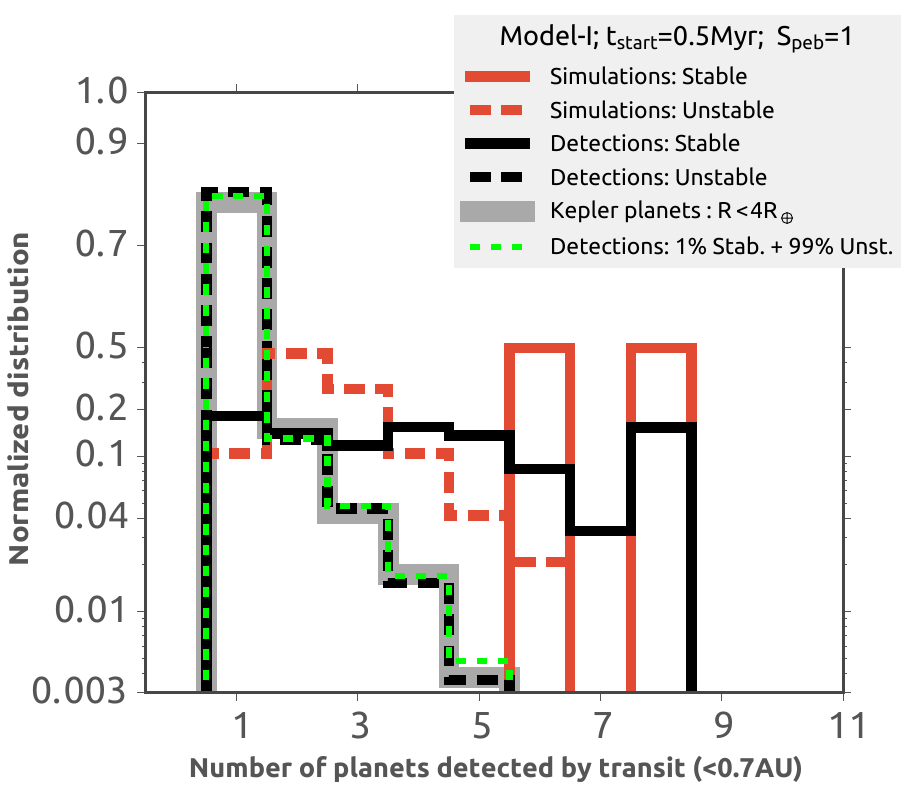}
\includegraphics[scale=.45]{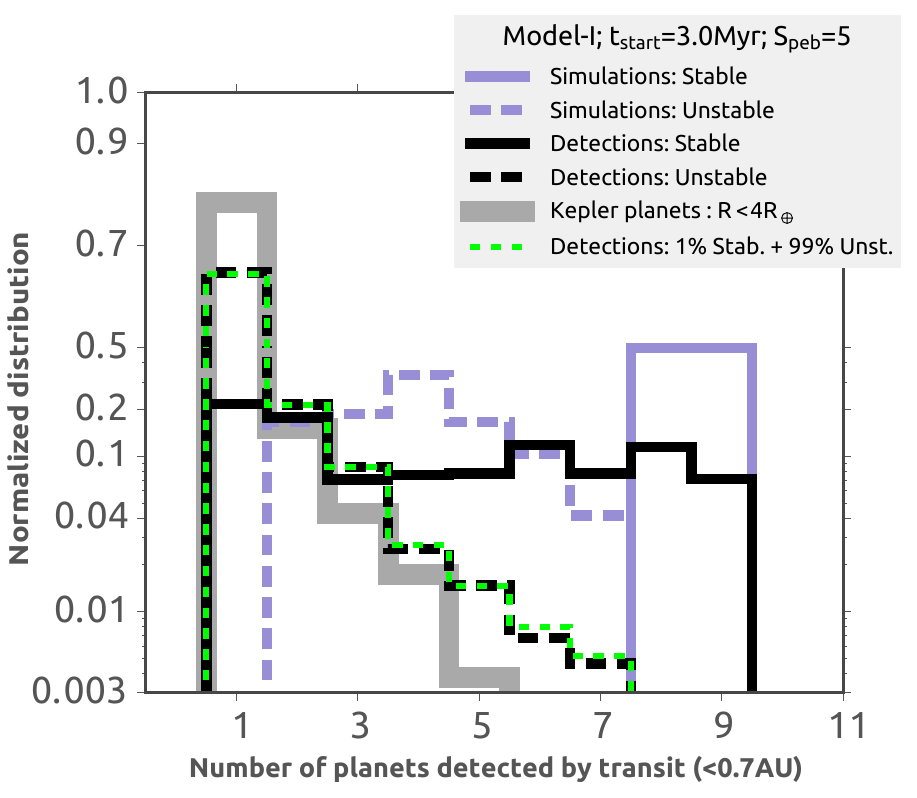}
\includegraphics[scale=.45]{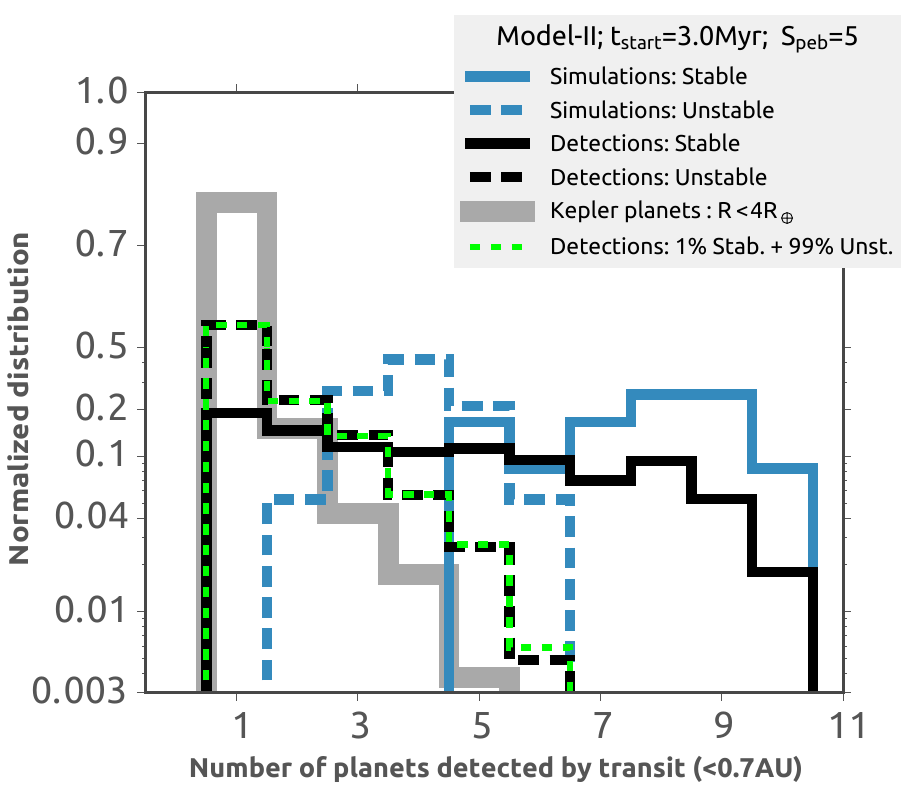}
\includegraphics[scale=.45]{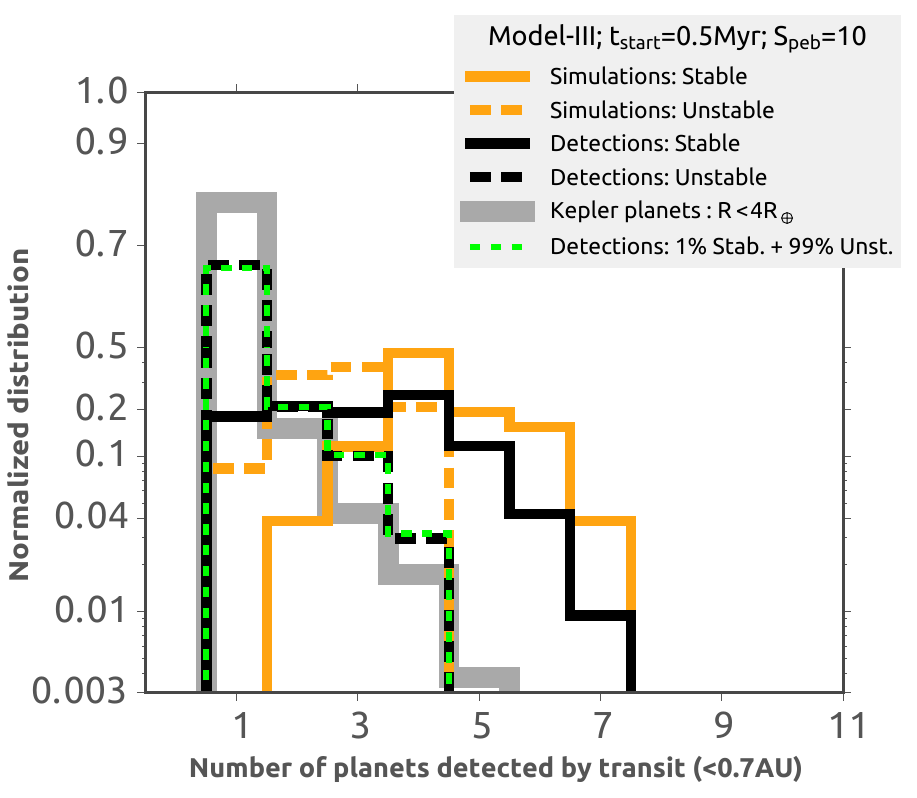}
\caption{Number of planets detected by transit in synthetic observations of planetary systems produced in our N-body simulations of Model I with ${\rm t_{start}= 0.5~Myr}$ and ${S_{{\rm peb}}= 1}$ (red), Model I with ${\rm t_{start}= 3.0~Myr}$ and ${S_{{\rm peb}}= 5}$  (purple), Model II with ${\rm t_{start}= 3.0~Myr}$ and ${S_{{\rm peb}}= 5}$ (blue), and Model III with ${\rm t_{start}= 0.5~Myr}$ and ${S_{{\rm peb}}= 10}$  (yellow). For all cases, ${\rm Rocky_{{\rm peb}}}$=0.1~cm. Solid lines show stable systems, and dashed  lines show unstable systems. The black lines show the results of our synthetic observations. The densely dashed green lines show  multiplicity distribution combining different real fractions of stable and unstable  systems. We note that the y-axes combine linear and log scaling.}
    \label{fig:dichotomy}
\end{figure*}

Figure \ref{fig:dichotomy} shows the multiplicity distributions of our  synthetically detected  systems compared with that of the Kepler sample.  Here we make use of the simulated detections produced earlier in this section.  We note that more than 90\% of the unstable planetary systems of Model I with ${\rm  t_{start}=0.5~Myr}$ have two or more planets  inside 0.7~AU (red dashed line in the left-hand panel of Figure \ref{fig:dichotomy}). The two stable systems of Model I with ${\rm t_{start}= 0.5~Myr}$ have six and eight planets in their chains.

Simulated detections of unstable systems of Model I with ${\rm t_{start}= 0.5~Myr}$ provide a very good  match to the Kepler multiplicity. Simulated observations of unstable systems rise the fraction of single planets from $\sim$10\%\footnote{in reality, these unstable systems have additional planets but they are beyond 0.7~AU and thus do not account in our analysis. We also stress that these systems formed with multiple planets inside 0.7~AU but dynamical instabilities after gas dispersal resulted in multiple scattering events and collisions resulting in a single planet inside 0.7~AU. } to about $\sim$75\%. This result is consistent with that of \citet{izidoroetal17}. Simulated observations of stable planetary systems result in a very spread and almost flat multiplicity distribution.

We show that our simulated observations match the period ratio distribution of the Kepler sample if one mixes  typically $\lesssim5\%$ of  stable systems with $\gtrsim95\%$ of unstable ones. Figure \ref{fig:dichotomy} shows that mixing 1\% stable and 99\% unstable  systems provides an almost perfect match to the Kepler dichotomy in Model I with ${\rm t_{start}=0.5~Myr}$. The middle and right-hand panels of Figure \ref{fig:dichotomy} show that Model I with ${\rm  t_{start}=3.0~Myr}$ and Model III with ${\rm  t_{start}=0.5~Myr}$ do not match the Kepler dichotomy equally well, even assuming that only 1\% of the  systems are stable. The detected multiplicity distribution in the middle and right-hand panels of Figure \ref{fig:dichotomy} show a  deficit at ${\rm N_{det}=1}$ (dashed green line) when compared to the Kepler population. Model
II provides a poorer match to the observed multiplicity distribution compared to other models (see bottom-left panel of Figure \ref{fig:dichotomy}). We also reiterate the fact that Model
II fails to match Kepler observations period ratio distribution (Figure \ref{fig:simulated_observations}).
\begin{figure*}
\centering
\includegraphics[scale=.5]{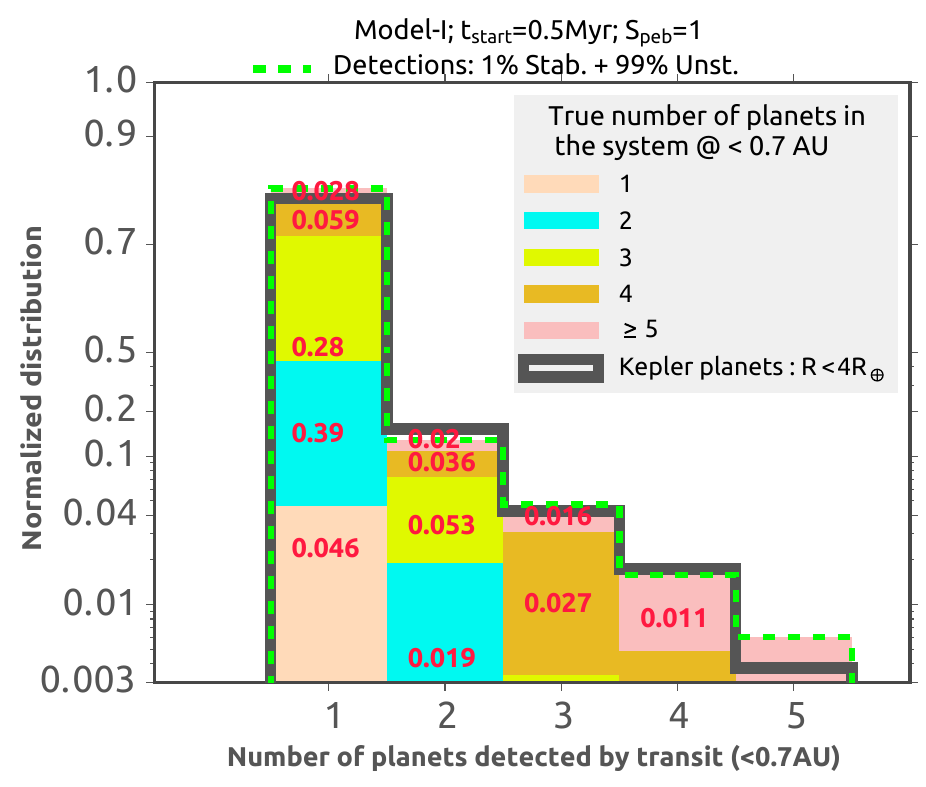}
\includegraphics[scale=.5]{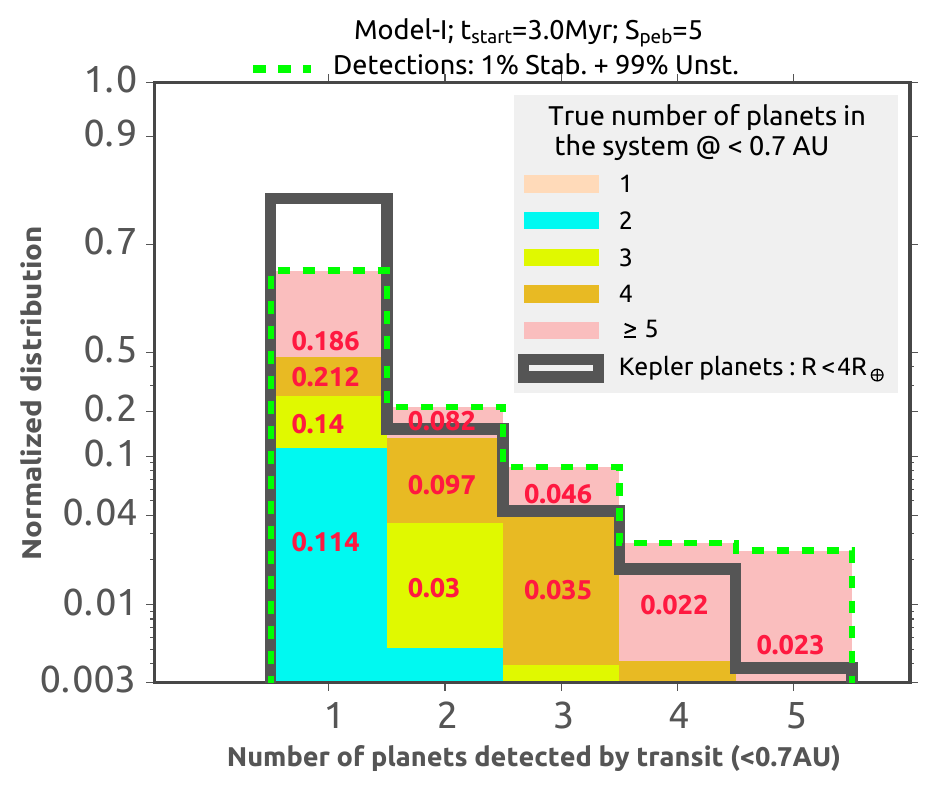}
\caption{Number of planets detected by transit in synthetic observations of planetary systems of two sets of simulations of Model I. The colors filling the dashed green histogram bins show the fractional contributions as a function of true system multiplicity. The left and right panels show simulations   with ${\rm t_{start}= 0.5~Myr}$, and  ${\rm t_{start}= 3.0~Myr}$, respectively. In both cases we use ${\rm Rocky_{{\rm peb}}}$=0.1~cm. The dashed green histograms show detection distributions considering different real mixing ratios of stable and unstable systems, as in Figure \ref{fig:dichotomy}. The over-plotted red numbers show the size of each colored sub-bar representing fractional contributions. We note that the y-axes combine linear and log scaling.}
    \label{fig:multiplicity_inclination}
\end{figure*}

To better understand how planetary systems with different (true-)numbers of planets contribute to the origin of the observed dichotomy, we selected simulations of Model I to conduct a deeper analysis. Figure \ref{fig:multiplicity_inclination}  shows again (as in Figure \ref{fig:dichotomy}) the observed planet multiplicity distribution in synthetic transit observations of a mix of 1\%  stable and 99\% unstable systems (dashed green histograms).
The colors filling each ${\rm N_{det}}$ dashed green histogram bin show the fractional contributions as a function of true system multiplicity.  In the left panel, where ${\rm t_{start}= 0.5~Myr}$, 80\% of all the observed systems show a single transit. However, only $\sim$5\% of these systems have only one planet inside 0.7~AU, $\sim$39\% of systems have two planets, $\sim$28\% have three planets, and $\sim$8\% have four or more planets. These results show that, in this case, the excess of single detected planets comes mainly from systems of intrinsically 2-3 super-Earths. In the right panel, where ${\rm t_{start}= 3.0~Myr}$, the peak at ${\rm N_{det}=1}$ constitutes $\sim$65\% of the systems, where $\sim$11\% of the systems have two planets, $\sim$14\% have three planets, $\sim$21\% have four planets, and 18\% of the systems have five or more planets. In this latter case, the relative contributions in terms of the systems' true multiplicity are not dramatically different and are within a factor of two.


Although Model I with ${\rm t_{start}= 3.0~Myr}$ and  ${S_{{\rm peb}}= 5}$ and Model
III with ${\rm t_{start}= 0.5~Myr}$ and  ${S_{{\rm peb}}= 10}$  do not perfectly match the dichotomy,  there may be several false positives in the ${\rm N_{det}=1}$ planetary systems of Kepler. The  rate of false positives for single Kepler planets is estimated to be of 20\%, at least two times higher than the false-positive rate in multi-planet systems \citep{mortonjohnson11,fressinetal13,coughlinetal14,desertetal15,mortonetal16}. Thus, the peak in the Kepler multiplicity distribution at ${\rm N_{det}=1}$ could be shorter in reality. Additionally, some  single-planet systems in the Kepler sample may be  truly singles formed by different mechanisms \citep{izidoroetal15a,izidoroetal17}. Therefore, Model I with ${\rm t_{start}= 3.0~Myr}$ and Model
III with ${\rm t_{start}= 0.5~Myr}$ might be good after all but with an extremely high instability fraction.

\subsection{Super-Earths: Rocky or Icy?}\label{sec:rockyicy}

As discussed above, the migration model has been shown to produce mostly ice-rich super-Earths~\citep[see][]{raymondetal18}.  This is in conflict with the results of atmospheric photoevaporation models, which suggest that the  super-Earths closest to the parent star -- and perhaps the majority of all super-Earths -- are mostly rocky~\citep{owenjackson12,lopezfortney13,owenwu17,lopezetla17,jinmordasinietal18}. The water-rich composition of super-Earths in our nominal simulations is probably more consistent with interior modeling of the composition of   super-Earths, which suggests that super-Earths larger than 2-3$\rearth$ are water-rich worlds \citep{zengetal19}.

In this section we first examine the compositions of close-in super-Earths from the simulations presented above. Then, inspired by the current debate on the true nature of  super-Earths, we perform additional sets of simulations with different assumptions to find the conditions needed to form rocky super-Earths.

\subsubsection{Compositions of super-Earths in Models I, II, and III}

The super-Earths that grew in our Model I and Model
II  simulations (see Table \ref{tab:2}) grew mainly from pebbles originating from beyond  the snow line. Thus, their final ice-mass fractions are as high as 50\% (e.g., Figures \ref{fig:panels_fluxes_model_1}, \ref{fig:panels_fluxes_model_2}). Also, as suggested by Figure \ref{fig:panels_fluxes_model_3_mm}, only Model
III simulations with large refractory pebbles (${\rm Rocky_{{\rm peb}}=1~cm}$) or extremely large pebble fluxes produced multiple inner planets with low water-ice contents.
\begin{figure*}
\centering
\includegraphics[scale=.25]{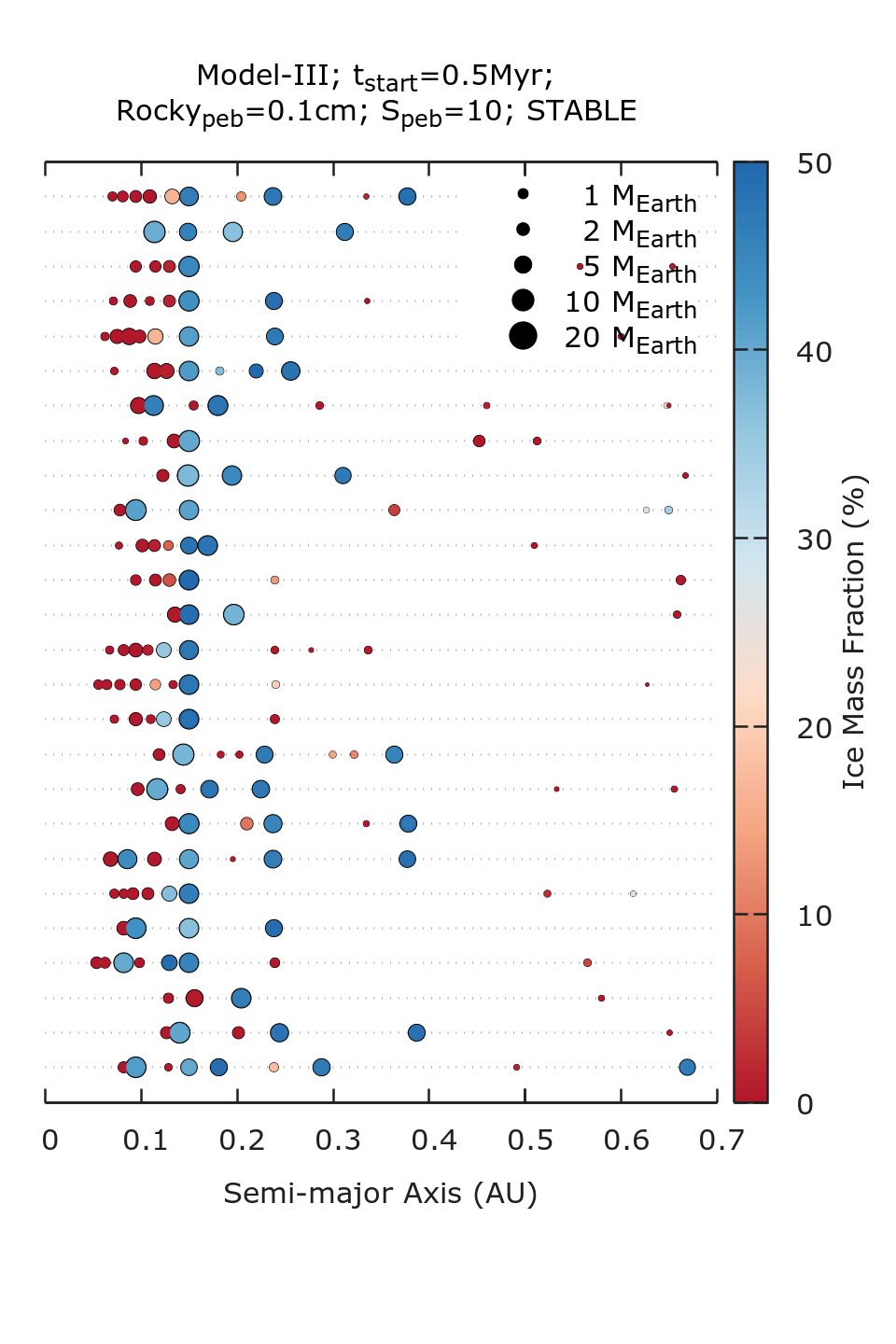}
\includegraphics[scale=.25]{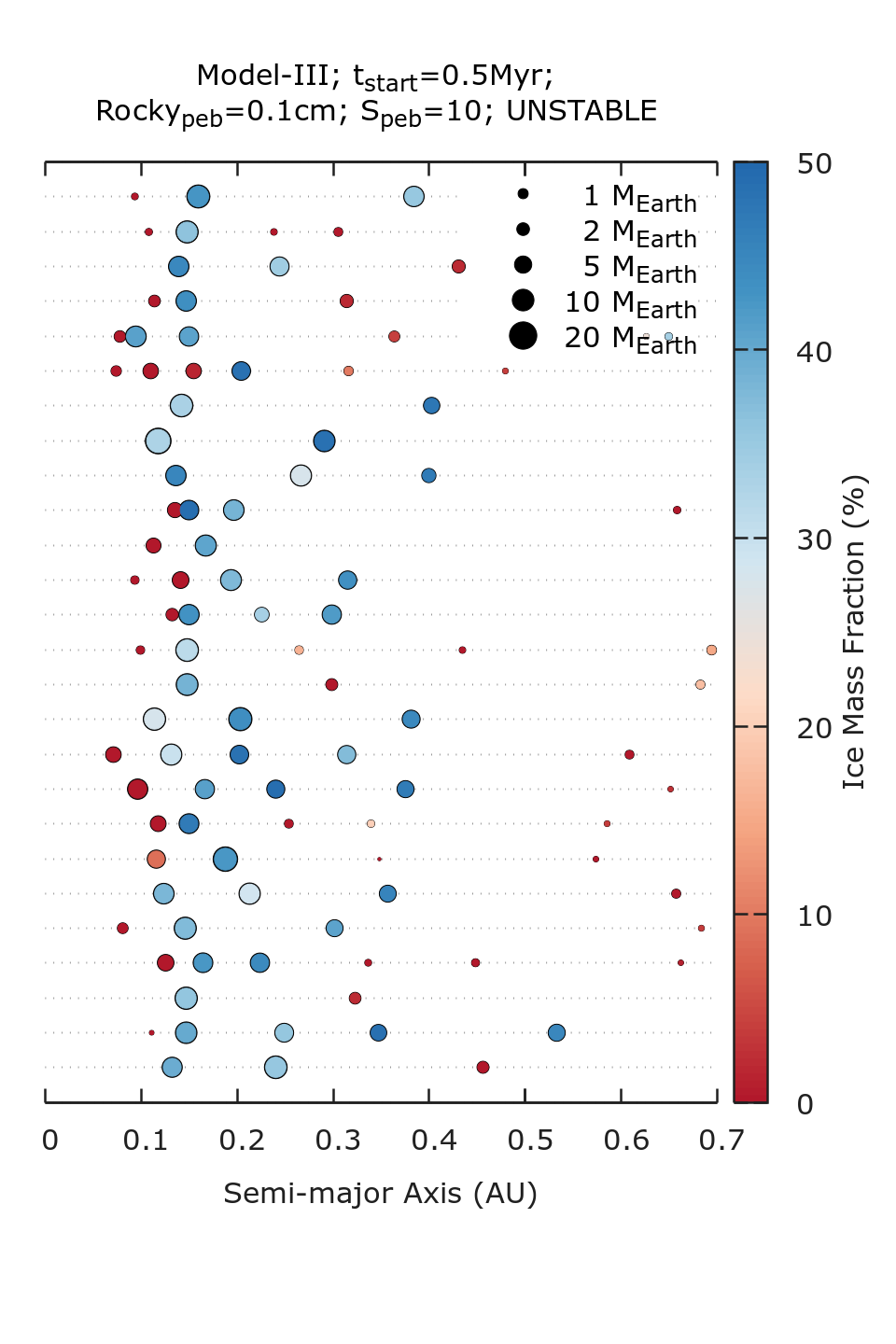}
\vspace{-1cm}
\caption{Planetary systems produced in  Model
III with  ${\rm t_{start}= 0.5~Myr}$,  ${S_{{\rm peb}}=10,}$ and ${\rm Rocky_{{\rm peb}}=0.1~cm}$ from 50 simulations at the end of the simulations (50~Myr). Two panels are shown. The left-hand panel shows the stable systems (50\%) and the right-hand panel shows the unstable ones (50\%). Planets are represented by dots. The sizes of the dots scale with mass as ${\rm m^{1/3}}$. The color of each dot indicates its ice-mass fraction.}
    \label{fig:panel_model_III_systems}
\end{figure*}

Figure \ref{fig:panel_model_III_systems} shows the final planetary systems produced in 50 simulations of Model
III with  $ {\rm t_{start}= 0.5~Myr}$,  ${S_{{\rm peb}}=10,}$ and ${\rm Rocky_{{\rm peb}}=0.1~cm}$. The results of these simulations are separated in two batches. The left-hand panel shows planetary systems that remained stable after the gas disk dispersal. The right-hand panel shows the final planetary systems that became dynamically unstable after the gas dispersal. It is clear that the former are much more compact than the latter for reasons explained above. More importantly, Figure \ref{fig:panel_model_III_systems}  shows a diversity of planetary system compositions. The left-hand panel shows that in dynamically stable systems, the innermost planets are typically rocky. This is a consequence of resonant shepherding.  As the snow line moves sufficiently inwards, it first sweeps the outermost planetary embryos in the disk; these objects grow faster and migrate inwards, and on their way inwards they encounter in resonance growing rocky planetary objects and shepherd them towards the inner edge of the disk. This result supports those of \citet{raymondetal18} who proposed that the migration model is consistent with a variety of super-Earth compositions.

Some planetary systems in  the left-hand panel of Figure \ref{fig:panel_model_III_systems} are particularly interesting, such as for example the second planetary system from the bottom. This planetary system shows five planets with masses ranging between 1.7 and ${\rm \sim9~M_{\oplus}}$ with bulk compositions that alternate radially from rocky to icy. This shows that adjacent planets may have drastically different feeding zones. Even after dynamical instabilities (right-hand panel of Figure \ref{fig:panel_model_III_systems}) the innermost planets in several systems are typically rocky. In some cases, rocky super-Earths end up on orbits that are immediately adjacent to ice-rich super-Earths that are reminiscent of the Kepler-36 system~\citep{carter12} and consistent with the simulations of \cite{raymondetal18}. It is also important to point out that in Model
III with ${\rm Rocky_{{\rm peb}}=0.1~cm}$ the largest rocky close-in super-Earths in stable systems have typical masses of ${\rm 2  M_{\oplus}}$ or lower. Observations of photoevaporated planets show rocky super-Earths up to at least 5${\rm  M_{\oplus}}$.  unstable systems of Model
III with ${\rm Rocky_{{\rm peb}}=0.1~cm}$ do produce some ${\rm   \sim 5M_{\oplus}}$ rocky super-Earths but we can produce them far more easily in Model
III with ${\rm Rocky_{{\rm peb}}=1~cm}$.

We  also note that in our unstable systems of Figure \ref{fig:panel_model_III_systems}, planets farther out are typically icy, resulting in an overall population of mixed compositions with most planets being icy. These icy planets are beyond 0.1~AU, and as  they are typically larger than ${\rm \sim1-2~M_{\oplus}}$ they are very unlikely to lose their entire water content by photo-evaporation \citep{kurosakietal13}. The co-existence of water-rich super-Earths and relatively less massive rocky super-Earths in the same system, in  Figure \ref{fig:panel_model_III_systems}, is at least qualitatively consistent with the findings of \cite{zengetal19}, which suggest that super-Earths larger than 2$\rearth$ are water-rich worlds. We note that \cite{zengetal19} also suggest that the innermost (hottest) super-Earths (shown as redish dots in their Figure 1) are all rocky. This is roughly consistent with the results of Figure \ref{fig:panel_model_III_systems}. Our results in  Figure \ref{fig:panel_model_III_systems} are also consistent with a few planetary systems tested against the photoevaporation model \citep[][e.g., Kepler-100, Kepler-142, and Kepler-36]{owencampos20} which suggest that adjacent planets in a system may have very distinct water contents.




We conclude this section by emphasizing that the absence of icy super-Earth systems in our companion paper \citep[Paper I;][]{lambrechtsetal18} is consequence of assuming an initial distribution of seeds only inside the snow line and without taking into account the snow line's evolution. The results of Paper I and this work agree on the fact that the formation of systems of rocky super-Earths requires special conditions. In the following section, we test if a different disk model or a putative lack of gas-driven planet migration can result in the formation of rocky super-Earths.

\subsubsection{How can we produce rocky super-Earths?}

In this section, we attempt to decipher whether or not we can produce close-in planetary systems dominated by rocky super-Earths by ignoring type-I migration, considering a nonevolving gas disk, or a combination of these two scenarios.


We performed three additional sets of simulations to answer this question.  In all of these, seeds are initially distributed from 0.5~AU  to $\sim$20~AU. The initial mutual radial separation of seeds are set by a geometric progression with common ratio equal to 1.06. Other orbital elements of our initial seed distribution were sampled as described in Section \ref{sec:setupmodels}. We name this scenario Model IV. We stop the numerical integration of all simulations of this section at the end of the gas disk phase. 
\begin{figure*}
\centering
\includegraphics[scale=.15]{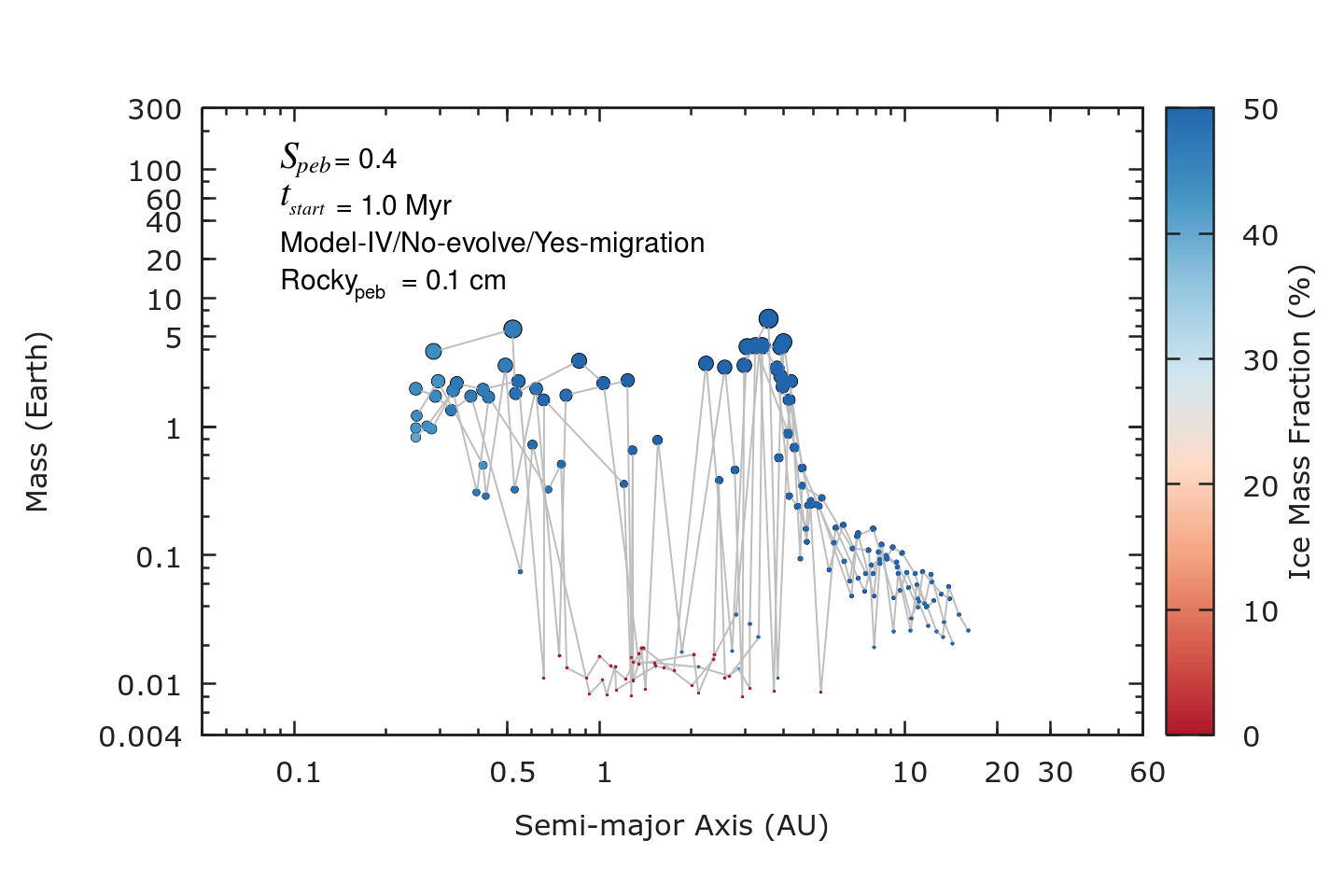}
\includegraphics[scale=.15]{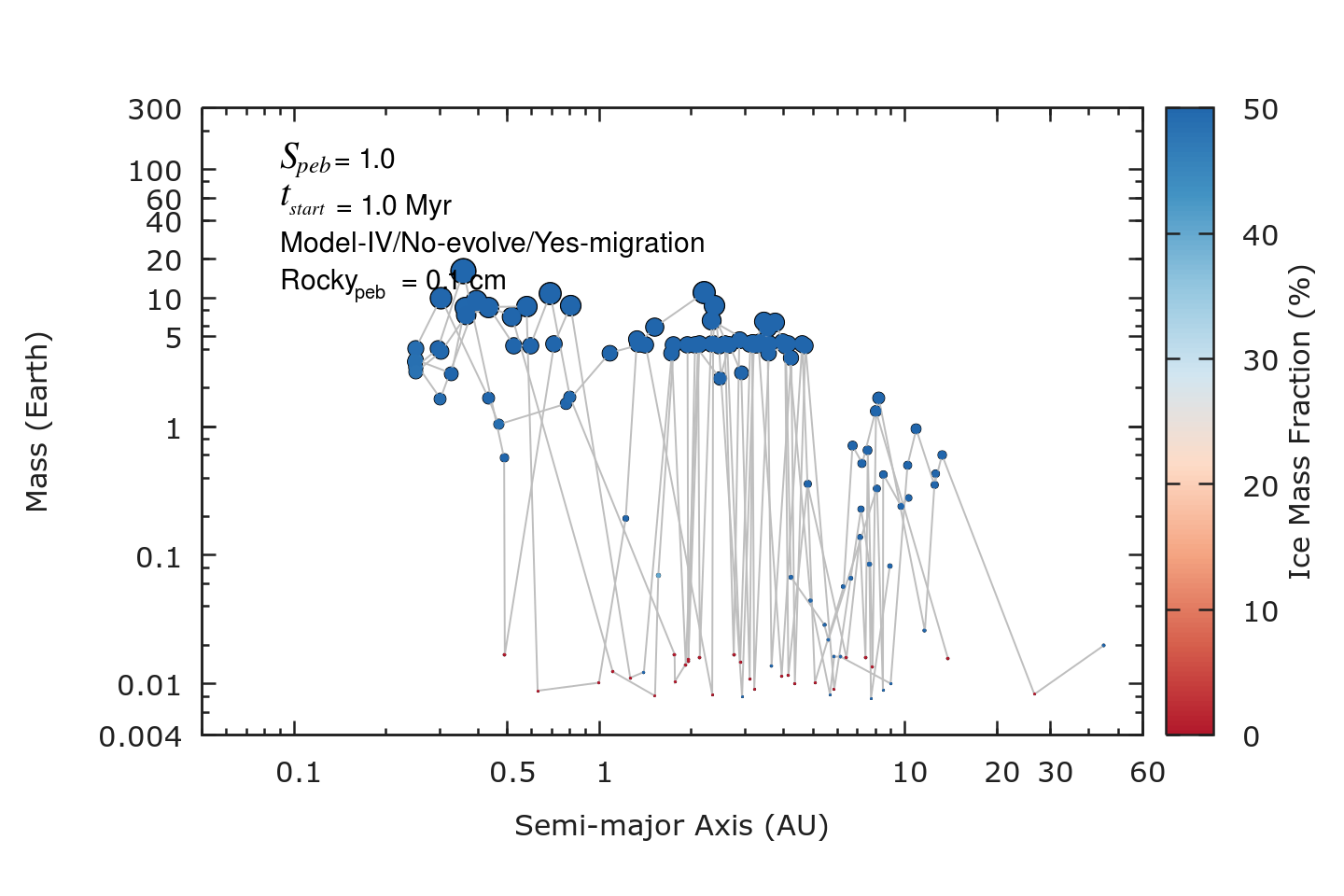}
\includegraphics[scale=.15]{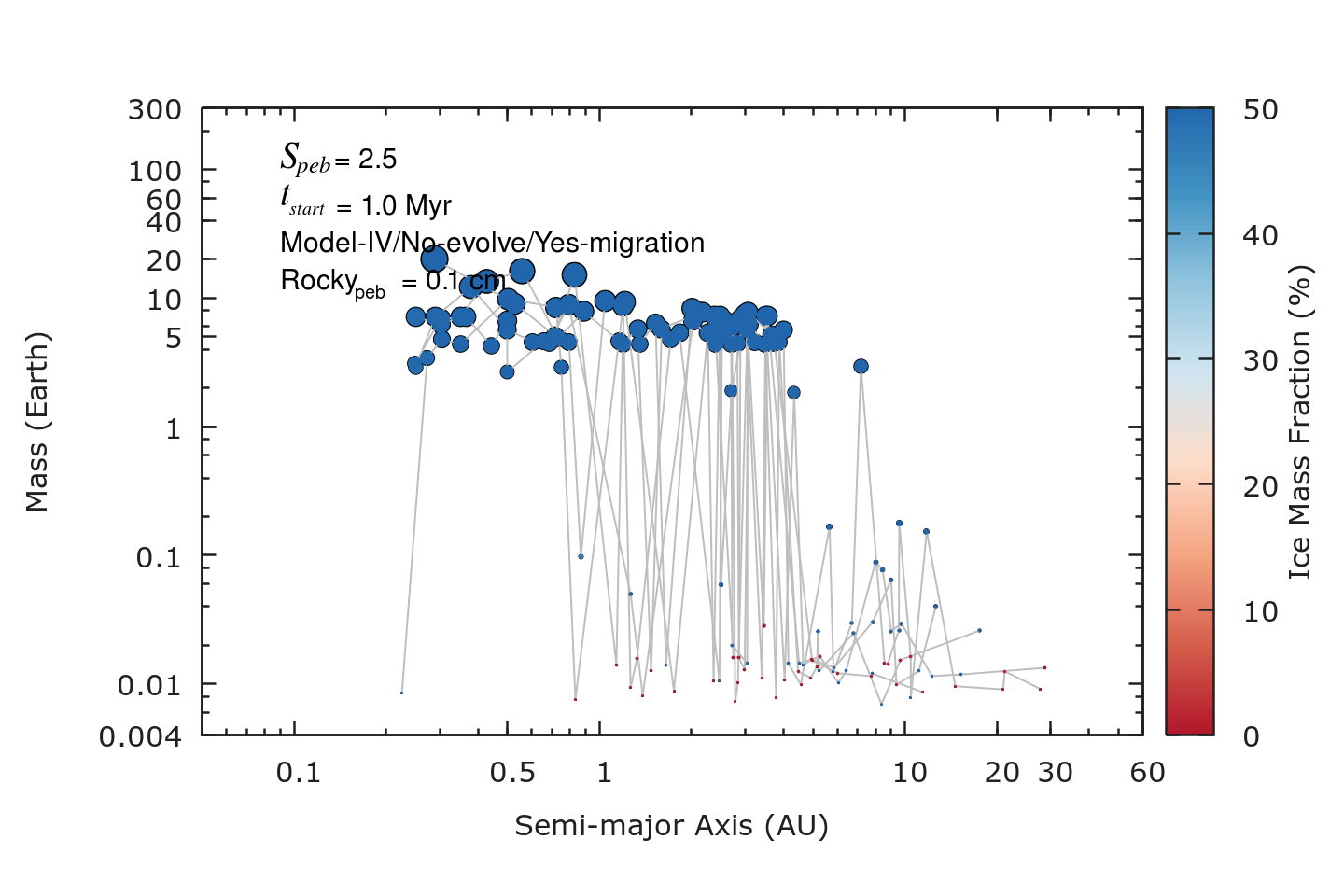}
\includegraphics[scale=.15]{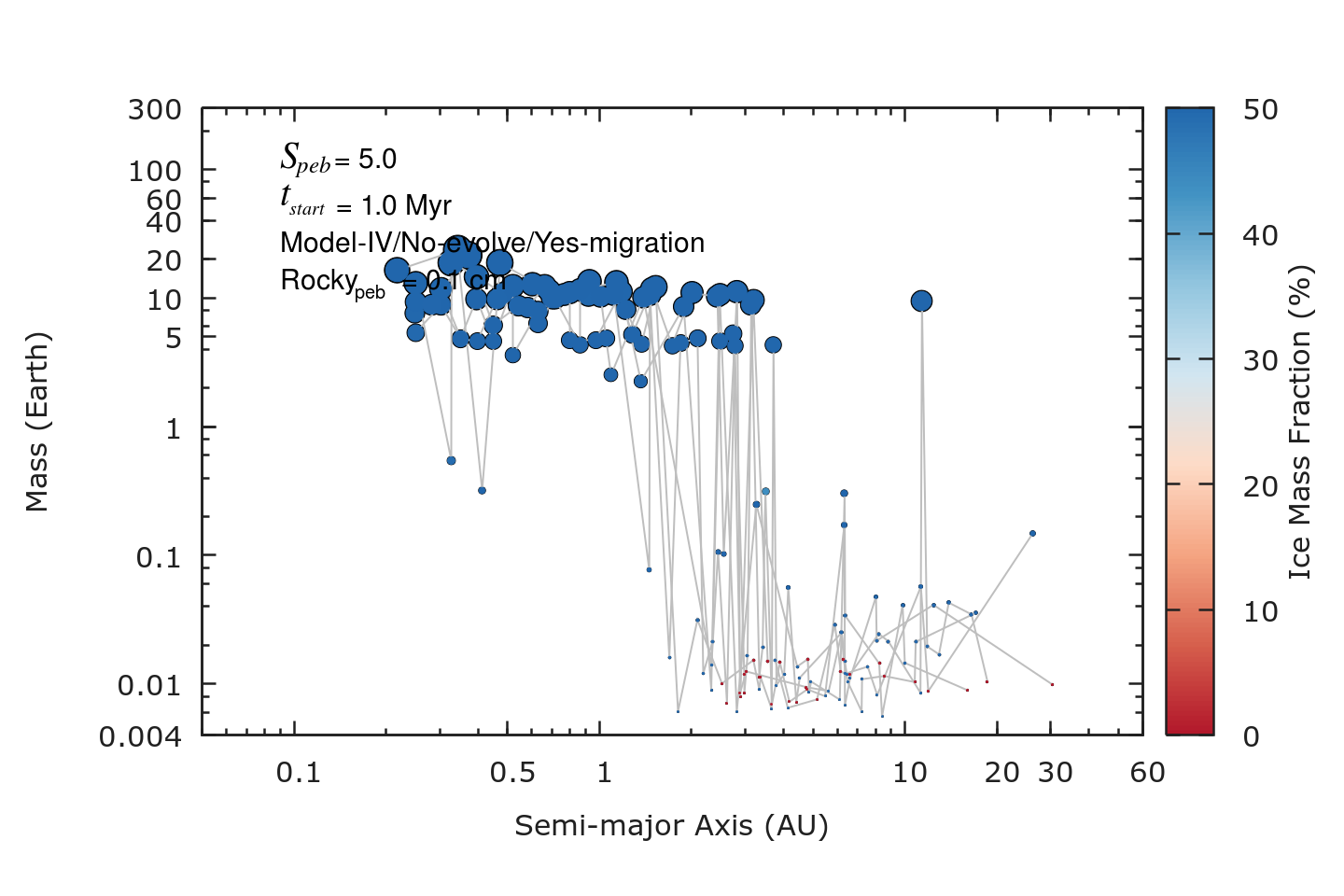}
\caption{Final masses of protoplanetary embryos in simulations where the gas disk is not evolving and the snow line is kept fixed. The size of the rocky pebbles is ${\rm Rocky_{{\rm peb}}=0.1~cm}$ for different pebble fluxes ${S_{{\rm peb}}}$. Each panel shows the outcome of five different simulations with slightly different initial conditions. Each final planetary object is represented by a colored dot where the color represents its final ice mass fraction. Planetary objects belonging to the same simulation are connected by lines.}
    \label{fig:noevolve}
\end{figure*}

Figure \ref{fig:noevolve} shows the final masses of planets growing by pebble accretion in a nonevolving disk (both gas surface density and temperature are kept constant)  with different pebble fluxes. The gas disk structure is set by the starting time of the simulation (${\rm t_{start}=0.5~Myr}$). We keep the initial disk structure throughout the entire simulation, and therefore the snow line is kept fixed at about ${\rm \sim3~AU}$. This allows us to test the effects of the movement of the snow line on our results. We recall that, in our nominal simulations,  at the end of the gas disk phase the snow line has shifted into the inner solar system down to ${\rm \sim1~AU}$. 

 Figure \ref{fig:noevolve} shows that this scenario also fails to produce rocky super-Earths for all considered pebble fluxes. Close-in Earth-mass planets are predominately icy. Seeds from the outer disk  grow faster and  regulate the pebble flux to the inner region, impairing the growth of rocky seeds. Also,  icy super-Earths eventually migrate into the inner disk, scattering or colliding with small rocky seeds.
\begin{figure*}
\centering
\includegraphics[scale=.15]{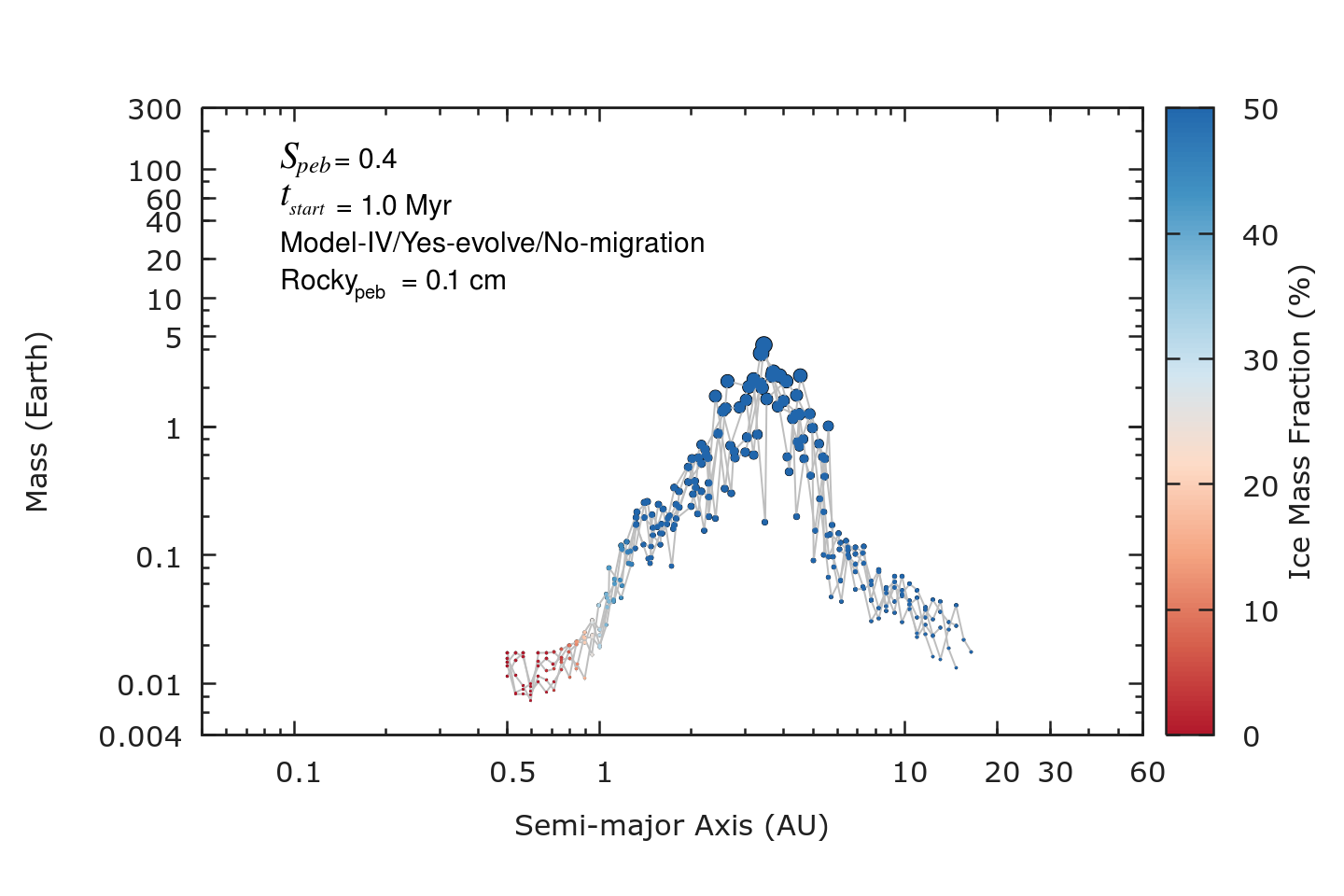}
\includegraphics[scale=.15]{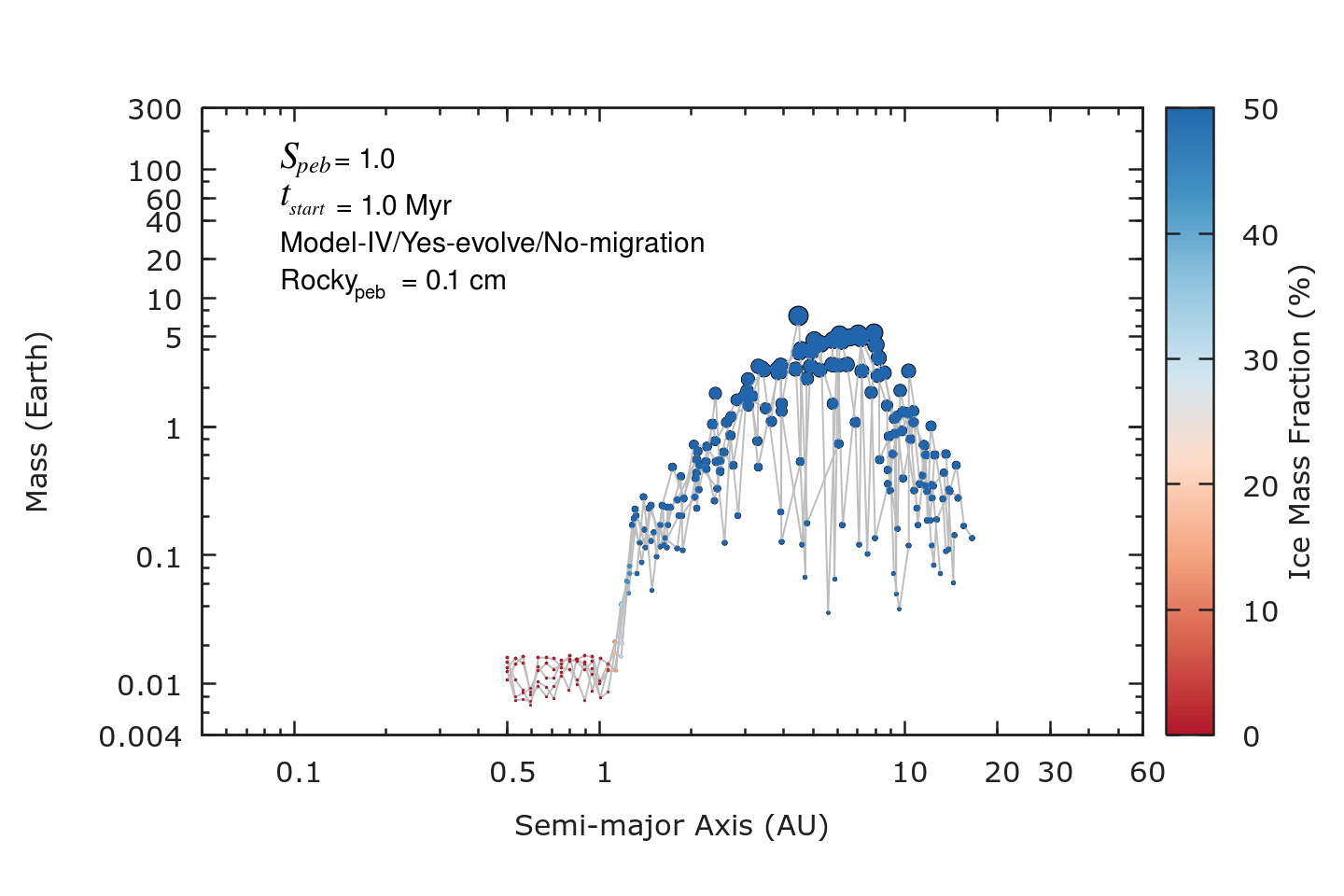}
\includegraphics[scale=.15]{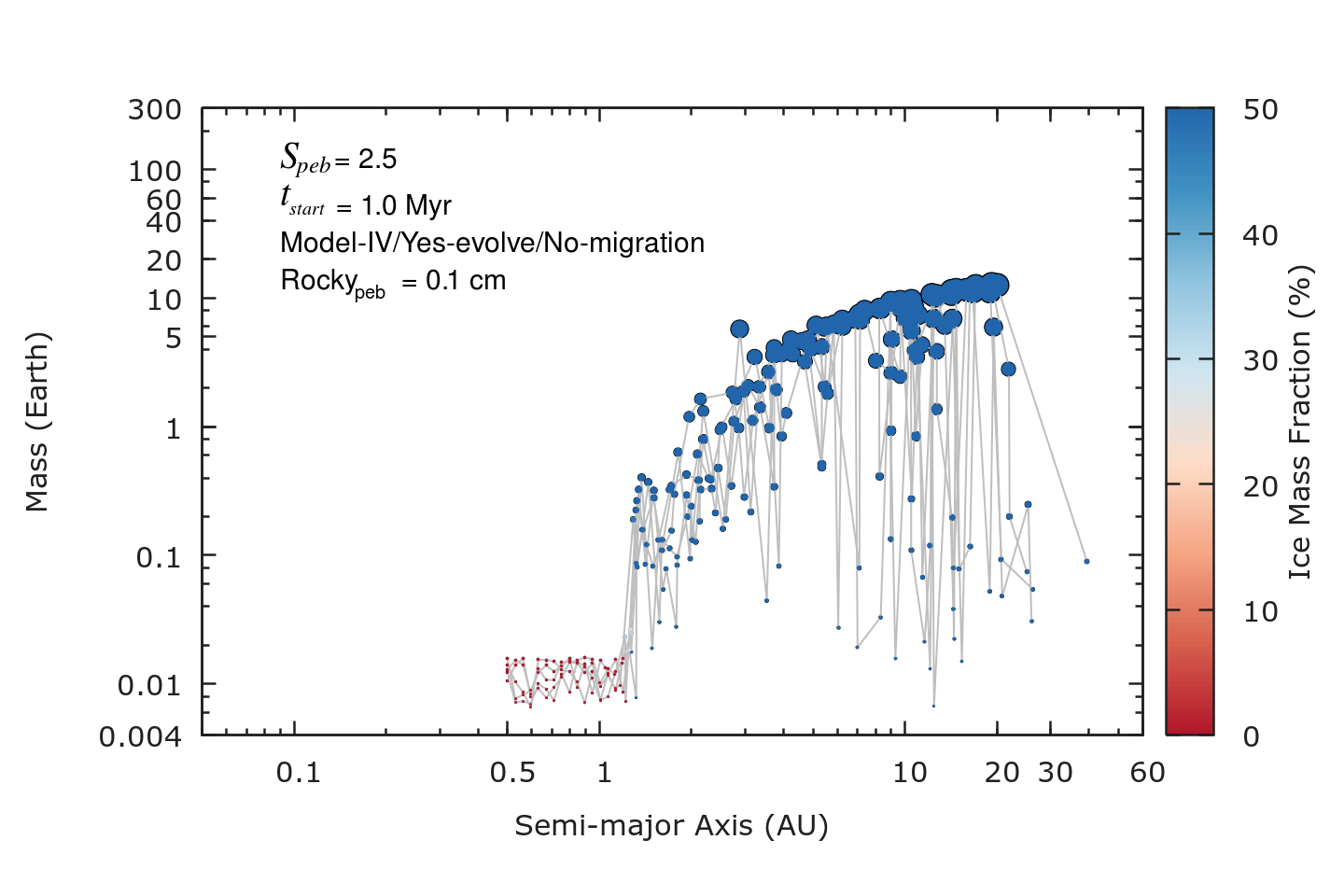}
\includegraphics[scale=.15]{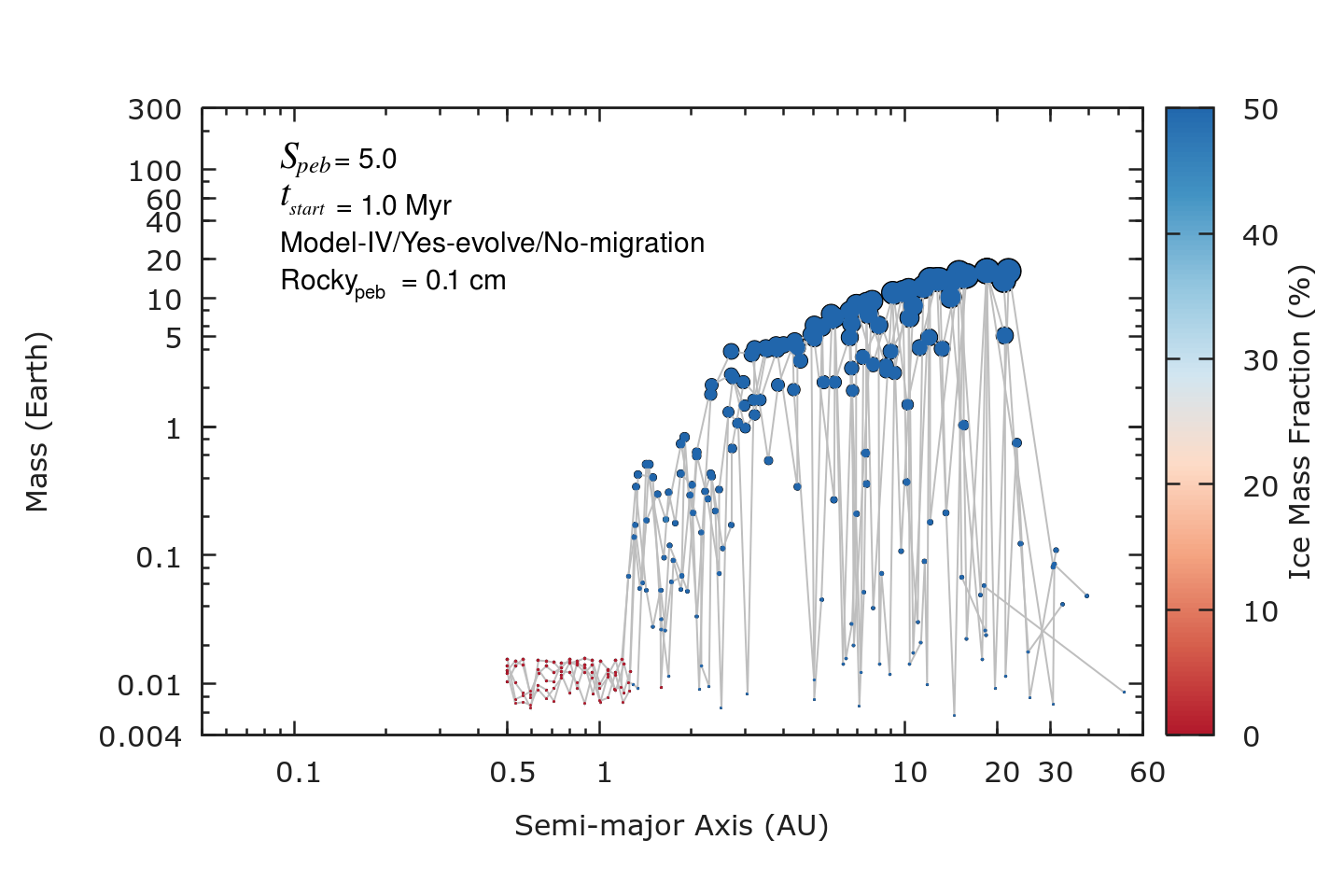}
\caption{Final masses of protoplanetary embryos in simulations where type-I migration is neglected but the disk evolves as in our nominal simulations. The size of rocky pebbles is ${\rm Rocky_{{\rm peb}}=0.1~cm}$ for different pebble fluxes ${S_{{\rm peb}}}$. Each panel shows the outcome of five different simulations with slightly different initial conditions. Each final planetary object is represented by a colored dot where the color represents its final ice mass fraction. Planetary objects belonging to the same simulation are connected by lines.}
    \label{fig:nomigrate}
\end{figure*}

Figure \ref{fig:nomigrate} shows the final masses of planets  in simulations where type-I migration is neglected but the disk evolves as in our nominal simulations.  The results of simulations with different pebble fluxes are shown. Unlike the previous scenario, seeds from the outer disk do not enter into the inner system but they  still grow quickly, starving the seeds in the inner disk.  Our results show that rocky seeds barely grow from their initial size because there is too much filtering of pebbles from outer, growing planets and pebbles are too small in the inner system, making accretion of rocky seeds very inefficient. For completeness, we also performed simulations where type-I migration is neglected and the gas disk is not evolving. The results of this scenario are qualitatively equivalent to those of Figure \ref{fig:nomigrate}.


We conclude this section by stressing that we did not succeed in producing rocky super-Earth systems in any of our simulations where the initial distribution of seeds extends up to distances reasonably  beyond the snow line (e.g., ${\rm \sim10~AU}$). If super-Earths are mostly rocky, our best match  to this constraint ---while still far from ideal--- comes from the results of the simulations of Model
III  with ${\rm Rocky_{{\rm peb}}=0.1~cm}$ and  ${S_{{\rm peb}}=10}$. In the following section we revisit  Model
III, invoking cm-sized silicate pebbles in the inner disk as a path for more efficient planetary growth in the inner disk.

\subsubsection{Making rocky super-Earths systems: A more successful case}

\begin{figure*}
\centering
\includegraphics[scale=.38]{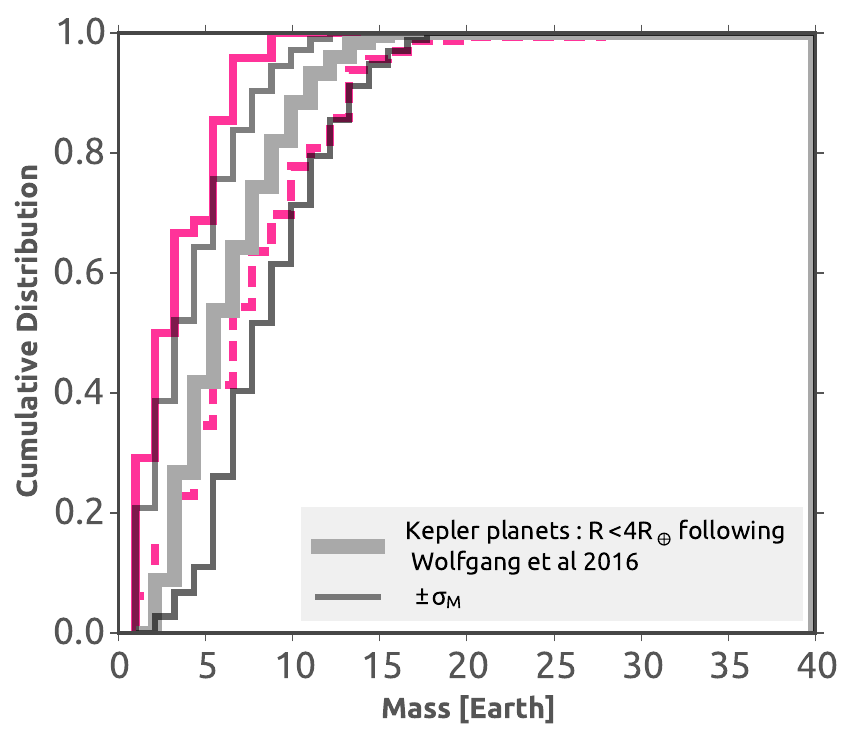}
\includegraphics[scale=.38]{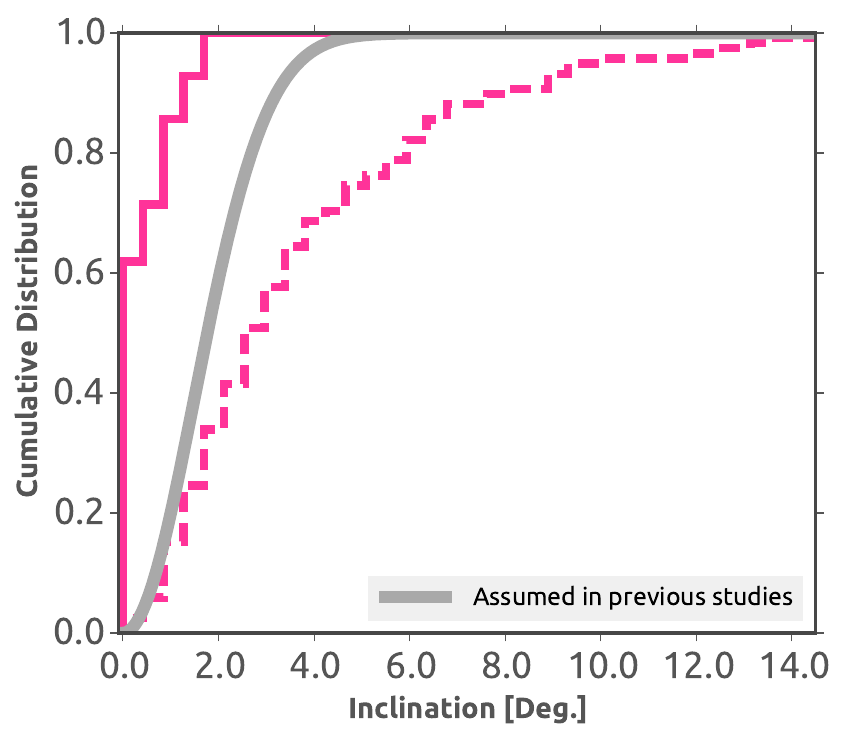}
\includegraphics[scale=.38]{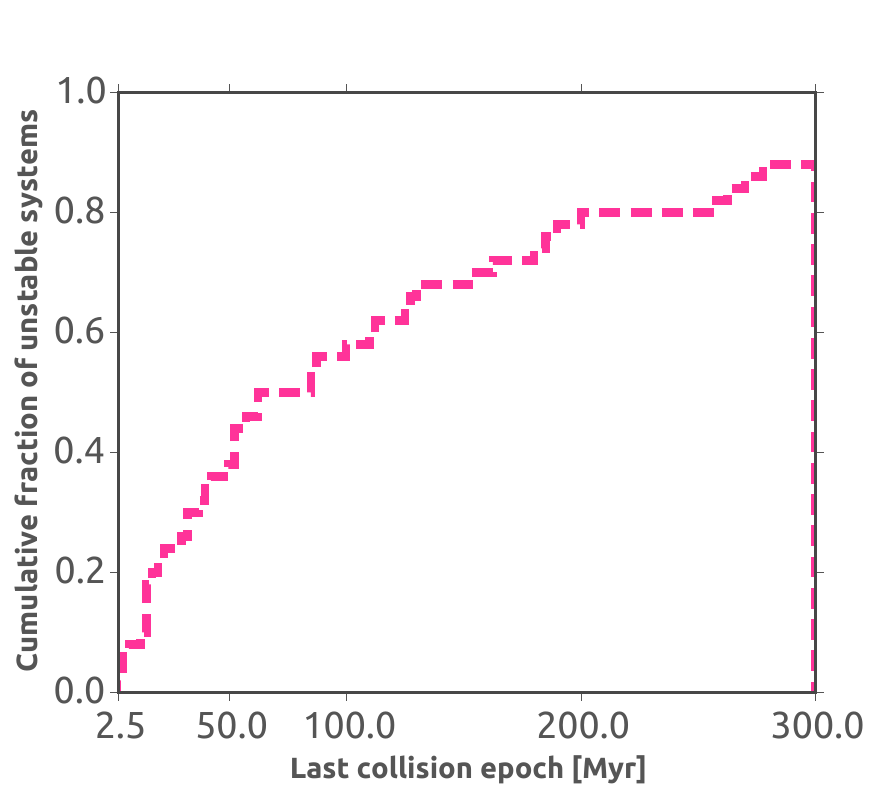}
\includegraphics[scale=.38]{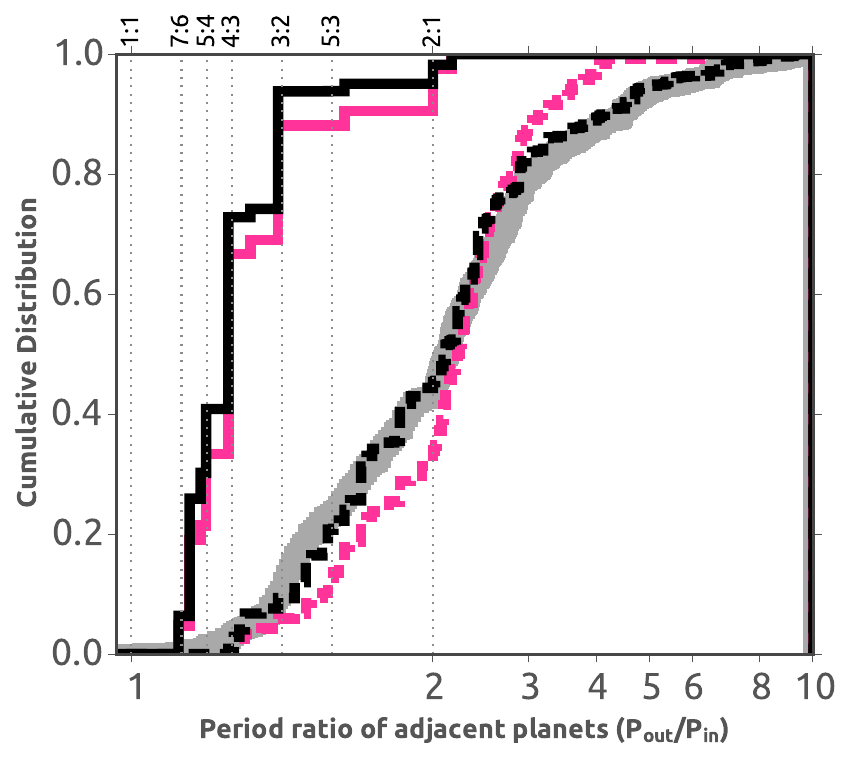} 
\includegraphics[scale=.38]{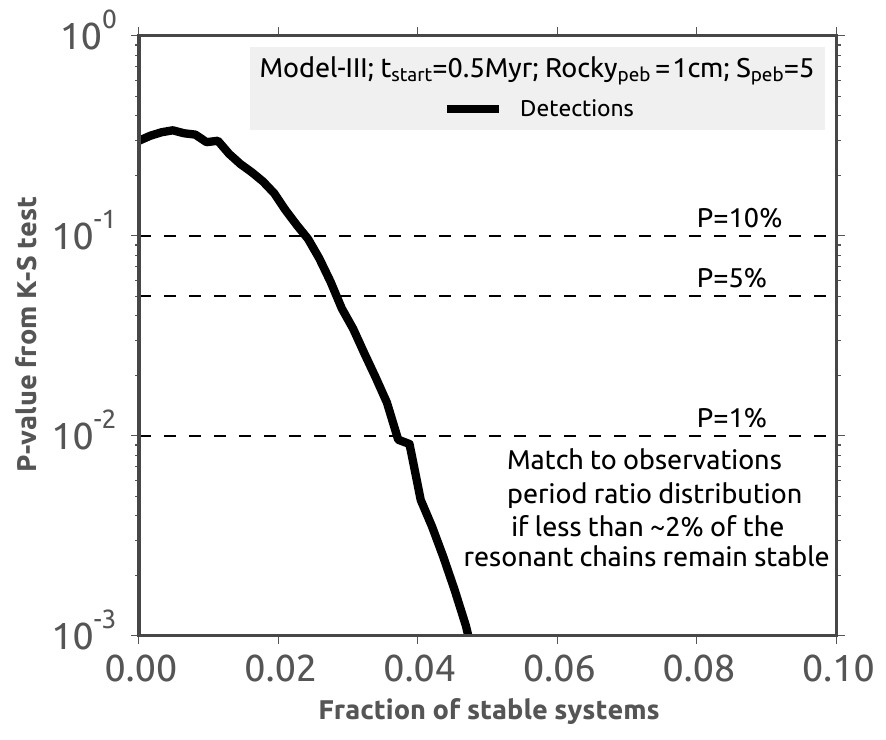}
\includegraphics[scale=.38]{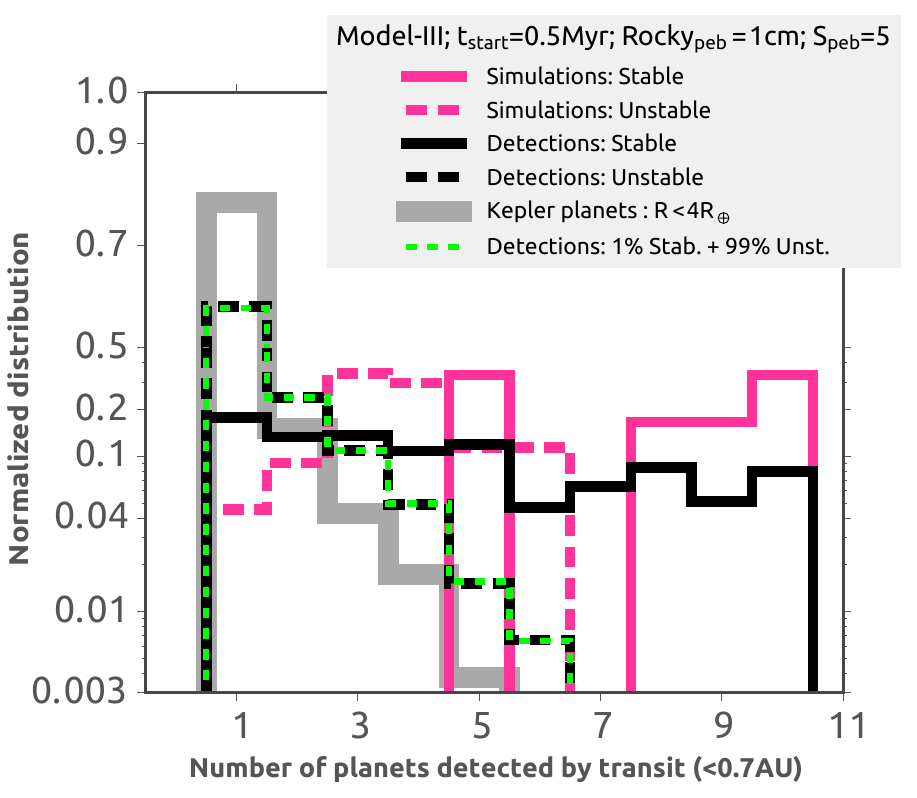}
\caption{Statistics of simulations of  Model
III with  ${\rm t_{start}=0.5~Myr}$,  ${S_{{\rm peb}}=5,}$ and ${\rm Rocky_{{\rm peb}}=1~cm}$. Solid lines show stable systems, and dashed  lines show unstable systems. The black lines show the results of our synthetic observations. The thick gray lines represent the Kepler planets. The upper-left panel shows the period ratio distribution of simulations and observations.  The
upper-right panel shows the KS-tests between the Kepler period ratio sample and our simulated detections of distributions mixing different real fractions of stable and unstable systems. The bottom-left panel shows the planet multiplicity distribution. The densely dashed green line shows the detected multiplicity distribution for a mix of 1\%  stable and  99\% unstable systems. The bottom-right panel shows the distribution of  the last collision epoch after gas dispersal. About 88\% of these systems became unstable after gas dispersal.}
    \label{fig:Model_III_1cm}
\end{figure*}

Our goal for this section is to modify the parameters of our simulations in order to produce predominantly rocky super-Earths as well as a high fraction of unstable systems. Figure \ref{fig:whyunstable} shows that dynamical instabilities after gas dispersal are more likely in systems that are very dynamically compact at the end of the gas disk phase and more importantly with a sufficiently larger number of planets  anchored at the disk inner edge (see also \citet{matsumotoetal12}). The simulations of Figure \ref{fig:panels_fluxes_model_3_cm}  were the only ones that produced predominantly rocky super-Earths inside of 0.7 AU. As a follow-up to that, we perform an additional set of simulations of Model
III considering cm-sized silicate pebbles in the inner regions (${\rm Rocky_{{\rm peb}}=1~cm}$) and  ${S_{{\rm peb}}=5}$. We do not re-use the results of Figure \ref{fig:panels_fluxes_model_3_cm} because there the disk inner edge was far from the star (at about 0.3~AU). Therefore, in order to compare our results with observations and for consistency with our previous analyses, we redo these simulations considering $r_{{\rm in}}=0.1~{\rm AU}$. We note that we stop the simulations of Figure \ref{fig:panels_fluxes_model_3_cm} at 5~Myr. We performed 50 new simulations. Because in this setup, most planets are fully formed during the first 2~Myr or less, and the disk lifetime in these simulations is set equal to 2.5~Myr, instead of the 5~Myr used for the simulations presented in Figure \ref{fig:panels_fluxes_model_3_cm} (this also allows us to save CPU time because these simulations are very computationally expensive).  We follow the long-term dynamical evolution of all these systems up to about 300~Myr after gas dispersal.
Figure \ref{fig:Model_III_1cm} shows our results.

The upper-panels of Figure \ref{fig:Model_III_1cm} show  planet mass, planet orbital inclination, and the last collision epoch distributions of stable and unstable systems. As expected, stable systems have lower mass planets and less mutually inclined planet pairs  compared to unstable systems. The mass distribution of planets in unstable systems agrees quite well with the inferred masses of Kepler planets from mass--radius relationship models \citep{wolfgangetal16}. Moreover, the left-hand panel of Figure \ref{fig:Model_III_1cm} shows that 88\% of these systems became dynamically unstable after gas dispersal. This is different from simulations of Model
III  with ${\rm Rocky_{{\rm peb}}=0.1~cm}$ where only ${\sim 50\%}$ of the systems became unstable\footnote{We also integrated the stable systems of Model
III with ${\rm Rocky_{{\rm peb}}=0.1~cm}$ up to 300~Myr but the fraction of unstable systems did not increase significantly, remaining around 50\%.}. This results in a much better though still imperfect match to observations (see bottom-right panel of \ref{fig:Model_III_1cm}).  

 The bottom-left panel of Figure  \ref{fig:Model_III_1cm} shows the period ratio distribution of planet pairs of stable and unstable systems and also their respective simulated detections. The bottom-middle panel shows statistics from a KS-test comparing Kepler observations with simulated detections of samples mixing different real fractions of stable and unstable systems. The bottom-left panel shows the planet multiplicity distributions from stable and unstable systems and also from our simulated detections. Overall, the results presented in this set of panels are very similar to those of Model
III with ${\rm t_{start}=0.5~Myr}$, ${\rm Rocky_{{\rm peb}}=0.1~cm}$, and ${S_{{\rm peb}}=10}$ (see bottom panels of Figure \ref{fig:simulated_observations} and  left-panel of   \ref{fig:dichotomy}). The period ratio distribution of our simulated planet pairs matches observations when one mixes about $<$2\% stable planet pairs with $>98$\%  unstable planet pairs. Our simulated detections do not perfectly match the Kepler dichotomy but there is nevertheless a prominent peak at ${\rm N_{det}=1}$ qualitatively similar to that seen in observations. The fraction of systems with single detected planets in our simulated detections is about 60\% compared to 80\% of observations (but see section \ref{sec:keplerdichotomy} for a discussion about the rate of false positives among single transiting planets in the Kepler data). Curiously, we also find that these simulations can provide a better match to the Kepler dichotomy if we rescale outwards the semi-major axis of planets in our simulations by a factor of 2-3. This is because close-in planets are more likely to be detected than those farther out. In our simulations, the disk inner edge is set at ${\rm \sim0.1~AU}$ and this essentially sets the typical location of the innermost planets in our simulations. If we had considered the disk inner edge to be slightly further out, our simulations would probably better match observations. This remains an interesting issue for future studies.

\begin{figure}
\centering
\includegraphics[scale=.25]{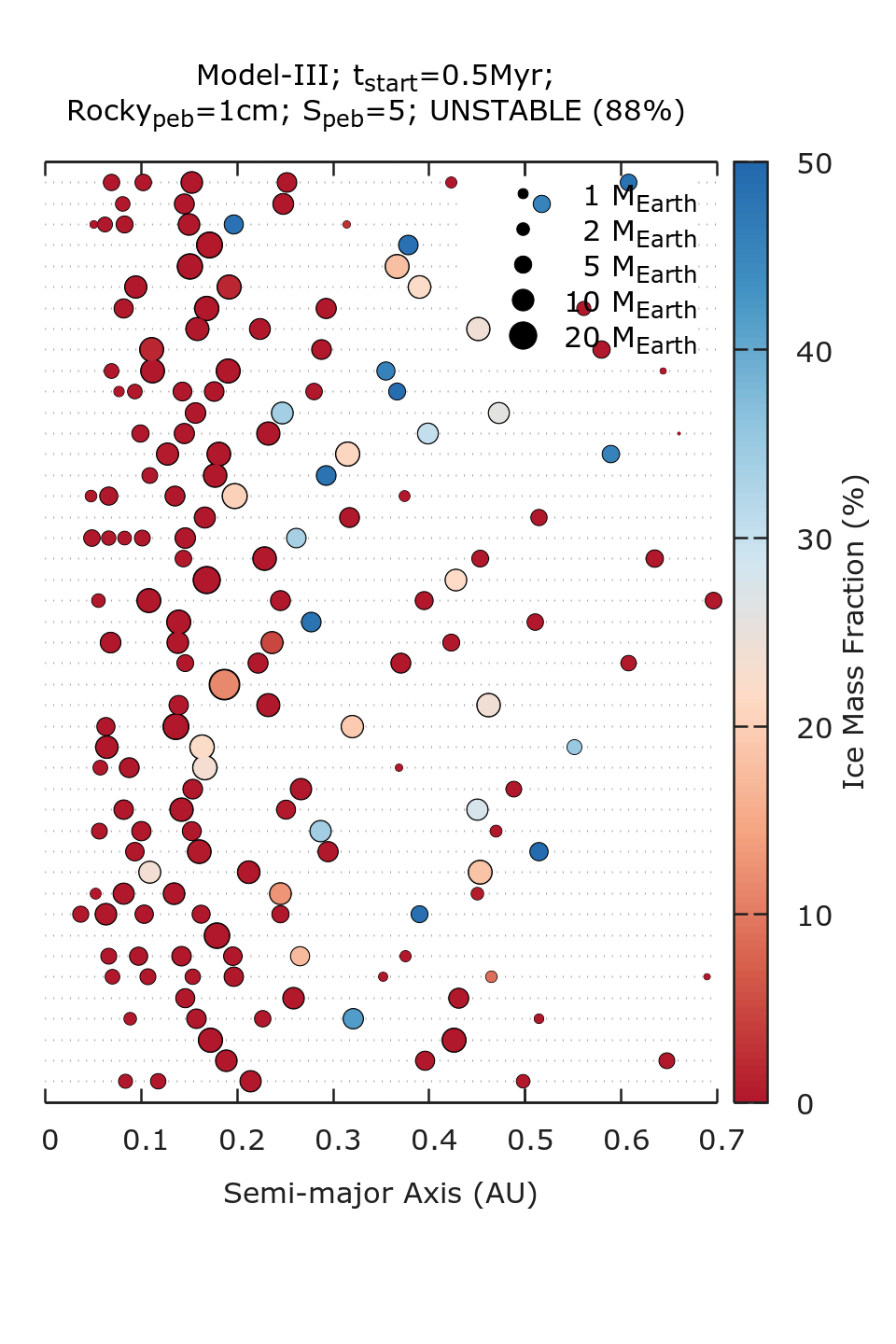}
\vspace{-1cm}
\caption{Planetary systems produced in  Model
III with  ${\rm t_{start}=0.5~Myr}$,  $S_{{\rm peb}}={\rm 5,}$ and ${\rm Rocky_{{\rm peb}}=1~cm}$ from 46 unstable systems at the end of the simulations (300~Myr). The sizes of the dots scale with mass as $m^{1/3}$. The color of each dot indicates its ice-mass fraction.}
    \label{fig:panel_model_III_1cm_systems_unstable}
\end{figure}

Finally, Figure \ref{fig:panel_model_III_1cm_systems_unstable} shows all unstable systems produced in simulations of  Model
III with ${\rm Rocky_{{\rm peb}}=1~cm}$ at the very end of our simulations (300~Myr). Most planets produced in these simulations are rocky instead of icy.

These simulations confirm the expectations from Figure \ref{fig:panels_fluxes_model_3_cm} that to form rocky super-Earths and achieve a final high instability fraction, growth inside the snow line has to be fast. Only in this case  can planets migrate faster than the snow line and avoid accreting icy pebbles at all times. To achieve such fast growth, we invoked large silicate pebbles in the inner disk. However, a similar growth mode could be equally achieved by invoking less stirring of the silicate pebble layer with mm-sized pebbles. Nevertheless,  we stress that any of these scenarios can only succeed in forming rocky super-Earths in our model if the initial distribution of seeds is restricted to regions well inside the snow line.


\section{The Solar System in context} \label{sec:solarsystem}

The Solar System is unusual among known exoplanet systems \citep[see recent review by][]{raymondetal18c}. It has been estimated that only $\sim$8\% of the planetary systems have the innermost planet (${\rm >1~R_{\oplus}}$) at an orbital period longer than that of Mercury \citep{muldersetal18}. The structure of the Solar System is also clearly segregated with low-mass rocky terrestrial planets residing in the inner regions and gaseous and/or icy giant planets in the outer region. The origin of this dichotomy in mass has been interpreted as a consequence of the process of pebble accretion and more precisely the presence of small silicate pebbles in the inner Solar System and larger icy pebbles in the outer Solar System \citep{morbidellietal15b}.


Jupiter and Saturn are the great architects of the Solar System. The cores of Jupiter and Saturn probably formed early and regulated the pebble flux to the terrestrial region. They first intercepted and consumed part of the pebbles drifting inwards but  eventually reached pebble isolation disconnecting the inner and outer Solar System \citep{kruijeretal17,lambrechtsetal18,weberetal18}. As a consequence of this process, terrestrial protoplanetary embryos were starved and only grew to about Mars-mass ---at most--- and did not migrate to the inner edge of the disk to become  close-in super-Earths (see Paper I and \citet{morbidellietal15b}). As Jupiter and Saturn grew, they probably migrated  into a resonant configuration ---also because of the interaction with sun's natal disk \citep[e.g.,][]{ward86,linpapaloizou86}. The exact gas-driven migration history of Jupiter and Saturn is not constrained, but their specific mass ratio prevented their migration into the Earth's zone \citep{massetsnellgrove01,morbidellicrida07,pierensraymond11,pierensetal14}. Jupiter and Saturn may have also blocked the inward gas-driven migration of Uranus and Neptune (and their precursors) to the inner Solar System, thus preventing the migration of icy super-Earths into the inner system \citep{izidoroetal15a,izidoroetal15c}. 
 
In light of our current view of the formation mechanism of the Solar System, none of our simulations comes close to matching our own system. Perhaps our closest approximation is produced in Model
III with ${\rm Rocky_{{\rm peb}}=0.1~cm}$ and ${S_{{\rm peb}}=1}$ (see top-left of Figure \ref{fig:panels_fluxes_model_3_mm}). Figure 8 suggests that if just a few seeds form near the snow line, they can grow much more than the inner planetary seeds. In some of these simulations, the more massive cores have a few Earth-masses and  sit around 2-3~AU while the innermost planetary embryos  all have masses below that of Mars. In reality, we need the two innermost   cores to grow more than in this simulation in order to become gas-giant planets before migrating too much in the type-I regime (once giant planets, their mutual interactions can prevent their type-II migration as in \citet{massetsnellgrove01}; but see also Paper III).  We can see some hope of this taking place in the narrow corner of parameter space explored in this paper but we are  far from coming to a firm conclusion.  Certainly, this issue requires further investigation.

\section{Conclusions}\label{sec:7}

We used N-body numerical simulations to model the growth and migration of planetary embryos in gaseous protoplanetary disks simultaneously. Our simulations start with a distribution of roughly subMoon-mass planetary seeds that grow predominantly by pebble accretion. Our results show that the  integrated pebble flux primarily sets the final planet masses. Planetary embryos growing in low-pebble-flux environments ---as in simulations where the integrated pebble flux is $\sim39~M_{\oplus}$--- have final typical masses of $\sim 2~M_{\oplus}$ or lower. Simulations where the total pebble reservoir is of $\sim194~M_{\oplus}$ produce multiple super-Earth-mass planets. Pebble accretion stops when planetary embryos reach pebble isolations mass. As the disk evolves, radial migration promotes mutual collisions and the delivery of multiple (super) Earth-mass planets to the disk inner edge. Our simulations also show that a simple increase by a factor of two in the total pebble flux from $\sim194~M_{\oplus}$ to about $\sim485~M_{\oplus}$ is enough to bifurcate the evolution of our planetary systems from one leading to  typical super-Earth-mass planets ($\lesssim 15~M_{\oplus}$) to another achieving systems of super-massive planetary embryos which are very likely to become gas giants. We dedicated two companion papers to modeling the formation of truly terrestrial planets  as opposed to rocky super-Earths \citep{lambrechtsetal18} and gas giant planets \citep{bitschetal18c}. The focus of this paper is to model the formation and long-term dynamical evolution of close-in super-Earths systems. Although some observed super-Earth systems host external gas giant planets, it is not clear whether this is predominant among such systems \citep{barbatoetal18,zhuwu18,bryanetal19}. In this work, we do not model the effects of very massive external perturbers on our close-in systems. This issue was recently addressed by \cite{bitschetal20}.


In our simulations we tested the effects of considering different silicate pebble sizes and also different initial radial distributions of seeds.
Some of our models are more successful than others in matching observations. In some models, up to $\sim95\%$ of the resonant chains become dynamically unstable after gas dispersal. This fraction of naturally unstable systems after gas dispersal is significantly higher than that found by \citet{izidoroetal17}, providing a better match to observations. Overall, our simulations match  the period ratio distribution of the Kepler sample if one combines a fraction of stable and unstable planetary systems, typically $\lesssim$2\%  stable with $\gtrsim$98\%  unstable. Supporting the results of \citet{izidoroetal17}, the results of our simulations also suggest that the excess of detected single-planet transiting systems compared to multiplanet  transiting  systems  arises from a dichotomy in orbital inclination rather than in planet multiplicity.

The Kepler period distribution  has been also used to constrain formation models. \cite{leechiang17} and \cite{choksichiang20} argue that super-Earths do not undergo significant orbital migration during their last doubling in mass which is consistent with our results. In the context of the migration model, the period distribution of planets with orbital periods shorter than 100 days has also been found to be relatively consistent with gas-driven migration and dynamical instabilities \citep{carreraetal19}. The migration model and dynamical instabilities can also account for the origin of nearly coplanar, high-planet-multiplicity, and nonresonant systems of close-in super-Earths, such as the Kepler-11 system~(\cite{estevesetal20}; see also \cite{mcnallyetal19}). Future studies  ---ideally including interior modeling and  effects of gas accretion on planets, stellar tides, fragmentation, and atmospheric loss generated by impacts--- should also aim to test how the migration model matches the inferred  `peas in a pod pattern'' found in Kepler's high-multiplicity systems,  which suggests that planets in a given system have very similar sizes \citep[e.g.,][]{weisspetigura20}.

Our simulations where planetary seeds are initially distributed in the inner and outer disk (inside and outside the snow line) produce systematically close-in icy super-Earths. This is in contrasts with the results of atmospheric photoevaporation models that suggest that super-Earths are mostly rocky \citep{owenetal12,jinmordasinietal18,vaneylen17}. Our finding is probably more aligned with the results of \cite{zengetal19}, which suggest that super-Earths larger than 2$\rearth$ are water-rich planets. Finally, we also show that the formation of close-in systems dominated by  rocky super-Earths requires special conditions such as the  formation of planetary seeds only inside the snow line (e.g., $<$1-2~AU) and vigorous growth by pebble accretion in the inner disk (suggesting the existence of large rocky pebbles or of a low-scale-height pebble layer in the inner disk). These same systems typically show that the innermost rocky super-Earths tend to be rocky whereas the outermost ones tend to be icy. This is consistent with the results of \cite{zengetal19} which suggest that the super-Earths exposed to the greatest stellar fluxes are mostly rocky.


Overall, the results presented here are in agreement with \citet{izidoroetal17} and suggest that pebble accretion, migration, and dynamical instabilities is a powerful combination of mechanisms to explain the bulk of the  orbital and mass properties of  super-Earth systems. However, if super-Earths are all rocky,  very special conditions would be required to promote their formation in our models. The required conditions seem to diverge from our current understanding of planetesimal formation, which indicates the snow line as the most favorable place to produce planetary seeds \citep{armitageetal16,drazkowskaalibert17}, and are also in contrast with the assumption that larger icy ---rather than rocky--- pebbles are necessary to explain the  structure of the Solar System.  Instead, our results suggest that at least some super-Earths should be water-rich planets.



\section*{Acknowledgements}
 A. I. thanks FAPESP for support via grants 16/19556-7 and 16/12686-2, and CNPq via process 313998/2018-3. Most simulations of this work were performed on the computer cluster at UNESP/FEG acquired with resources from FAPESP. S.~N.~R and A.~M.  thank  the  Agence  Nationale  pour la Recherche for support via grant ANR-13-BS05-0003- 01  (project  MOJO). B.B., thanks the European Research Council (ERC Starting Grant 757448-PAMDORA) for their financial support. A.J. was supported by the European Research Council under ERC Consolidator Grant agreement 724687-PLANETESYS, the Swedish Research Council (grant 2014-5775), and the Knut and Alice Wallenberg Foundation (grants 2012.0150, 2014.0017, and 2014.0048). Computer time for this study was also provided by the computing facilities MCIA (Mésocentre de Calcul Intensif Aquitain) of the Université de Bordeaux and of the Université de Pau et des Pays de l'Adour. This research has made use of the NASA Exoplanet Archive, which is operated by the California Institute of Technology, under contract with the National Aeronautics and Space Administration under the Exoplanet Exploration Program.

\begin{appendix}
\section{Planet migration prescription}\label{sec:AppendixA} 

Growing planets interact gravitationally with the protoplanetary gas disk, which exerts a torque on the planet, and can thus migrate through the disk (see \citealt{baruteauetal14} for a review). Low-mass planets migrate by type-I migration, whereas massive planets migrate by type-II migration. The torques responsible for type-I migration are the Lindblad torques, exerted from the spiral waves of the planet \citep{ward86,ward97a}, and the corotation torques, originating from the horseshoe region around the planet \citep{goldreichtremaine79,ward92}. The Lindblad torque results in inward migration for most disk parameters, where the migration timescale is much shorter than the disk lifetime for bodies more massive than the Earth \citep{tanakaetal02}.

However, the corotation torque can, depending on the local disk properties, overcompensate the negative Lindblad torque, resulting in outward migration 
\citep{paardekoopermellema06,paardekoopermellema08,baruteaumasset08,paardekooperpapaloizou08,kleyetal09,paardekooperetal11}. This outward migration depends strongly on the radial profile of the gas surface density and temperature and thus entropy in the disk \citep{baruteaumasset08,bitschkley11b}. So-called regions of outward migration, where the total torque is positive and can be sustained by the entropy-driven corotation torque in the disk, are associated with shadowed regions, where the disk's aspect ratio H/r decreases with orbital distance \citep{bitschetal13b,bitschetal14,bitschetal15,baillieetal15}. Additionally, outward migration can exist if the radial gas surface density gradient increases with orbital distance \citep{massetetal06}, as at the inner edge of the disk.

For the migration of planets, we follow the torque formula given by \citet{paardekooperetal11}, which has been extensively tested against 3D hydrodynamical simulations \citep{bitschkley11b,legaetal14,legaetal15}.

In the torque formula by \citet{paardekooperetal11}, the total torque $\Gamma_{\rm tot}$ is the sum of the Lindblad torque $\Gamma_{\rm L}$ and the corotation torque $\Gamma_{\rm C}$. The total type-I torque  also depends on the planet's orbital eccentricity and inclination. In order to account for the eccentricity and inclination effects, the total torque formula of \cite{paardekooperetal11} is rewritten as 

\begin{equation}
{\Gamma_\text{tot} = \Gamma_\text{L}\Delta_\text{L} + \Gamma_\text{C}\Delta_\text{C}},
\end{equation}
where $\Delta_\text{L}$ and $\Delta_\text{C}$ are rescaling factors accounting for torque reduction due to the orbital eccentricity and inclination of the planet. The Lindblad torque is
\begin{equation}
{ \Gamma_\text{L}= (-2.5 -1.7\beta + 0.1x)\frac{\Gamma_\text{0}}{\gamma_{\text{eff}}}},
\end{equation}
where $\beta$ and x  are the negative of the  gas surface density and temperature gradients at the location of the planet \citep[e.g.,][]{lyraetal10}:
\begin{equation}
{ x = - \frac{\partial ln ~\Sigma_{gas}}{\partial ln~r},~~~\beta = - \frac{\partial ln ~T}{\partial ln~r} }.
\end{equation}
The Lindblad torque reduction $\Delta_\text{L}$ \citep{cresswellnelson08} is
\begin{multline}
 \Delta_\text{L}  = \left[   P_\text{e} + \frac{P_\text{e}}{|P_\text{e}|} \times \left\lbrace 0.07 \left( \frac{i}{h}\right)  + 0.085\left( \frac{i}{h}\right)^4 \right. \right. \\  \left. \left. - 0.08\left(  \frac{e}{h} \right) \left( \frac{i}{h} \right)^2 \right \rbrace \right] ^{-1}  ,
\end{multline}
where
\begin{equation}
{P_\text{e} = \frac{1+\left( \frac{e}{2.25h}\right)^{1.2} +\left( \frac{e}{2.84h}\right)^6}{1-\left( \frac{e}{2.02h}\right)^4}},
\end{equation}
where \textit{e}  and \textit{i} are the planet's orbital eccentricity and inclination, respectively, and \textit{h} is the gas disk aspect ratio.

The co-orbital torque  is written as
\begin{align}
\Gamma_{\text{C}} ~=~\Gamma_{\text{c,hs,baro}} F(p_{\rm{\nu}}) G(p_{\rm\nu}) +  (1 - K(p_{\rm\nu}))\Gamma_{\text{c,lin,baro}} ~~+  \nonumber\\   \Gamma_{\text{c,hs,ent}}F(p_{{\rm \nu}})F(p_{\rm \chi})\sqrt{G(p_{\rm \nu})G(p_{\rm \chi})} \nonumber  +  \\ \sqrt{(1 - K(p_{\rm \nu}))(1 - K(p_{\rm \chi})}\Gamma_{\text{c,lin,ent}}~,
\end{align}
where 
\begin{equation}
{\Gamma_{\text{c,hs,baro}}= 1.1\left( \frac{3}{2}-x\right) \frac{\Gamma_\text{0}}{\gamma_{\text{eff}}}},
\end{equation}
\begin{equation}
{ \Gamma_{\text{c,lin,baro}}= 0.7\left( \frac{3}{2}-x\right) \frac{\Gamma_\text{0}}{\gamma_{\text{eff}}}},
\end{equation}
\begin{equation}
{\Gamma_{\text{c,hs,ent}}= 7.9\xi\frac{\Gamma_\text{0}}{\gamma_{\text{eff}}^2}},
\end{equation}
and
\begin{equation}
 \Gamma_{\text{c,lin,ent}}= \left( 2.2 - \frac{1.4}{\gamma_{\text{eff}}}\right)\xi \frac{\Gamma_0}{\gamma_{\text{eff}}}.
\end{equation}
In the previous equations,  ${ \xi=\beta - (\gamma -1)x}$ is the negative of the entropy gradient and  $\gamma=1.4$ is the adiabatic index. The scaling torque ${ \Gamma_0=(q/h)^2\Sigma_{\text{gas}} r^4 \Omega_k^2}$ is calculated at the planet's location. Here,  q is the planet--star mass ratio, and ${\Omega_k}$ is the planet's Keplerian orbital frequency. The functions F, G, and K control the torque saturation due to viscous and thermal effects. They depend on

\begin{equation}
{p_{\rm \nu} = \frac{2}{3}\sqrt{\frac{r^2\Omega_k}{2\pi\nu}x_s^3}},
\end{equation}
where ${x_{\rm s}}$ is the nondimensional half width of the horseshoe region,
\begin{equation}
{ x_{\rm s}=\frac{1.1}{{\gamma_\text{eff}}^{1/4}}\sqrt{\frac{q}{h}}},
\end{equation}
and 
\begin{equation}
{ p_{\chi} = \frac{2}{3}\sqrt{\frac{r^2\Omega_k}{2\pi\chi}x_s^3}},
\end{equation}
where ${\rm \chi}$ is the thermal diffusion coefficient which reads as
\begin{equation}
{ \chi = \frac{16\gamma(\gamma -1)\sigma T^4}{3 \kappa \rho^2(hr)^2 \Omega_k^2}}.
\label{eq:chi}
\end{equation}
In Eq. \ref{eq:chi}, $\rho$ is the gas volume density, $\kappa$ is the gas disk opacity, and $\sigma$ is the Stefan-Boltzmann constant. To set the protoplanetary disk opacity we follow \cite{belllin94}.

Different diffusion timescales (e.g., viscosities) influence the corotation torque as well, where lower levels of viscosity will not allow outward migration even when the radial gradients in entropy are steep enough to promote outward migration \citep[e.g.,][]{baruteaumasset08,bitschetal13c}. Different viscosities can therefore result in different migration speeds, which might influence the structure of planetary systems in the inner disk \citep[e.g.,][]{bitschetal13c}.

$\gamma_\text{eff}$ depends on
\begin{equation}
Q  = \frac{2 \chi}{3 h^3 r^2 \Omega_k},
\end{equation}
as
\begin{equation}
\gamma_\text{eff} =  \frac{2Q \gamma}{\gamma Q + \frac{1}{2}\sqrt{2 \sqrt{( \gamma^2Q^2+1)^2-16 Q^2(\gamma-1)}+2 \gamma^2 Q^2 - 2}}.
\end{equation}

F, G, and K take the form 
\begin{equation}
F(p) = \frac{1}{1+ \left( \frac{p}{1.3}\right)^2, }
\end{equation}
\begin{equation}
  G(p)=\begin{cases}
    \frac{16}{25}\left( \frac{45\pi}{8}\right)^{\frac{3}{4}}p^{\frac{3}{2}}, & \text{if $p< \sqrt{\frac{8}{45\pi}}$}\\
    1 - \frac{9}{25}\left( \frac{8}{45\pi}\right)^{\frac{4}{3}}p^{-\frac{8}{3}}, & \text{otherwise}.
  \end{cases},
\end{equation}
and
\begin{equation}
  K(p)=\begin{cases}
    \frac{16}{25}\left( \frac{45\pi}{8}\right)^{\frac{3}{4}}p^{\frac{3}{2}}, & \text{if $p< \sqrt{\frac{28}{45\pi}}$}\\
    1 - \frac{9}{25}\left( \frac{28}{45\pi}\right)^{\frac{4}{3}}p^{-\frac{8}{3}}, & \text{otherwise}.
  \end{cases}.
\end{equation}
Here, p is either $p_{\nu}$ or $p_{\chi}$ as defined above.

The co-orbital torque reduction is
\begin{equation}
{ \Delta_\text{C}=\exp\left(\frac{e}{e_\text{f}} \right)\left\lbrace 1-\tanh\left(\frac{i}{h} \right)\right\rbrace   },
\end{equation}
where ${e_\text{f}}$ is  given by \cite{fendykenelson14} as
\begin{equation}
{e_\text{f} = 0.5h + 0.01}.
\end{equation}
We nte that for sufficiently large eccentricities, $\Delta_\text{L}$ may assume negative values favoring unrealistic reversal of migration. In extreme situations, where $\Delta_\text{L} < 0,$ we impose  $\Delta_\text{L} $ = $\Delta_\text{C} $.  This approach is slightly different from that in \citet{izidoroetal17}, but both approaches produce qualitatively equivalent  results. 

We follow \cite{papaloizoularwood00} and \cite{cresswellnelson08}  to write the migration timescale  used in our N-body code:

\begin{equation}
{ t_m =- \frac{L}{\Gamma_{tot}}},
\end{equation}
where L and ${\rm \Gamma_{tot}}$  are the planet's orbital angular momentum and  the total type-I torque, respectively.

The damping of orbital eccentricity and inclination due to gas tidal interaction follow the prescriptions of  \cite{papaloizoularwood00} and \cite{tanakaward04} reformulated by \citet{cresswellnelson06,cresswellnelson08}. ${\rm t_e}$ and ${\rm t_i}$ are the orbital eccentricity and inclination damping timescales, respectively:

\begin{multline}
t_e = \frac{t_{wave}}{0.780} \left(1-0.14\left(\frac{e}{h}\right)^2 + 0.06\left(\frac{e}{h}\right)^3  \right. \\ \left. + 0.18\left(\frac{e}{h}\right)\left(\frac{i}{h}\right)^2\right),
\label{eq:te}
\end{multline}

\begin{multline}
 t_i = \frac{t_{wave}}{0.544} \left(1-0.3\left(\frac{i}{h}\right)^2 + 0.24\left(\frac{i}{h}\right)^3 \right.  \\  \left.  + 0.14\left(\frac{e}{h}\right)^2\left(\frac{i}{h}\right)\right).
 \label{eq:ti}
\end{multline}

In Eq. \ref{eq:te} and \ref{eq:ti}, 
\begin{equation}
 t_{wave} = \left(\frac{M_{\odot}}{m_p}\right)  \left(\frac{M_{\odot}}{\Sigma_{gas} a^2}\right)h^4 \Omega_k^{-1}
,\end{equation}
with ${\rm M_{\odot}}$, ${\rm a_p}$, ${\rm m_p}$, ${\rm i}$, and ${\rm  ~e}$ being the solar mass, the planet's semimajor axis, its mass, its orbital inclination, and its eccentricity, respectively. 

Finally, in order to incorporate all these effects into our simulations, we follow  \cite{papaloizoularwood00} and \cite{cresswellnelson06,cresswellnelson08} and implement the following artificial accelerations  into FLINTSTONE accounting for migration ($\mathbf{a}_m$),  orbital eccentricity ($\mathbf{a}_e$),  and inclination damping ($\mathbf{a}_i$):
\begin{equation}
{\mathbf{a}_m = -\frac{\mathbf{v}}{t_m}},
\end{equation}

\begin{equation}
{\mathbf{a}_e = -2\frac{(\mathbf{v.r})\mathbf{r}}{r^2 t_e}},
\end{equation}
and
\begin{equation}
{\mathbf{a}_i = -2\frac{v_z}{t_i}\mathbf{k},}
\end{equation}
where  ${ \mathbf{r}}$, ${ \mathbf{v}}$, and ${ \mathbf{k}}$ are the planet's heliocentric distance, velocity, and the unit vector in the z-direction. The formula of $\mathbf{a}_i$ includes a factor of two correcting for a typo in \citep{cresswellnelson06,cresswellnelson08}.

\end{appendix}

\bibliographystyle{aa}
\bibliography{library} 


\end{document}